\documentclass[apj,revtex4,numberedappendix]{emulateapj}

\usepackage{graphics}
\usepackage{subfigure}
\usepackage{hyperref}
\usepackage{graphicx}
\usepackage{color}

\usepackage{amsmath}
\usepackage{natbib}
\usepackage{float} 
\newcommand{\CH}[1]{\colhead{#1}}

\newcommand\ii{{\sc ii}}
\newcommand\iii{{\sc iii}}
\newcommand\iv{{\sc iv}}
\newcommand\W{{$\lambda$}}

\begin{document}

\shortauthors{Berg et al.}
\title{A Window On The Earliest Star Formation: Extreme Photoionization Conditions of a High-Ionization, Low-Metallicity Lensed Galaxy at z$\sim$2
\footnote{This work was supported by a NASA Keck PI Data Award, administered by the NASA Exoplanet Science Institute. Data presented 
herein were obtained at the W. M. Keck Observatory from telescope time allocated to the National Aeronautics and Space Administration 
through the agency's scientific partnership with the California Institute of Technology and the University of California. The Observatory was 
made possible by the generous financial support of the W. M. Keck Foundation.}}

\author{Danielle A. Berg\altaffilmark{1,2}, 
	    Dawn K. Erb\altaffilmark{1}, 
	    Matthew W. Auger\altaffilmark{3},
	    Max Pettini\altaffilmark{3}, 
	    Gabriel B. Brammer\altaffilmark{4}}

\altaffiltext{1}{Center for Gravitation, Cosmology and Astrophysics, Department of Physics, University of Wisconsin Milwaukee,
3135 N. Maryland Avenue, Milwaukee, WI 53211; bergda@uwm.edu; erbd@uwm.edu}
\altaffiltext{2}{Department of Astronomy, The Ohio State University, 140 W. 18th Avenue, Columbus, OH 43202; berg.249@osu.edu}
\altaffiltext{3}{Institute of Astronomy, Madingley Road, Cambridge, CB3 0HA, United Kingdom; mauger@ast.cam.ac.uk; pettini@ast.cam.ac.uk}
\altaffiltext{4}{Space Telescope Science Institute, 3700 San Martin Dr, Baltimore, MD 21211; brammer@stsci.edu}

%----------------------------------------------------------------------------------------------------------------------
%----------------------------------------------------------------------------------------------------------------------

\begin{abstract}

We report new observations of SL2SJ021737-051329, a lens system consisting of a bright arc at $z=1.84435$, 
magnified $\sim17\times$ by a massive galaxy at $z=0.65$. 
SL2SJ0217 is a low-mass (M $<10^9$ M$_{\odot}$), 
low-metallicity (Z$\sim1/20$ Z$_{\odot}$) galaxy, with extreme star-forming conditions
that produce strong nebular UV emission lines in the absence of any apparent outflows. 
Here we present several notable features from rest-frame UV Keck/LRIS spectroscopy:
(1) Very strong narrow emission lines are measured for C~\iv~\W\W1548,1550, He~\ii~\W 1640, 
O~\iii]~\W\W1661,1666, Si~\iii]~\W\W1883,1892, and C~\iii]~\W\W1907,1909.
(2) Double-peaked Ly$\alpha$ emission is observed with a dominant blue peak and centered near the systemic velocity.
(3) The low- and high-ionization absorption features indicate very little or no outflowing gas along
the sightline to the lensed galaxy.
The relative emission line strengths can be reproduced with a very high-ionization, 
low-metallicity starburst with binaries, with the exception of He~\ii, which indicates an additional ionization source is needed.
We rule out large contributions from AGN and shocks to the photoionization budget, 
suggesting that the emission features requiring the hardest radiation field likely result from 
extreme stellar populations that are beyond the capabilities of current models. 
Therefore, SL2S0217 serves as a template for the extreme conditions that are important for reionization 
and thought to be more common in the early Universe.

\end{abstract}


%----------------------------------------------------------------------------------------------------------------------
%----------------------------------------------------------------------------------------------------------------------
%%%% 1 %%%%
\section{INTRODUCTION}\label{sec1}
%----------------------------------------------------------------------------------------------------------------------

Determining the precise epoch and source of reionization is currently 
a major focus of observational cosmology. 
Yet, the relative contributions to ionizing radiation from stellar and nuclear 
activity are still uncertain \citep[e.g.,][]{fontanot14}.
However, there is a general consensus that metal-poor, low-mass galaxies 
host a substantial fraction of the star formation in the high-redshift Universe 
and are likely the key contributors to reionization \citep[e.g.,][]{wise14, madau15}. 
Depending on the redshift, luminosity function, intergalactic medium (IGM) clumping factor, and ionizing photon production efficiency assumed, the estimated escape fraction 
of ionizing radiation needed to sustain cosmic reionization ranges from $<13$\% up to 50\% \citep{finkelstein12}.
Therefore, detailed studies of these chemically unevolved galaxies at early epochs are 
necessary to probe the initial stages of galaxy evolution and assess whether the physical conditions
allow sufficient Ly continuum radiation to escape.

Surveys of galaxies at $2 \lesssim z \lesssim 4$ reveal a rich diversity of galaxy 
properties and mark a critical stage of galaxy evolution at which the peak of both 
star formation and black hole accretion activity occur (see \citet{shapley11} for a review). 
These surveys have provided us with a broad understanding of how galaxies 
assemble and evolve, but the spatial and spectral limitations inherent in 
observing faint, distant objects mean that many of the physical processes 
regulating this dynamic evolution are poorly constrained. 
Clearly, a more detailed understanding of this key epoch is needed.

Until recently, very few high-redshift galaxies had been observed spectroscopically in detail,
as they are generally too faint to obtain high signal-to-noise observations with even 8$-$10m class telescopes.
Previously, this challenge has been overcome by building composite spectra of high-redshift 
galaxies \citep[e.g.,][]{steidel01, shapley03, steidel16, rigby17}, 
the properties of which generally indicate the sources are young ($\sim10^8$ yrs), 
metal-poor (Z$_{neb} < 0.5$Z$_{\odot}$) galaxies 
with high specific star formation rates and strong outflows of ionized gas. 
Currently, our most detailed information on the physical conditions in galaxies at moderate 
redshift comes from careful, high S/N spectroscopic studies of gravitationally lensed objects
\citep[e.g.,][]{pettini00, pettini02b, fosbury03, hainline09, quider09, yuan09, 
quider10, dessauges-zavadsky10, rigby11, jones13, amorin14, vanzella16}.
However, samples of such galaxies are necessarily obtained one at a time, and so remain 
small and do not yet adequately sample the diversity among high redshift galaxy populations.
In particular, there are few detailed studies of low mass, low metallicity (Z$_{neb} < 0.1$Z$_{\odot}$) 
sources, and only a handful of targets with direct determinations of their physical conditions
such as electron temperature, density, and gas phase oxygen abundance 
\citep[e.g.,][]{hainline09, bayliss14, christensen14}.
 
In order to understand how the physical conditions in galaxies have evolved over cosmic time,
we need detailed studies of galaxies at early epochs that have undergone little chemical evolution
in comparison with nearby counterparts. 
The Strong Lensing Legacy Survey \citep{tu09} has discovered several such targets,
including the $z \sim 1.85$ gravitationally lensed arc SL2SJ0217-0513, hereafter SL2S0217.
Low-resolution rest-frame UV and optical spectra of this arc
display very strong, high-ionization nebular emission lines, suggesting SL2S0217 
is both very metal poor and highly ionized \citep{tu09, brammer12}.

In preparation for the next frontier of high-redshift studies made possible 
by the forthcoming generation of space-based (i.e., JWST) and 30m ground-based 
observatories (i.e., GMT, ELT, and TMT), we seek to understand the conditions 
producing the extreme UV emission lines seen in SL2S0217 and even higher-redshift sources. 
Because the intergalactic medium becomes increasingly neutral at $z>6$, 
Ly$\alpha$ emission, which is typically the strongest emission feature in 
UV spectra of high redshift Ly$\alpha$-emitting galaxies\footnotemark[2], 
is increasingly scattered and therefore suppressed \citep{stark16}.
Instead, the next strongest UV emission lines, such as C~\iii] must be used to spectroscopically confirm redshifts.
Additionally, these lines will provide useful constraints on the physical properties of the emitting 
source and may be used as important diagnostics for characterizing high-redshift galaxies.
Recently, these strong UV nebular emission lines have been observed in high redshift ($z > 6$) galaxies
\citep[e.g.,][]{stark15,stark16}, indicative of low-metallicity and high-ionization \citep[e.g.,][]{erb10, stark14}.
Because high-ionization targets may also have large escape fractions of H-ionizing photons \citep[e.g.,][]{brinchmann08,jaskot13,stark14},
SL2S0217 is a good candidate for studying the chemically unevolved, extreme physical 
conditions that are expected in the early galaxies responsible for reionizing the Universe. 

Here we present the analysis of new Keck LRIS spectroscopy of SL2S0217. 
Section~\ref{sec2} describes the previous observations and derived properties of SL2S0217 presented in \citet{brammer12}.
In Section~\ref{sec3} we present our analysis of the existing photometry and an improved lensing model for SL2S0217.
The spectral observing plan and data reduction for the new LRIS spectrum are laid out in Section~\ref{sec4}. 
We examine the remarkable features of the resulting rest-frame UV spectrum in Section~\ref{sec5}, paying particular
attention to the nebular emission lines (\S~\ref{sec5.2}), the double-peaked Ly$\alpha$ profile (\S~\ref{sec5.3}),
and the interstellar absorption line profiles (\S~\ref{sec5.4}).
The physical conditions of the gas and chemical abundances of oxygen, carbon, nitrogen, and silicon 
are estimated from the nebular emission lines in Section~\ref{sec7}.
Finally, in Section~\ref{sec8}, we attempt to reproduce our spectra and constrain the 
ionization source by inspecting a large grid of photoionization models, considering contributions from stars, 
shocks (\S~\ref{sec8.3.1}), and active galactic nuclei (AGN; \S~\ref{sec8.3.2}).
The photoionization budget is discussed further in the context of the strong 
C~\iv\ (\S~\ref{sec8.4.1}) and He~\ii\ (\S~\ref{sec8.4.2}) emission.
Throughout this paper we adopt a flat FRW metric with $\Omega_{\mbox{m}}$ = 0.3, 
$\Omega_{\Lambda}$ = 0.7, and H$_0$ = 70 km s$^{-1}$ Mpc$^{-1}$, and the
solar metallicity scale of \citet{asplund09}, where 12 + log(O/H)$_{\odot} = 8.69$.

\footnotetext[2]{Typical, moderately massive $z\sim2-3$ galaxies actually have net Ly$\alpha$ absorption.}

%----------------------------------------------------------------------------------------------------------------------
%----------------------------------------------------------------------------------------------------------------------

% Table1: Observations
\begin{deluxetable}{rl}
\tabletypesize{\footnotesize}
\tablewidth{0pt}
\tablecaption{Observations of SL2SJ0217-0513}
\tablehead{
\CH{Observing Program} 		& \CH{Band / Wavelength}}
\startdata
{\sc{Photometry:}}			& {}	\\												
{{\it HST} CANDELS}		& {ACS F606W (V), F814W (i)}		\\
{}						& {WFC3 F125W (J), F160W (H)}	\\
{3D-HST}					& {F140W (H$_{\mbox{wide}}$)}	\\	
{$Spitzer$}				& {MIPS 24 $\mu$m (mJy)}		\\
{HST Ly$\alpha$}			& {WFC3 F343N, F390M}			\\
{}						& {}							\\
{\sc{Spectroscopy:}}			& {}	\\
{3D-{\it HST} Optical}		& {WFC3 G141; $1.10-1.65\ \mu$m} \\
{Keck/LRIS UV}			& {600/4000; $\sim3100-9000$ \AA}	
\enddata
\tablecomments{Existing imaging and spectral observations of SL2SJ0217-0513,
including the new Keck/LRIS spectra presented here. }
\label{tbl1}
\end{deluxetable}

%----------------------------------------------------------------------------------------------------------------------
%----------------------------------------------------------------------------------------------------------------------
%%%% 2 %%%%
\section{GALAXY PROPERTIES AND PAST OBSERVATIONS}\label{sec2}
%----------------------------------------------------------------------------------------------------------------------

SL2SJ0217 is a gravitationally lensed galaxy
at a redshift $z\sim1.85$ that is lensed by a massive galaxy at $z=0.6459$
as discovered by the Strong Lensing Legacy Survey \citep{tu09}.
The general properties of SL2S0217 are well characterized due to its location within both the
{\it HST} CANDELS imaging survey 
\citep[ACS F606W and F814W and WFC3 F125W and F160W][]{grogin11, koekemoer11},
and 3D-{\it HST} spectroscopic survey \citep[WFC3 G141 grism spectroscopy;][hereafter B12]{brammer12}.
Additionally, a low-resolution Keck LRIS spectrum \citep{tu09} exists for the arc.
All existing imaging and spectral observations for SL2S0217 are summarized in Table~\ref{tbl1}.

B12 report that the arc is very blue with a UV slope of $\beta = -1.7\pm0.2$ 
($f_{\lambda} \propto \lambda^{\beta}$) for $\lambda_{\mbox{\footnotesize rest}} = 2100-2800$ \AA,
but is rather red in the F160W band due to strong [O~\iii] \W\W4959,5007 emission. 
The low resolution WFC3 G141 grism spectrum of the lens system presented in B12 shows extremely
high equivalent width rest-frame optical emission lines that are atypical of $z\sim0$ star-forming 
galaxies (namely, rest-frame EW(H$\beta$) = 1470 \AA\ and EW([O~\iii] \W\W4959,5007) = 5690 \AA).
The emission lines are all spatially extended, indicating that strong nebular 
emission is coming from multiple clumps along the arc.
From their analysis of this grism spectrum and modeling of the broadband spectral energy distribution, 
B12 concluded that the arc is a young, low-mass, low-metallicity (12+log(O/H) $\sim$ 7.5) star-bursting 
(specific star formation rate, sSFR, $\sim 100$ Gyr$^{-1}$) galaxy, 
with similar characteristics as local blue compact dwarf galaxies. 
SL2S0217 is thus one of the lowest metallicity star-forming galaxies yet identified at $z > 1$.
Derived quantities from B12 are given in the top of Table~\ref{tbl2}.

%----------------------------------------------------------------------------------------------------------------------

% Table 2: Galaxy Properties
\begin{deluxetable}{lc}
\tabletypesize{\footnotesize}
\tablewidth{0pt}
\tablecaption{Properties of SL2SJ0217-0513}
\startdata
\hline\hline \\
[-1.0 ex]
{Parameter}													& {Value} 				\\
\hline \\ 
[-1.0 ex]
{R.A.}														& {02:17:37.237}		\\
{Dec.}														& {-05:13:29.78}		\\
{$z$}															& {1.84435 $\pm$ 0.00066}	\\
\hline \\
[-1.0 ex]
\multicolumn{2}{c}{From \cite{brammer12}:}  \\
[1.0 ex]
\hline \\
[-1.0 ex]
{log (age/yr)}													& {7.2 $\pm$ 0.2}		\\
{$\mu_{\star}$}													& {25 $\pm$ $1^a$}		\\
{$\mu'$}														& {$\sim1.4^a$}			\\
{$A_V$ (continuum)}		 										& {0.09 $\pm$ 0.15}		\\
{log ($\mu\ \cdot M / M_{\odot}$)}									& {9.5 $\pm$ 0.1}		\\
{$\beta$ (2000$-$2800 \AA)}										& {-1.7 $\pm$ 0.2}		\\
{$\mu\ \cdot$ SFR$_{\mbox{H}\beta}$ ($M_{\odot}$ yr$^{-1}$)} 				& {390 $\pm$ 9} 		\\
{$\sqrt{\mu'} \cdot r_e$}											& {350 pc}				\\
\hline \\
[-1.0 ex]
\multicolumn{2}{c}{Using Updated Lens Model:}  \\
[1.0 ex]
\hline \\
[-1.0 ex]
$\mu_{\mbox{\it tot}}$											& $17.3\pm1.2$		\\
$\mu_{\mbox{\it eff}}$											& $19\pm1.5$			\\
{log($M_{\star}/M_{\odot}$)}										& {$8.26\pm0.10$}		\\
{SFR ($M_{\odot}$ yr$^{-1}$)$^b$}									& {$22.5\pm1.6$} 	
\enddata
\tablecomments{
Observed and derived quantities for SL2SJ0217-0513.
The {\it HST} CANDELS \citep{koekemoer11} and MIPS \citep{labbe06} photometry,
as well as properties derived from the 3D-{\it HST} grism spectrum and SED fit, are taken the from \citet{brammer12}.
Updated parameters incorporating the lensing model from this work are also listed.
RA and Dec specify the optical center in units of
hours, minutes, and seconds, and degrees, arcminutes, and arcseconds respectively. \\
$^a$ $\mu_{\star}$ is the relative magnification between the integrated arc and the counter image.
$\mu'$ is the brightness magnification of the counter image. The total lens magnification is
$\mu = \mu_{\star} \cdot \mu'$. \\
$^b$ Determined from the H$\beta$ line luminosity.}
\label{tbl2}
\end{deluxetable}

%----------------------------------------------------------------------------------------------------------------------
%----------------------------------------------------------------------------------------------------------------------

% Figure 1: Lens Model:
\begin{figure*}[t]
\epsscale{1.0}
\includegraphics[scale=0.525,  trim=0mm 0mm 0 0mm, clip]{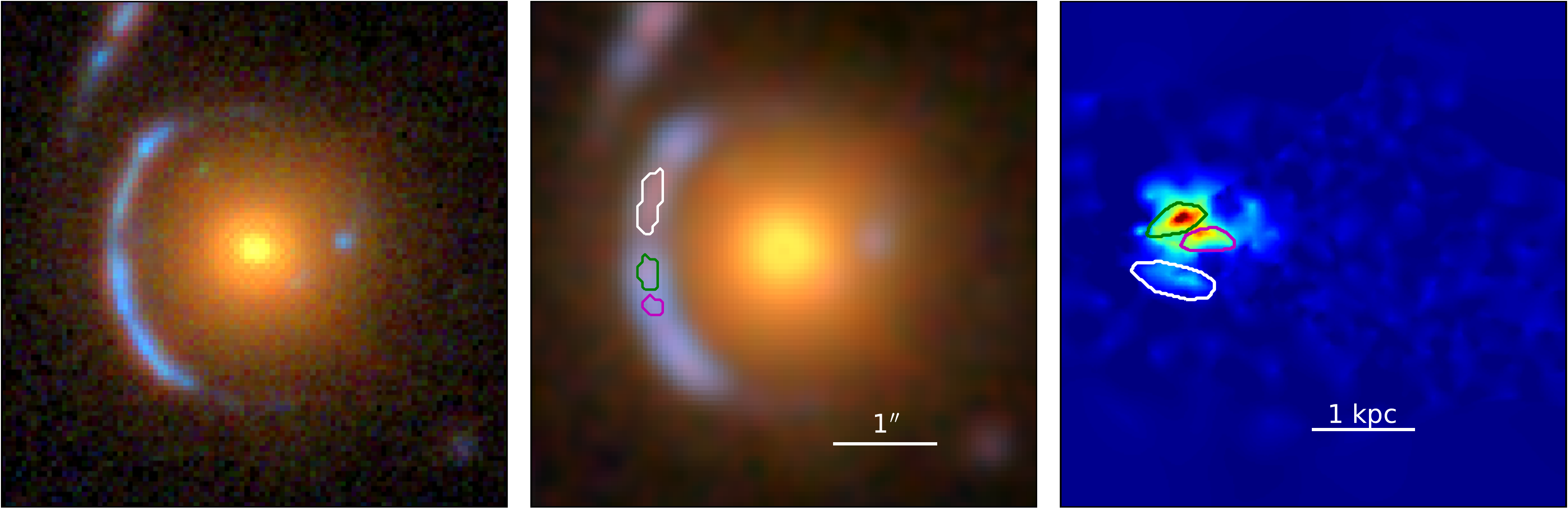}
\caption{(Left) {\it HST} F606W, F814W, and F125W image of the lens. 
(Middle) Same, but with the resolution degraded to the resolution of the F160W image 
to perform aperture photometry within the shown apertures. 
(Right) Reconstructed source in the F606W filter, showing the intrinsic three regions 
corresponding to the apertures in the middle panel. \\}
\label{fig1}
\end{figure*}

%----------------------------------------------------------------------------------------------------------------------

%%%% 3 %%%%
\section{IMPROVED PHOTOMETRY AND LENSING MODEL}\label{sec3}
Using the available 5-band {\it HST} imaging, we quantify the resolved photometric properties of SL2S0217. 
These data include optical F606W and F814W images observed with ACS/WFC (4,856 s and 11,300 s respectively)
and near infrared F125W, F140W, and F160W images observed with WFC3/IR (1,912 s, 812 s, and 3,512 s respectively). 
All of the data were reduced with the AstroDrizzle package \citep{fruchter10} 
and were resampled to a pixel scale of 0.05$^{\prime\prime}$. 
The absolute astrometry from the telescope differs between the different bands, 
but we ensured that each band was well registered by matching the \textit{apparent} coordinates of the 
central lensing galaxy in the output resampled images. 
We used an unsaturated bright star 21$^{\prime\prime}$ from the lens as a model for the 
PSF in our subsequent surface brightness and lens modeling.

The surface brightness distribution of the foreground lensing galaxy overlaps with the 
lensed arc features, and we therefore modeled and removed this light from each image
using the following process. 
First, we manually masked several interloping objects that lie both within and around the lens system. 
Next, we constructed a mask for the arc and counter image and simultaneously fit the foreground 
lensing galaxy light distribution in each band with a model that allowed three concentric S\'{e}rsic profiles. 
As a test of this procedure, we have also fit the foreground light in the F606W band while 
simultaneously modeling the light of the background source and found consistent results. 
Finally, the foreground surface brightness model was subtracted from the data to give an 
uncontaminated view of the background source structure.

The structure of the bright, lensed arc indicates that the background galaxy has a complex morphology.
We therefore used the adaptive pixellated source modeling technique described in \citet{vegetti09} 
to delens and `reconstruct' the lensed source. 
The intrinsic source surface brightness distribution was described on an irregular grid of pixels that 
approximately follows the magnification of the lens, with a PSF-deconvolved intensity at each pixel  
determined from the lens mass model and observed data. 
We modeled the lensing potential as an elliptical powerlaw mass distribution with an empirical external 
shear, and constrained it with the F606W data, which have excellent spatial resolution and S/N. 
The reconstructed source surface brightness distribution is shown in the right panel of Figure~\ref{fig1}.

The total observed (magnified) magnitudes are given in Table~\ref{tbl3}, with systematics-dominated 
uncertainties of $\sim0.03$ magnitudes. 
While the increased freedom in our lens model generally increases the uncertainties on the magnification, 
our flexible source model allows a better fit to the data, decreasing the systematic uncertainty. 
Therefore, the lens modeling yields a total magnification of $\mu=17.3\pm1.2$, significantly lower than the value of 
$\sim$35 determined by \citet{cooray11} and reported in B12.
This discrepancy in magnification is largely driven by the choice of source model. 
In our analysis, we find that there is a significant amount of low-surface-brightness (and low-magnification) flux
that is not encapsulated by the \citet{cooray11} sources, and for that reason the magnification 
(observed light)/(modeled source light) is larger for their model.\footnotemark[3]

\footnotetext[3]{
\citet{cooray11} used two exponential profiles for the source and a singular isothermal elliptical (SIE) lens,
perturbed by a distant singular isothermal sphere (SIS) halo. 
This model fixes the external shear field and the radial distribution. 
The subsequent lack of structure in the source surface brightness model results in significant residuals 
(see Figure 2(h) of \citet{cooray11}). }

This low-surface-brightness flux is readily seen in the bluer filters, where the light from the lensing galaxy 
is less prominent, as extensions of the bright arc immediately north and south of the lensing galaxy. 
Our flexible source model allows magnification to vary across the source, 
and results in the intrinsically higher-surface-brightness regions (highlighted in the right panel of Figure~\ref{fig1}) 
having somewhat higher magnifications of $\sim$25.
In other words, there is clearly \textit{differential magnification} across the broad-band image of the source. 
We simulated slit losses of $\sim10$\% by convolving the light profile from the {\it HST} F606W image 
with the seeing disk and integrating through the 1\arcsec\ slit of the LRIS spectrum. 
Accounting for the lensing magnification and applying the slit loss correction, we find that the flux 
within the spectroscopic aperture has an \textit{effective mean amplification} of 19$\pm$1.5.

As in B12, we find significant broad-band color differences across the galaxy. 
To quantify this, we convolved each of the four bluer images with a gaussian filter to match the resolution 
of the F160W image and determined aperture colors within the three regions shown in Figure~\ref{fig1}. 
The colors measured for the magenta, green, and white apertures are listed in Table~\ref{tbl3}. 
The white region, which is clearly redder, is also much lower surface brightness 
and lower magnification than the other regions,
and so contributes less to the observed LRIS spectrum. 
Thus, the UV spectrum of SL2S0217 is largely produced by the neighboring magenta and green star-forming regions.

%----------------------------------------------------------------------------------------------------------------------

% Table 3: Lens Photometry
\begin{deluxetable}{lccc}
\tabletypesize{\footnotesize}
\tablewidth{0pt}
\tablecaption{Lens Model Photometry}
\tablehead{
\CH{Parameter}		& \multicolumn{3}{c}{Region} }
\startdata
{}					& \multicolumn{3}{c}{Total}  \\ 
\cline{2-4} \\
[-1.0 ex]
{F606W}				& \multicolumn{3}{c}{22.15} \\
{F814W}				& \multicolumn{3}{c}{22.03} \\
{F125W}				& \multicolumn{3}{c}{21.80} \\
{F140W}				& \multicolumn{3}{c}{21.14} \\
{F160W}				& \multicolumn{3}{c}{20.98} \\
\hline \\
[-1.0 ex]
{}					& {Magenta}		& {Green}			& {White}	\\ 
\cline{2-4} \\
[-1.0 ex]
{F606W$-$F160W}		& {$1.03\pm0.10$}	& {$1.35\pm0.19$}	& {$1.79\pm0.27$}	\\
{F814W$-$F160W}		& {$1.01\pm0.10$}	& {$1.24\pm0.18$}	& {$1.53\pm0.25$}	\\
{F125W$-$F160W}		& {$0.79\pm0.08$}	& {$0.93\pm0.14$}	& {$1.01\pm0.21$}	\\
{F140W$-$F160W}		& {$0.25\pm0.08$}	& {$0.28\pm0.13$}	& {$0.26\pm0.20$}	\\
\enddata
\tablecomments{
Top: The total observed (magnified) magnitudes, with systematics-dominated 
uncertainties of $\sim0.03$ magnitudes. 
Bottom: The colors measured for the magenta, green, and white apertures show
significant broad-band color differences across the galaxy.}
\label{tbl3}
\end{deluxetable}

%----------------------------------------------------------------------------------------------------------------------
%----------------------------------------------------------------------------------------------------------------------

%%%% 4 %%%%
\section{NEW KECK/LRIS SPECTROSCOPIC OBSERVATIONS AND DATA REDUCTION}\label{sec4}

%----------------------------------------------------------------------------------------------------------------------

%%%% 4.1 %%%%
\subsection{Observations}\label{sec4.1}

We obtained spectra of SL2S0217 using the Low Resolution Imaging Spectrograph 
\citep[LRIS;][]{oke95} on the Keck I telescope on the UT date of 2015 October 12.
The 600/4000 grism and 600/7500 grating were used with the blue and red detectors respectively,
with the dichroic at 560 nm, resulting in an observed-wavelength coverage of approximately $3100-9000$ \AA\ 
and an average resolution at full width half maximum (FWHM) of roughly 4.0 \AA\ (275 km s$^{-1}$) 
in the blue and 4.7 \AA\ (190 km s$^{-1}$) in the red. 
This corresponds to a rest-wavelength coverage of approximately $1100-3000$ \AA\ and 
resolutions of 1.4 (1.7) \AA\ in the blue (red). 

\begin{figure}
\epsscale{1.0}
\includegraphics[scale=0.45, trim=0mm 10mm 0mm 5mm, clip]{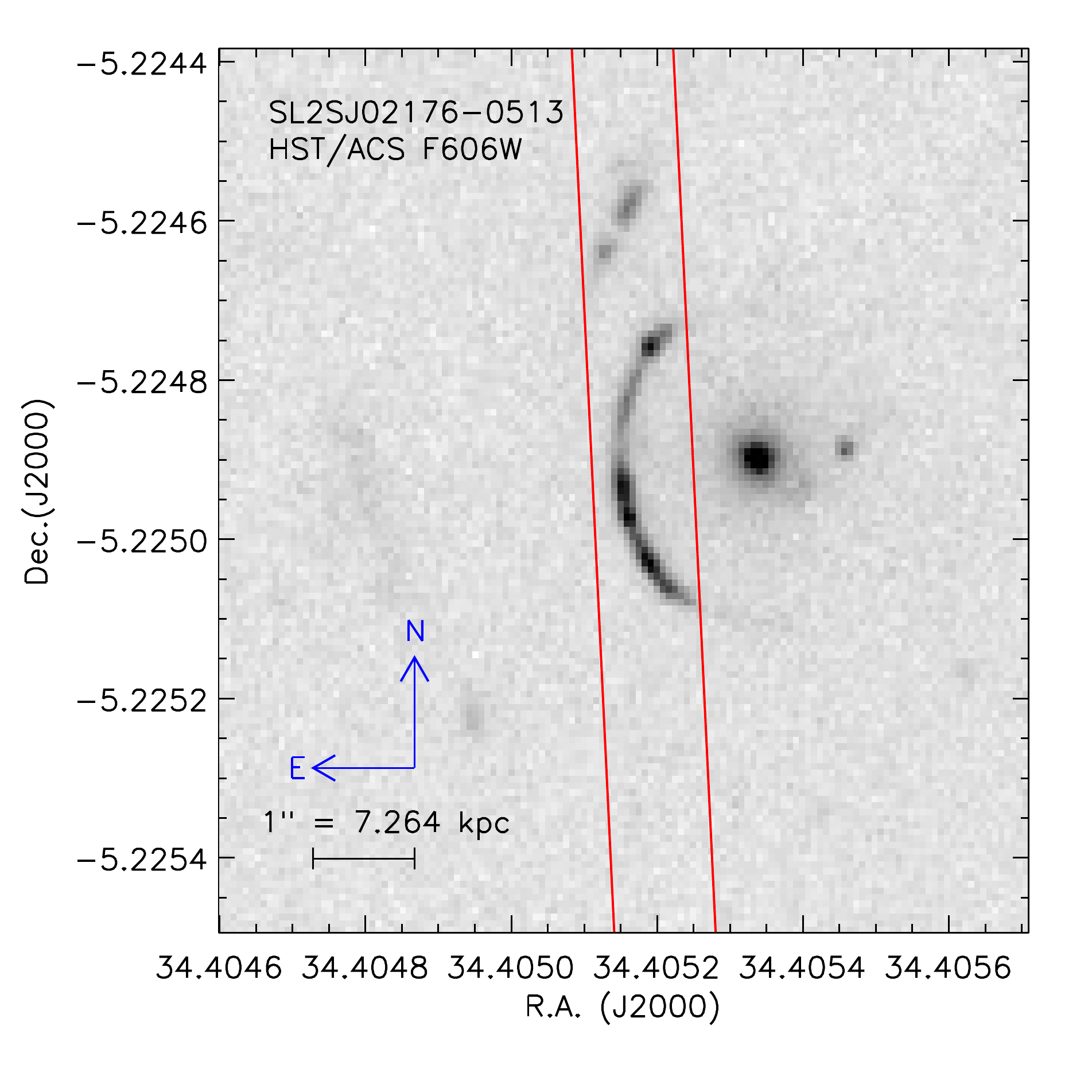}
\caption{CANDELS {\it HST} ACS F606W image of SL2SJ0217-0513.
The 1.0\arcsec\ LRIS slit is overlaid in red, demonstrating that the majority
of the arc light was captured in the slit. 
A second lensed source is visible in the top part of the slit, but it is spatially distinct 
(both along the slit and in redshift) and so does not affect the spectrum of SL2SJ0217-0513.}
\label{fig2}
\end{figure}

Internal and twilight calibration flats were obtained at the beginning of the night to account for
differences between the chip illumination patterns on the sky and the internal lamps.
Two standard stars, Feige~24 and Feige~110, with spectral energy distributions peaking in the blue 
were observed over a range of hour angles throughout the night, allowing the flux calibration to be 
determined as a function of airmass.
SL2S0217 was observed for $13\times1800$ s or 6.5 hours in clear, dark conditions with 
0.7$-$0.8\arcsec\ seeing, over an airmass range of $1.0 - 1.7$.
A slit of 1.0\arcsec$\times$168\arcsec\ at an angle of $3^{\circ}$ was chosen 
in order to encompass the maximum flux from the arc, as shown in Figure~\ref{fig2}.
HgNeArCdZn arc lamps were observed after every other exposure to mitigate the effects of 
instrument flexure in the wavelength calibration.

%----------------------------------------------------------------------------------------------------------------------
%----------------------------------------------------------------------------------------------------------------------

% Figure 2: SL2S HST Image:

%----------------------------------------------------------------------------------------------------------------------
%----------------------------------------------------------------------------------------------------------------------

% Figure 3: Blue and Red LRIS Spectra:
\begin{figure*}
\begin{center}
        \includegraphics[scale=0.85, trim=0mm 8mm 5mm 0mm, clip]{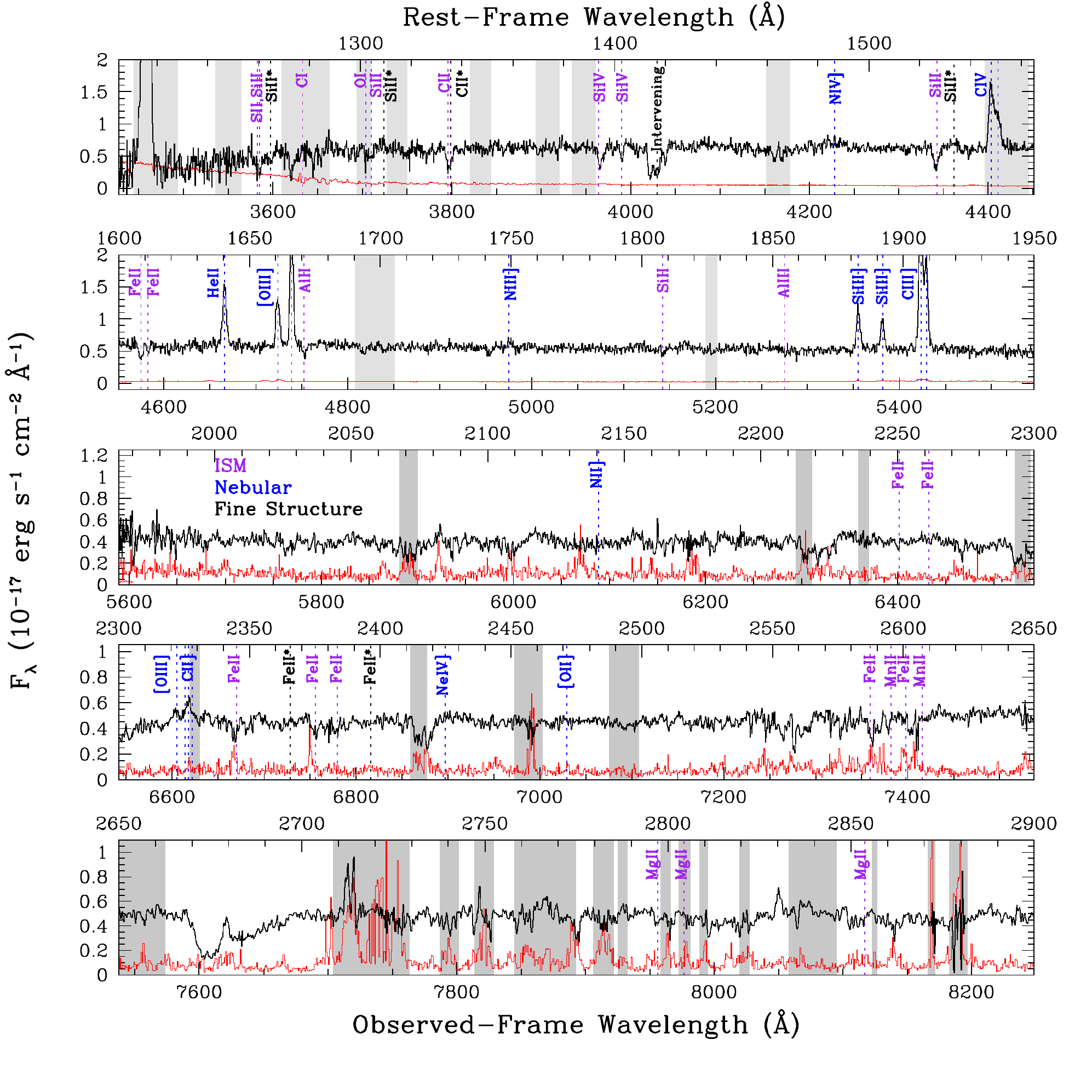} 
	\caption{
	The absorption and emission features observed in the LRIS spectra of SL2S0217,
	with the error spectrum shown in red. 
	Regions of strong sky line emission are shaded grey.
	Note that the sky features in the red spectrum are stronger than in the blue,
	and so are shaded darker gray accordingly.
	The locations of nebular emission lines are identified with dashed blue lines.
	 Absorption features due to the interstellar medium are identified by dashed purple lines,
	 and corresponding fine structure emission lines are labeled in black.
	 Not all labels correspond to detections, but identify the location of features discussed in the text.}
   \label{fig3}
\end{center} 
\end{figure*}     

%----------------------------------------------------------------------------------------------------------------------
%----------------------------------------------------------------------------------------------------------------------

%%%% 4.2 %%%%
\subsection{Spectra Reduction}

The LRIS spectra were processed using ISPEC2D \citep{moustakas06}, 
a long-slit spectroscopy data reduction package written in IDL. 
Master sky and internal flats were constructed by taking the median 
at each pixel after normalizing the counts in the individual images. 
These calibration files were then used to bias-subtract, flat-field, and illumination-correct the raw science data frames. 
Misalignment between the trace of the light in the dispersion direction and the orientation of the CCD detector 
was  rectified via the mean trace of the standard stars, providing alignment to within a pixel across the detector. 
A median sky subtraction was performed at each column along the dispersion, followed by a wavelength calibration 
applied from the HgNeArCdZn comparison lamps taken at the nearest airmass. 
The one-dimensional spectra were produced using a box-car aperture that encompassed roughly 
99\%\ of the light in the Ly$\alpha$ feature.
We note that while a second lensed source is with in the top part of the slit (see Figure~\ref{fig2}), 
it is distinct spatially, as well as in redshift \citep[$z\sim2.32$;][]{cooray11}, 
and so does not affect the extracted spectrum of SL2SJ0217.
Individual exposures were then flux calibrated using the sensitivity curve derived from the standard star
observations taken throughout the night.
Finally, the 13 sub-exposures were median combined, eliminating cosmic rays in the process. 

The resulting 1D spectra were then compared as a check on the flux calibration:
the overlapping regions of flux-calibrated blue and red spectra were found to be in good agreement
within the dispersion of their continua.
We then checked the absolute flux calibration of the combined blue$+$red LRIS spectrum using newly obtained 
{\it HST} images in the WFC3 F390M (covering rest-frame $\sim1300-1400$ \AA) and 
F343N (covering the Ly$\alpha$ emission line) filters; these images will be presented in a future paper. 
We used the PYSYNPHOT package in Python to predict the F390M AB magnitude from the spectrum, 
and then scaled the spectrum to match the observed F390M AB magnitude ($22.68\pm0.07$; Erb et al.\ 2018, in prep.). 
Note that we used the bluest possible continuum band so that contamination of the lens is negligible.
The resulting 1D spectrum and error spectrum are shown in Figure~\ref{fig3}.
Regions of strong sky line emission are shaded grey. 
Potential emission and absorption features are labeled accordingly. 

%----------------------------------------------------------------------------------------------------------------------
%----------------------------------------------------------------------------------------------------------------------

% Table 4:
\begin{deluxetable}{cccc}[H]
\tabletypesize{\scriptsize}
\tablewidth{0pt}
\setlength{\tabcolsep}{3pt}
\tablecaption{Emission Line Redshift Determinations}
\tablehead{
\CH{Ion}	& \CH{\W$_{\mbox{\scriptsize lab}}$ (\AA)$^a$}	& \CH{\W$_{\mbox{\scriptsize obs}}$ (\AA)} & \CH{z} }
\startdata
O~\iii]		& 1660.81		& 4723.97	   	   & 1.84438 \\
O~\iii]		& 1666.15		& 4739.03		   & 1.84430 \\
Si~\iii]		& 1882.71		& 5355.23		   & 1.84443 \\
Si~\iii]		& 1892.03		& 5381.48		   & 1.84429 \\
{[C~\iii]}		& 1906.68		& 5423.25		   & 1.84434 \\
C~\iii]		& 1908.73		& 5429.13		   & 1.84437 \\
\hline \\
Average:		& \multicolumn{3}{l}{1.84435$\pm$0.00066} 
\enddata 
\tablecomments{
Systemic redshift of SL2S0217 calculated from the strong nebular emission lines
in the LRIS spectrum. \\
$^a$ Vacuum wavelengths. }
\label{tbl4}
\end{deluxetable}

%----------------------------------------------------------------------------------------------------------------------
%----------------------------------------------------------------------------------------------------------------------

% Figure 4: Comparison to Composite Spectrum:
\begin{figure*}
\begin{center}
\includegraphics[scale=0.85, trim=0mm 5mm 5mm 0mm, clip]{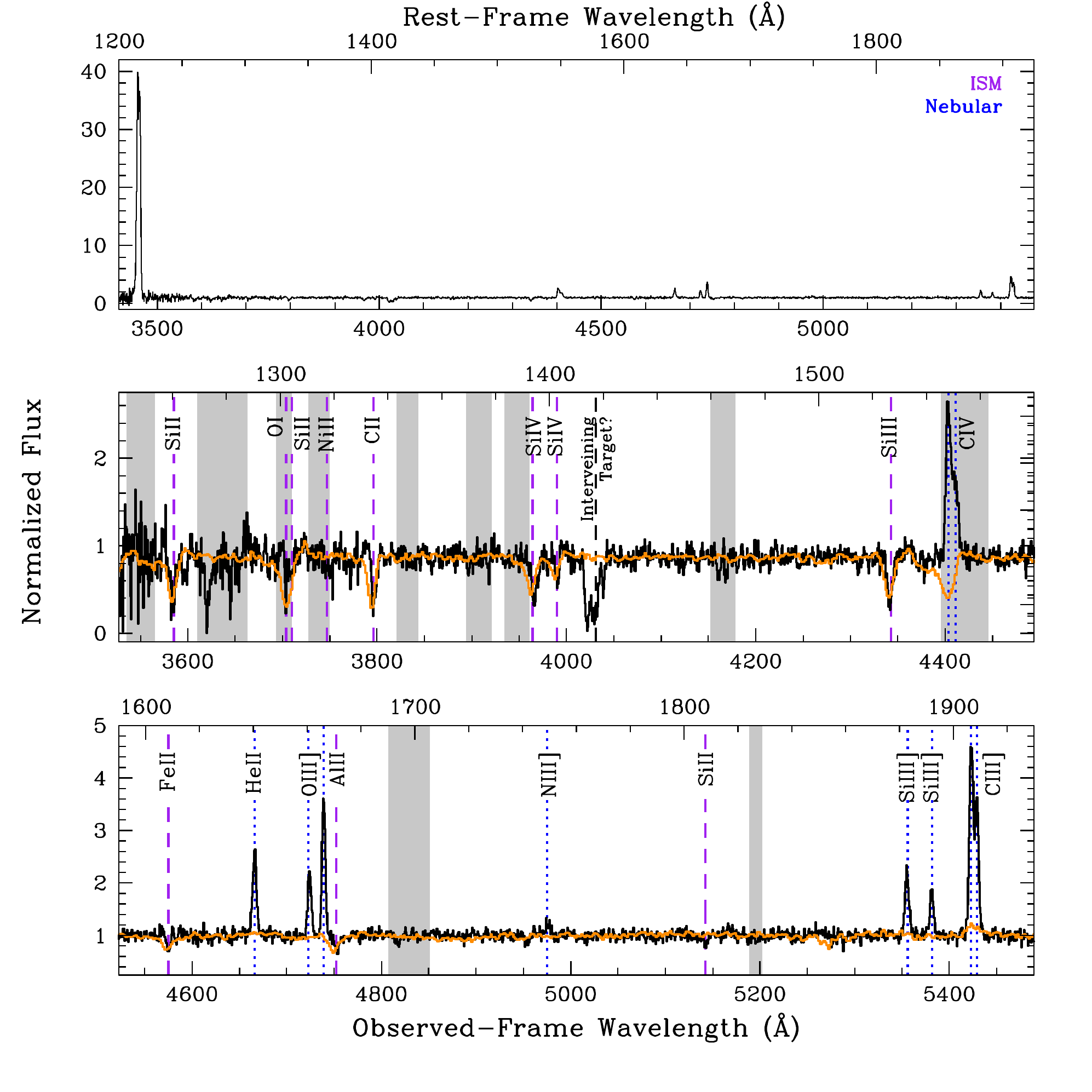}
\caption{
Blue arm of the Keck LRIS spectrum of SL2S0217, where regions of significant sky contamination 
are designated by the grey shading.
The top panel shows the full extent of the spectrum, whereas the bottom two panels are 
zoomed in to depict the numerous interstellar medium (ISM) absorption and nebular emission features. 
The $z\sim2$ composite spectrum from \citet{erb10} is over plotted in orange, demonstrating 
the staggering difference in the strength of the emission features and widths of the absorption 
features between the two spectra.}
\label{fig4}
\end{center}
\end{figure*}

%----------------------------------------------------------------------------------------------------------------------
%----------------------------------------------------------------------------------------------------------------------

%%%% 5 %%%%
\section{The Rest-Frame UV Spectrum}\label{sec5}

%----------------------------------------------------------------------------------------------------------------------
%----------------------------------------------------------------------------------------------------------------------

The blue portion of the rest-frame UV LRIS spectrum of SL2S0217 is shown in Figure~\ref{fig4}, 
where the middle and bottom panels are zoomed in on the vertical scale to highlight the numerous 
absorption and emission features observed. 
In comparison we plot the composite spectrum of $\sim$1,000 $z\sim2$ galaxies from \citet{erb10},
and note two defining properties of the SL2S0217 spectrum:
1) The interstellar absorption features typical of $z~\sim2$ galaxies appear to have velocity profiles 
that are roughly 3 times narrower in SL2S0217, and
2) significant, high-ionization nebular emission is present in SL2S0217 that is atypical of $z~\sim2$ galaxies.
We further discuss some of the most striking features of the LRIS spectrum below. 

%%%% 5.1 %%%%
\subsection{Source Redshift}\label{sec5.1}
In order to analyze the velocity structure of the absorption and emission features seen
in the SL2S0217 UV spectrum, it is necessary to first establish a reliable determination
of the systemic redshift as the reference velocity zero point.
\citet{tu09} used all of the strong emission lines present in their low-resolution Keck/LRIS spectrum
(Ly$\alpha$, C~\iv, He~\ii, O~\iii], and C~\iii]) to estimate $z_{arc} = 1.84691\pm0.0024$.
However, our LRIS spectrum shows that the Ly$\alpha$ emission for SL2S0217 is double peaked
(see Figure~\ref{fig6}).
Additionally, both the He~\ii\ and C~\iv\ emission can have complex profiles complicated
by combinations of stellar and nebular contributions. 
Therefore, we choose not to use these lines in our redshift measurement.
With the higher resolution of our LRIS spectrum, we are able to use the 
O~\iii] \W\W1660,1666, Si~\iii] \W\W1883,1892, and C~\iii] \W\W1907,1909 
emission lines for our systemic redshift determination. 

The line centers used to calculate the redshift are given in Table~\ref{tbl4},
resulting in a systemic redshift of $z_{sys} = 1.84435 \pm 0.00066$ for SL2S0217.
Since this value is within the uncertainty of the redshift measured by \citet{tu09},
the line centers of the observed Ly$\alpha$, He~\ii, and C~\iv\ profiles must lie close to the systemic velocity.

%----------------------------------------------------------------------------------------------------------------------
%----------------------------------------------------------------------------------------------------------------------

% Figure 5:  Emission Line Spectra:
\begin{figure*}
\begin{center}
\includegraphics[scale=0.75, trim=2mm 4mm 0mm 0mm, clip]{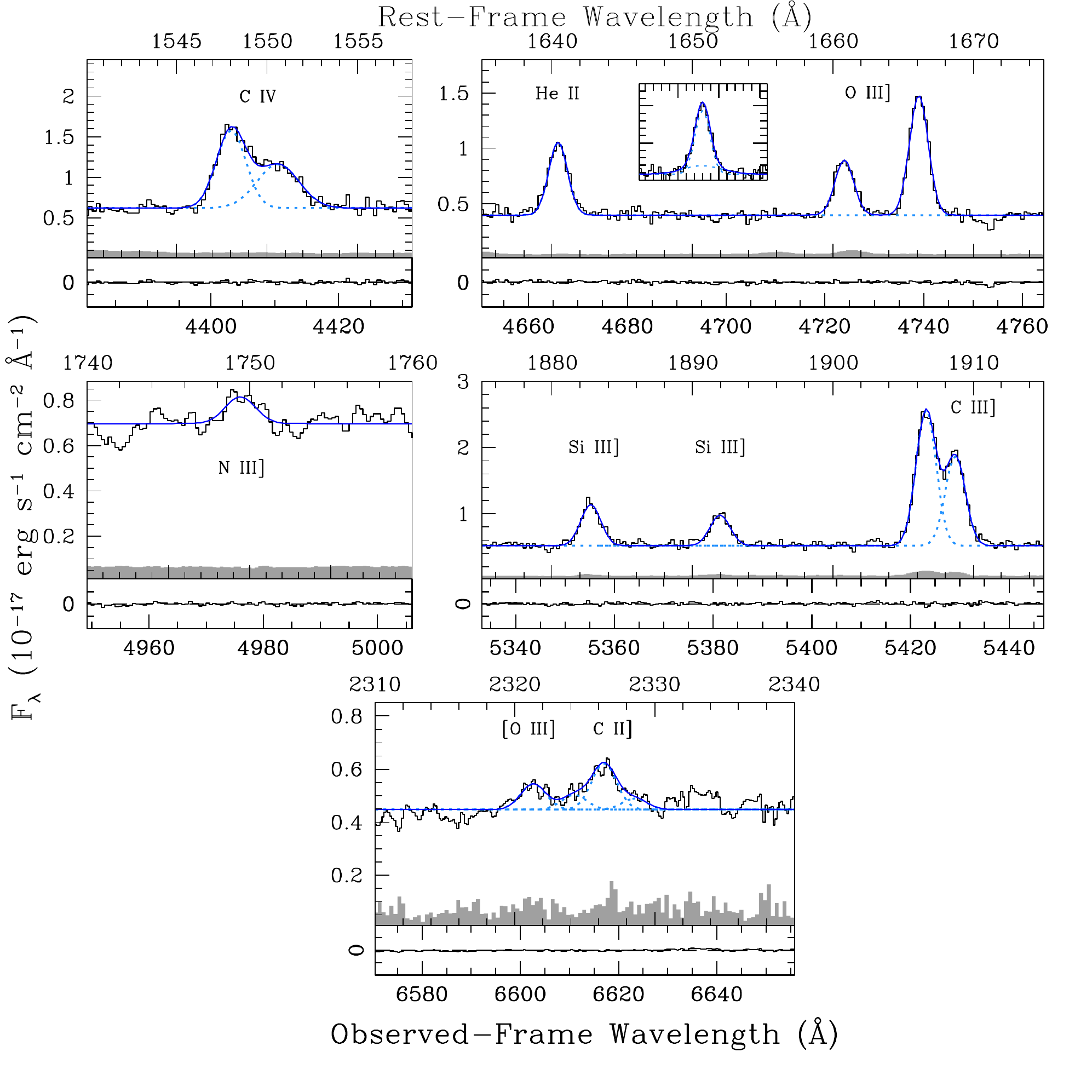}
\caption{
The unusually strong nebular emission lines of SL2S0217.
Lines are fit by Gaussian profiles (blue), 
with FWHMs and wavelength shifts tied together for nearby lines.
When multiple components are fit to a single blended structure, the components
are shown by dashed light blue lines.
In the case of He~\ii, we show both a single Gaussian fit and a two-component fit
(inset window) allowing for narrow ($\sim280$ km s$^{-1}$) nebular and 
broad ($\sim710$ km s$^{-1}$) stellar contributions. 
Two features are particularly interestingly:
(1) the C~\iv\ emission is surprisingly strong and
(2) the He~\ii\ emission is both strong and narrow, appearing mostly nebular in origin.
The $\pm1\sigma$ error spectrum (shaded grey) is shown for comparison.
Residuals to the fit are plotted below the corresponding spectral windows on a scale of -1 to 1.}
\label{fig5}
\end{center}
\end{figure*}

%----------------------------------------------------------------------------------------------------------------------
%----------------------------------------------------------------------------------------------------------------------

% Table 5:
\begin{deluxetable*}{lccccc}[H]
\tabletypesize{\scriptsize}
\tablewidth{0pt}
\tablecaption{Nebular Emission Lines}
\tablehead{
\CH{Ion}									& 
\CH{\W$_{\mbox{\scriptsize lab}}$ (\AA)$^a$}		& 
\CH{$F_{\lambda}^b$}						& 
\CH{$W$ (\AA)$^c$}							& 
\CH{$I_{\lambda}^b$} 						& 
\CH{$I_{\lambda}/I_{\lambda1909}$}				} 
\startdata	
Ly$\alpha_b$ 	& 1215.67		& 58.7$\pm$0.6	& 66.0	& 71.3$\pm$1.2	& 8.94	\\ 
Ly$\alpha_r$	& 1215.67		& 39.0$\pm$0.4	& 43.2	& 47.3$\pm$0.8	& 5.94	\\
Ly$\alpha_{tot}$& 1215.67	& 100.8$\pm$1.0	& 113.0	& 122.4$\pm$2.1	& 15.3	\\
C~\iv\ 		& 1548.19		& 5.54$\pm$0.07	& 3.1		& 6.40$\pm$0.12	& 0.80	\\ 
			& 1550.77		& 4.47$\pm$0.06	& 2.5		& 5.17$\pm$0.10	& 0.65	\\
He~\ii		& 1640.42		& 4.62$\pm$0.07	& 2.8		& 5.32$\pm$0.10	& 0.67	\\ 
He~\ii$_{n}$	& 1640.42		& 3.98$\pm$0.06	& 2.4		& 4.58$\pm$0.09	& 0.57	\\ 
He~\ii$_{w}$	& 1640.42		& 1.33$\pm$0.05	& 0.8		& 1.53$\pm$0.07	& 0.19	\\ 
O~\iii]		& 1660.81		& 3.50$\pm$0.06	& 2.1		& 4.03$\pm$0.09	& 0.51	\\
			& 1666.15		& 7.60$\pm$0.09	& 4.5		& 8.74$\pm$0.15	& 1.10	\\
N~\iii]$^d$		& 1749		& 0.52$\pm$0.10	& 0.4		& 0.60$\pm$0.12	& 0.08	\\
Si~\iii] 		& 1882.71		& 3.17$\pm$0.06	& 2.1		& 3.66$\pm$0.08	& 0.46	\\ 
	 		& 1892.03		& 2.35$\pm$0.05	& 1.6		& 2.72$\pm$0.07	& 0.34	\\ 
{[C~\iii]} 		& 1906.68		& 10.4$\pm$0.12	& 7.0		& 12.1$\pm$0.19	& 1.51	\\ 
C~\iii] 		& 1908.73		& 6.90$\pm$0.08	& 4.7		& 7.98$\pm$0.13	& 1.00	\\ 
{[O~\iii]}		& 2321.66		& 0.64$\pm$0.03	& 0.6		& 0.74$\pm$0.08	& 0.09	\\ 
C~\ii]			& 2325.40		& 0.36$\pm$0.03	& 0.4		& 0.42$\pm$0.08	& 0.05	\\
			& 2326.93		& 1.11$\pm$0.03	& 0.9		& 1.29$\pm$0.09	& 0.16	\\
			& 2328.12		& 0.27$\pm$0.03	& 0.4		& 0.32$\pm$0.08	& 0.04	\\
\hline 
{[Ne~\iii]} 		& 3870.16		& 30$\pm$4		& \nodata	&   32.6$\pm$4.4	& \nodata \\
H$\delta$ 		& 4101.00		& 19$\pm$2		& \nodata	&   21.1$\pm$2.2	& \nodata	\\
H$\gamma^e$ 	& 4341.69		& 35$\pm$2		& \nodata	&   37.8$\pm$2.2	& \nodata	\\ 
{[O~\iii]$^e$}	& 4364.44		& 5$\pm$3		& \nodata	&     5.3$\pm$3.2	& \nodata	\\  
H$\beta$ 		& 4862.69		& 74$\pm$2		& 517 	&   78.9$\pm$2.3	& \nodata	\\ 
{[O~\iii]}		& 4960.30		& 79$\pm$2		& \nodata	&   84.6$\pm$3.3	& \nodata	\\
			& 5008.24		& 230$\pm$2		& 2095$^f$ & 244.2$\pm$4.1	& \nodata		
\enddata
\tablecomments{
Rest-Frame UV emission lines from this work and optical emission lines 
from the 3D-{\it HST} spectrum of B12.
Emission features were fit with Gaussian profiles, where nearby lines were constrained to
the same FWHMs and a single wavelength offset. 
The exceptions to this rule are the Ly$\alpha$, C~\iv, and He~\ii\ profile fits, which have
additional contributions from resonant scattering (Ly$\alpha$ and C~\iv) and stars (C~\iv\ and He~\ii).
The two fits listed for Ly$\alpha_b$ and Ly$\alpha_r$ correspond to blue and red Gaussian components, 
 whereas Ly$\alpha_{tot}$ was determined by integrating the profile as a whole.
For He~\ii, a single Gaussian fit is reported, as well as a two-component fit with narrow and (potential) wide parts.
Following B12, who found negligible reddening in SL2S0217,
we simply corrected for the galactic extinction \citep[$E(B-V)=0.0194$;][]{schlafly11}. 
Line fluxes (observed-frame) and equivalent widths (rest-frame) are listed in Columns 3 and 4, followed by the extinction 
corrected intensities in Column 5. \\
$^a$Vacuum wavelengths. \\
$^b$ Units are $10^{-17}$ erg s$^{-1}$ cm$^{-2}$, uncorrected for lens magnification.  \\
$^c$ Equivalent widths are measured in the rest-frame. \\
$^d$N~\iii] is the blend of N~\iii] \W\W1748,1749. \\
$^e$Lines were completely blended in the 3D-{\it HST} grism spectrum.  \\
$^f$Total EW of the blended [O~\iii] \W\W4959,5007 doublet. }
\label{tbl5}
\end{deluxetable*}

%----------------------------------------------------------------------------------------------------------------------
%----------------------------------------------------------------------------------------------------------------------

%%%% 5.2 %%%%
\subsection{Nebular Emission Lines}\label{sec5.2}
The rest-frame UV spectrum of SL2S0217 contains numerous emission line
features that are sensitive to the ionizing stellar population, physical properties 
of the emitting gas, and kinematics of outflows.
In particular, SL2S0217 shows double-peaked Ly$\alpha$ emission and exceptionally strong,
nebular-like emission from high-ionization species such as C~\iv\ \W\W1548,1550, He~\ii\ \W1640, 
O~\iii] \W\W1661,1666, Si~\iii] \W\W1883,1892, and C~\iii] \W\W1907,1909, 
with additional weak detections of the low-ionization [O~\iii] \W2322 and C~\ii] \W\W\W2325,2327,2328 lines. 
While UV emission lines have rarely been observed at strengths comparable to SL2S0217 
in star-forming galaxies at any epoch,
these features are especially uncommon for the typically older, more massive 
star-forming galaxies studied at $z\sim2$ \citep[see, e.g.,][]{jones12, reddy12, whitaker12}.
Figures~\ref{fig3} and \ref{fig4} highlight these atypical emission features of SL2S0217.
However, extreme emission-line features may be more common at higher redshifts 
where we expect to find harder radiation fields associated with less evolved, more metal-poor galaxies.
Therefore, the magnified emission-line spectrum of SL2S0217 provides a unique window to examine the 
conditions driving powerful photoionization in distant galaxies.

To characterize the emission line spectrum, all emission line strengths for our LRIS spectrum 
were measured using the {\tt SPLOT} routine within IRAF\footnotemark[4]. 
With the exceptions of Ly$\alpha$, C~\iv, and He~\ii\, which have multiple components, 
groups of nearby lines were constrained to a single Gaussian FWHM and a single 
shift in wavelength from vacuum when possible\footnotemark[5].
Note that the nebular emission lines are unresolved, 
and so subsequent discussions of line widths do not account for the instrumental resolution.
Following B12, who found that the H$\gamma$/H$\beta$ ratio from the grism spectrum was consistent 
with no reddening, we ignored the reddening from dust and simply corrected the line fluxes for the galactic 
extinction along the line of sight to SL2S0217 \citep[$E(B-V)=0.0194$;][]{schlafly11}\footnotemark[6]
using the \citet{cardelli89} reddening law.
The adopted emission line fits are depicted in Figure~\ref{fig5}, with
with the resulting line strengths and intensities given in Table~\ref{tbl5}.
We examine a few of the interesting spectral features of SL2S0217 individually in the next sections 
and in the discussion (Section~\ref{sec8}).

\subsubsection{Blended Emission Features}\label{sec5.2.1}
Neither the components of the double Ly$\alpha$ emission feature nor the individual lines of the C~\iii] \W\W1907,1909
or C~\iv\ \W\W1548,1550 doublets are resolved, but are still well fit by two blended Gaussians.
C~\iv\ and He~\ii\ are further complicated by the potential for their profiles to be modified by multiple components.
In the case of He~\ii, which can have both nebular and stellar emission, we used a two-component fit with narrow 
and wide profiles respectively.
Both single and multi-component fits to the He~\ii\ profile are demonstrated in Figure~\ref{fig5}.
While the two-component model may offer a more visually appeasing fit to the He~\ii\ profile, it does
not significantly reduce the fit residuals, and so the significance of the wide ($\sim700$ km s$^{-1}$), stellar component is 
difficult to assess with the current S/N and resolution. 
We therefore suggest that the reported fits are upper/lower limits for the stellar/nebular components.
Regardless of the correct fit, the He~\ii\ emission in SL2S0217 is clearly dominated by a narrow
component whose width is consistent with nebular emission ($\sim280$ km s$^{-1}$). 

In the case of C~\iv, which is typically an interstellar medium (ISM) absorption feature or a P-Cygni profile from 
the stellar winds of massive stars, it is clearly seen in SL2S0217 as an emission doublet.
Because the C~\iv\ doublet, like Ly$\alpha$, can also be resonantly scattered, its profile widths were not constrained to the pure nebular line widths.
The resulting best fit had an observed-frame Gaussian FWHM 6.0 \AA\ or about 410 km s$^{-1}$.
Compared to the nearby profiles of the O~\iii] nebular lines (285 km s$^{-1}$)
this 45\%\ increase in profile width indicates that C~\iv\ is likely dominated by nebular emission,
but is broadened by resonant scattering. 
Additionally, strong emission may be masking an absorption component, and,
therefore, we report the C~\iv\ emission line strengths as lower limits.

\footnotetext[4]{IRAF is distributed by the National Optical Astronomical Observatories.}
\footnotetext[5]{For the current spectrum, the emission line shape is largely determined by the instrument resolution, 
which is wavelength dependent, and so the FWHM changes slightly across the spectrum ($\sim1.5-1.7$ \AA\ in the rest frame).}
\footnotetext[6]{B12 presented further evidence for minimal dust.
They determined the extinction of the arc continuum to be $A_V = 0.09$ from their 
2D model fit to the observed 3D-{\it HST} grism spectrum using a \citet{calzetti00} reddening law. 
Assuming $R_V = 4.05$, this corresponds to a very small reddening value of $E(B-V) = 0.022$,
and so was concluded to be negligible.}

%----------------------------------------------------------------------------------------------------------------------
%----------------------------------------------------------------------------------------------------------------------

%%%% 5.3 %%%%
\subsection{Ly$\alpha$}\label{sec5.3}

Ly$\alpha$ emission is predominantly produced by recombination in the 
ionized gas surrounding star-formation, but also by collisionally excited 
neutral H gas \citep[cooling radiation;][]{dijkstra14}.
However, the emission feature is increasingly obscured and scattered with 
increasing H{\sc i} column density and dust attenuation.
Therefore, the shape of the Ly$\alpha$ profile depends on the kinematics 
and distribution of the outflowing gas and dust in star-forming galaxies 
\citep[e.g.,][]{verhamme06, steidel10, kornei10}.

\begin{figure}
\begin{center}
\includegraphics[scale=0.425, trim=0mm 0mm 10mm 5mm, clip]{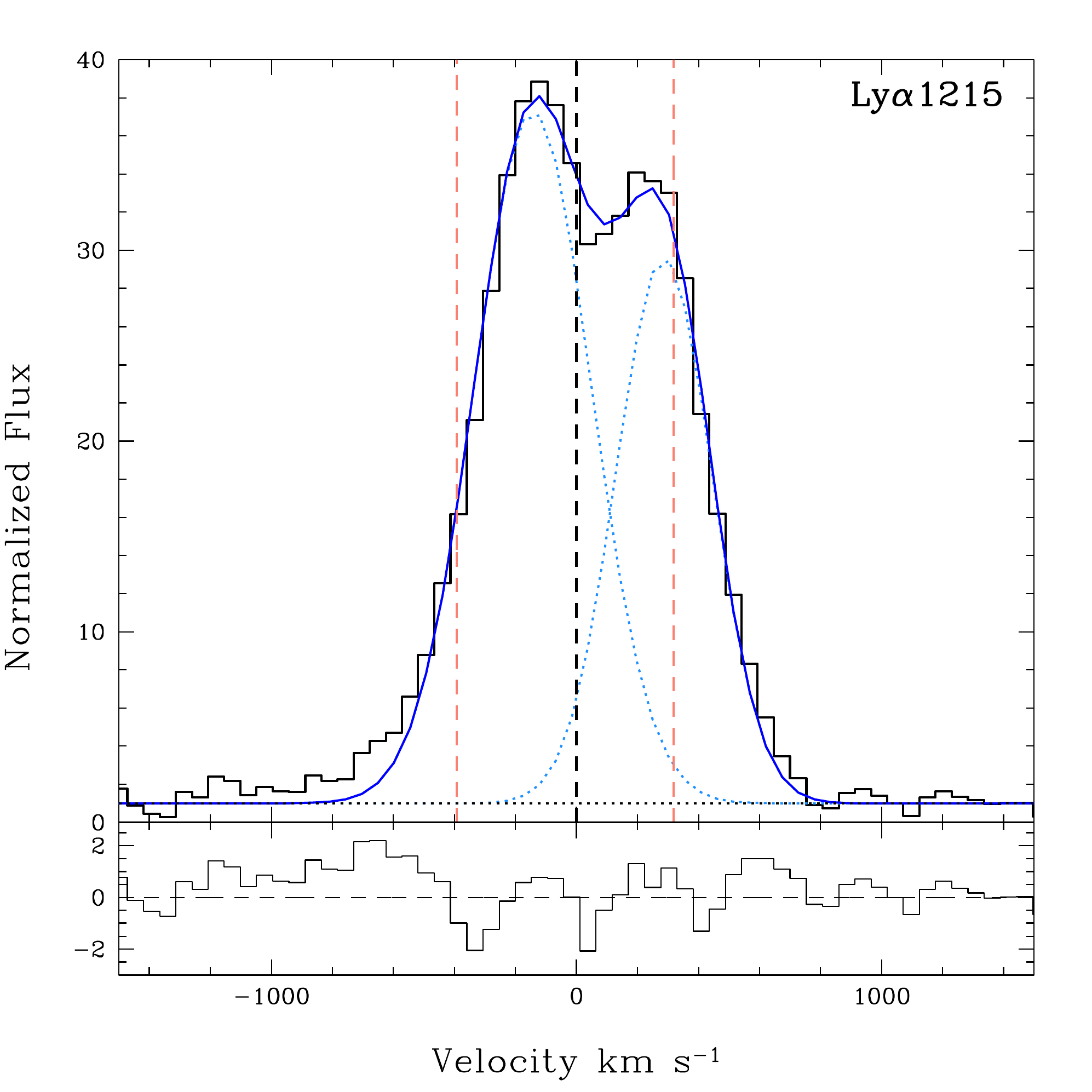}
\caption{
Velocity profile of the double-peaked Ly$\alpha$ emission in SL2S0217.
The overall profile is reasonably well-fit using two Gaussians (blue lines), 
but a residual blue tail remains (shown in the bottom panel), indicating a more complex fit is needed to 
fully describe the kinematic properties of the Ly$\alpha$-emitting gas.
Interestingly, the blue component is stronger than the red, and has a slightly larger velocity width.
The Ly$\alpha$ peak separation is relatively small, shown here to be narrower than the velocity
range of the average absorption feature (red dashed lines), with a nearly systemic central velocity. }
\label{fig6}
\end{center}
\end{figure}

The Ly$\alpha$ velocity profile from the LRIS spectrum of SL2S0217 is 
shown in Figure~\ref{fig6}, where we observe strong, double-peaked 
emission with a dominant blue peak and nearly systemic central velocity. 
The total, integrated Ly$\alpha$ emission has a large equivalent width of 
$W_{Ly\alpha}=113$ \AA, consistent with values for other young, 
metal-poor galaxies \citep[e.g.,][]{cowie11, trainor16}.
Fitting the emission with two independent Gaussian profiles, 
we measure the relative blue to red flux to be $\sim1.5$.
The peak separation of this profile is $\Delta_{\mbox{\scriptsize peaks}} = 371$ km s$^{-1}$. 
From Figure~\ref{fig6} we can see that this model fits the profile fairly well,
however, some of the flux at the peaks is missed, and a residual blue tail extends to $\sim -1300$ km s$^{-1}$.

Multi-peaked Ly$\alpha$ emission (multiplicity) is commonly associated with star-forming galaxies.
In fact, recent studies have found multiplicity rates of $30-50$\% amongst large 
samples of $z\sim2-3$ star-forming galaxies \citep[e.g.,][]{kulas12} and Ly$\alpha$
emitters \citep[LAEs; e.g.][]{yamada12, trainor15}.
Targets with Ly$\alpha$ multiplicity are typically grouped according to whether
they are blue- or red-peak dominant.
While targets with red dominant peaks are much more common than 
blue dominant peaks, \citet{trainor15} find that 26\% of their 129 multi-peak 
LAEs are blue dominant, with an average peak separation of 
$\langle\Delta v_{\mbox{\scriptsize peaks}}\rangle = 660\pm300$ km s$^{-1}$. 
Additionally, for blue dominant profiles in star-forming galaxies, \citet{kulas12} 
found an average peak separation of 
$\langle\Delta v_{\mbox{\scriptsize peaks}}\rangle \sim 800$ km s$^{-1}$, where the 
blueshifted peak is located at $v_{\mbox{\scriptsize peak}} \sim -200$ km $s^{-1}$.
Note, however, that these results are based on low resolution spectra ($\Delta v\sim200-600$ km s$^{-1}$),
which are more sensitive to larger peak separations, and so may be biased to a higher average value.
Still, the peak separation we observe for SL2S0217  
($\Delta v_{\mbox{\scriptsize peaks}} \sim 371$ km s$^{-1}$)
is small relative to these samples of star-forming galaxies at similar redshifts.

%----------------------------------------------------------------------------------------------------------------------
%----------------------------------------------------------------------------------------------------------------------

% Figure:  Lya Profile:

%----------------------------------------------------------------------------------------------------------------------
%----------------------------------------------------------------------------------------------------------------------

The simplest explanation for a double-peaked Ly$\alpha$ emission profile is
scattering by a static, homogeneous, spherical or shell gas cloud.
Given such a medium, the Ly$\alpha$ peak separation is set by the total H{\sc i} column density.
For the peak separation and doppler parameter (thermal velocity dispersion)
of $b = \sqrt{2k_{\scriptsize{B}}T/m_{\mbox{\scriptsize{H}}}} = 16.3$ km s$^{-1}$ 
(for $T= 16100$ K) measured in SL2S0217, the radiative transfer models of \citet{verhamme15}
predict a column density of neutral gas of $N_{\mbox{\scriptsize HI}} \lesssim 10^{20}$ cm$^{-2}$.
Observationally, the trend in Ly$\alpha$ peak separation with H{\sc i} column density found for 12 Ly$\alpha$ emitters at $z \sim 2$
by \citet{hashimoto15} also suggests $N_{\mbox{\scriptsize HI}} \approx 10^{20}$ cm$^{-2}$.
However, static models produce red and blue emission components of equal strengths and widths, 
and so a more complex picture is needed to explain the larger intensity and wing of the blue 
Ly$\alpha$ peak in SL2S0217.

Additional explanations of the Ly$\alpha$ emission profile exist, including fluorescence,
where ionizing photons that escape from galaxies can produce double-peaked 
fluorescent Ly$\alpha$ emission in cold clouds \citep[e.g.,][]{mas-ribas16}.
Like the Ly$\alpha$ profile of SL2S0217, fluorescence tends to produce a small 
peak separation \citep[e.g.,][]{gould96}.
Alternatively, \citet{vanzella17} used MUSE integral field spectroscopy of two extended 
Ly$\alpha$ systems to measure spatial-dependent emission and varying sub-structures, 
suggesting radiative transfer through clumpy/multiphase media as the source 
of blue-dominant Ly$\alpha$ emission \citep[e.g.,][]{gronke16}. 
Unfortunately, our observations of SL2S0217 lack sufficient resolution to 
assess the spatial extent of or variations in emission, and are further complicated by lensing.

Even in a simple static model, viewing Ly$\alpha$ emission from different inclinations 
/ orientations affects the relative blue and red Ly$\alpha$ emission strengths \citep{behrens14},
especially in a differentially lensed system such as SL2S0217.
Alternatively, in radiation transfer models, blue- and red-peak dominant Ly$\alpha$ morphologies are 
associated with inflowing and outflowing gas respectively \citep[e.g.,][]{dijkstra06, verhamme06}.
In the case of infalling gas towards a central source, the red side of a double-peaked 
profile is depressed as the red-shifted Ly$\alpha$ photons within the galaxy see higher 
optical depth due to the line-of-sight infalling gas, resulting in dominant blue-peak 
Ly$\alpha$ emission \citep[see, e.g.,][]{yang12}.
The Ly$\alpha$ profile of SL2S0217 is consistent with this infalling halo model, which 
predicts a peak separation of $\langle\Delta v_{\mbox{\scriptsize peaks}}\rangle \sim 
400$ km s$^{-1}$ \citep[][see their Figure~5]{verhamme06}. 
However, such a model does not eliminate the possibility of outflows, as 
we could have a restricted sight line along which the gas is static or cold pristine gas 
is flowing into the galaxy, and such that outflows are not visible to our viewing angle.

If we are in fact observing the effects of gas inflowing towards the star-forming regions in 
SL2S0217, then we may be witnessing an early episode of star formation during which 
we might have expected relatively low $N_{\mbox{\scriptsize HI}}$ and negligible dust attenuation 
(inferred from the Balmer decrement of the grism spectrum) to favor a leakage of ionizing radiation.
The fraction of escaping Ly$\alpha$ can be estimated by comparing the intrinsic and observed Ly$\alpha$ luminosities.
We determine the intrinsic Ly$\alpha$ luminosity by multiplying the theoretical Ly$\alpha$/H$\beta$ ratio of 23.3
\citep[assuming $T_e = 1.5\times10^4$ K and $n_e = 10^2$ cm$^{-3}$;][]{osterbrock89} with the H$\beta$ intensity 
(only corrected for galactic extinction; see Section~\ref{sec5.2}).
Note, however, that the theoretical Ly$\alpha$/H$\beta$ ratio is density sensitive, and so we also consider
the $n_e = 10^3$ cm$^{-3}$ case in which Ly$\alpha$/H$\beta = 25.7$.
Finally, we use the total magnification ($\mu_{tot} = 17.3$) to correct the observed H$\beta$ intensity and the
effective mean magnification ($\mu_{eff} = 19$) to correct the observed Ly$\alpha$ intensity.
Interestingly, the Ly$\alpha$ escape fraction, $f\substack{Ly\alpha \\ esc}$, is $L\substack{obs \\ Ly\alpha}$/$L\substack{int \\ Ly\alpha}
= 0.061 (0.055)$ for $n_e = 10^2 (10^3)$ cm$^{-3}$, or only 6\%\ of Ly$\alpha$ emission escapes along the line of sight.

For a sample of local Ly continuum (LyC) emitters, \cite{verhamme17} found that 
both Ly$\alpha$ and LyC escape fractions increased with increasing Ly$\alpha$ EWs, 
while Ly$\alpha$ peak separations decreased.
Along these lines, the small Ly$\alpha$ peak separation (371 km s$^{-1}$) and large 
EW in SL2S0217 (113 \AA) argues for significant Ly$\alpha$ leakage \citep[$\sim40-60\%$;][]{verhamme17},
discordant with the observed estimated escape fraction. 
Note, however, that the Ly$\alpha$ escape fraction may be underestimated due to slit losses.
A preliminary analysis of the continuum-subtracted {\it HST} WFC3 F343N image suggests significant 
Ly$\alpha$ slit losses of roughly $30-40\%$.
Correcting the Ly$\alpha$ flux for this factor increases the Ly$\alpha$ escape fraction to 10\%,
still surprisingly small given the large observed Ly$\alpha$ EW.
A complete analysis will be discussed in a future paper (Erb et al.\ 2018, in prep.).

While SL2S0217 is a clear outlier to the trends found for the sample of LyC emitters in \citet{verhamme17}, 
\citet{jaskot17} recently reported an extreme Green Pea galaxy, J1608, with properties similar to SL2S0217.
In particular, these authors found J1608 to be a very high ionization (from [O~\iii]/[O~\ii]) galaxy, 
with no evidence of outflows, and measured a Ly$\alpha$ escape fraction of 0.16,
despite its strong Ly$\alpha$ EW of 163 \AA\ and narrow peak separation of $\Delta_{\mbox{\scriptsize peaks}} = 214$ km s$^{-1}$.
They suggest that multiple mechanisms for LyC escape exist, where such extreme targets may 
have suppressed superwinds, with radiation dominated feedback driving Ly$\alpha$ escape,
and likely escaping LyC emission.
Similarly, despite the low Ly$\alpha$ escape fraction measured for SL2S0217,
it has several characteristics, namely strong [O~\iii]/[O~\ii], large Ly$\alpha$ EW, and
small Ly$\alpha$ peak separation, that are indicative of LyC leakage \citep{henry15,izotov16,verhamme17}.
SL2S0217 is, therefore, a unique template to probe the conditions of extreme 
galaxies that may have played a critical role in the reionization of the Universe.

%----------------------------------------------------------------------------------------------------------------------
%----------------------------------------------------------------------------------------------------------------------

%%%% 5.4 %%%%
\subsection{Interstellar Absorption Lines}\label{sec5.4}
Outflowing gas, if present, can be directly probed by examining the interstellar absorption line profiles.  
Given sufficient spectral resolution and signal-to-noise in these features, 
a map of the covering fraction of the absorbing gas as a function of velocity 
can be inferred for both high- and low-ionization states. 
To better understand the gas kinematics in SL2S0217, the UV spectrum was 
normalized using the \citet{rix04} continuum windows as a guide, with further 
continuum designated by eye.

The normalized profiles of the significant absorption line detections 
are depicted in Figure~\ref{fig7}, arranged by ion.
Many of the strong, low-ionization 
absorption lines that are characteristic of $z\sim2-3$ galaxies \citep[e.g.,][]{pettini02b, shapley03} are 
also present in the SL2S0217 spectrum, but with ($\sim3\times$) narrower velocity ranges.
We can place constraints on the absorbing gas 
by measuring the properties of these profiles for various transitions.
Using a bootstrap Monte Carlo simulation in which the 1$\sigma$ uncertainty was used 
to perturb a given absorption profile and generate 1000 artificial spectra, we measured 
the flux-weighted centroid, velocity range, and integrated equivalent width for each line.
These values, along with the oscillator strengths for each line \citep{morton03}, are given 
in Table~\ref{tbl6}.
The integrated profiles are depicted by the filled absorption features in Figure~\ref{fig7}.

Measurements of two different lines from the same ion can be used to assess the optical depth quantitatively.
On the linear part of the curve of growth, $W\propto Nf\lambda^2$, where $W$ is the equivalent width,
$N$ is the column density, $f$ is the oscillator strength, and \W\ is the wavelength of a given transition.
Then, the theoretical ratio of the Si~\ii\ lines is $W_{\lambda1527}/W_{\lambda1808}$ = 45.2,
such that Si~\ii\ \W1527 is expected to be about 45 times stronger than Si~\ii\ \W1808 in the optically thin case.
Instead, the observed ratio of $W_{\lambda1527} / W_{\lambda1808} = 1.136$ \AA/ 0.271 \AA\ = 4.2
indicates line saturation and optically thick gas. 
Similar results are found for the low-ionization Fe~\ii\ \W\W1608,1611 lines
(theoretical $W_{\lambda1608} / W_{\lambda1611} = 41.7$ versus 2.7 observed), 
suggesting the Fe~\ii\ \W1608 line is strongly saturated. 
Deviations are also found for the high-ionization Si~\iv\ \W\W1394,1403 lines, but to a smaller degree,
as the observed ratio is about half of the theoretical ratio. 
Therefore, all but perhaps the weakest absorption lines in the UV SL2S0217 spectrum are saturated to varying degrees.
In this case, the equivalent width is dependent on both the velocity range and covering 
fraction of the absorbing gas, but  high resolution is needed to model the covering fraction 
as a function of velocity.
Given the low resolution of the UV SL2S0217 spectrum (FWHM$\sim$280 km s$^{-1}$ at 1500\AA),
a more detailed kinematic analysis of individual lines is unfeasible at this time.

\subsection{Average Absorption Profiles}\label{sec5.4.1}
\begin{figure}
\begin{center}
\includegraphics[scale=0.65, trim=5mm 5mm 65mm 0mm, clip]{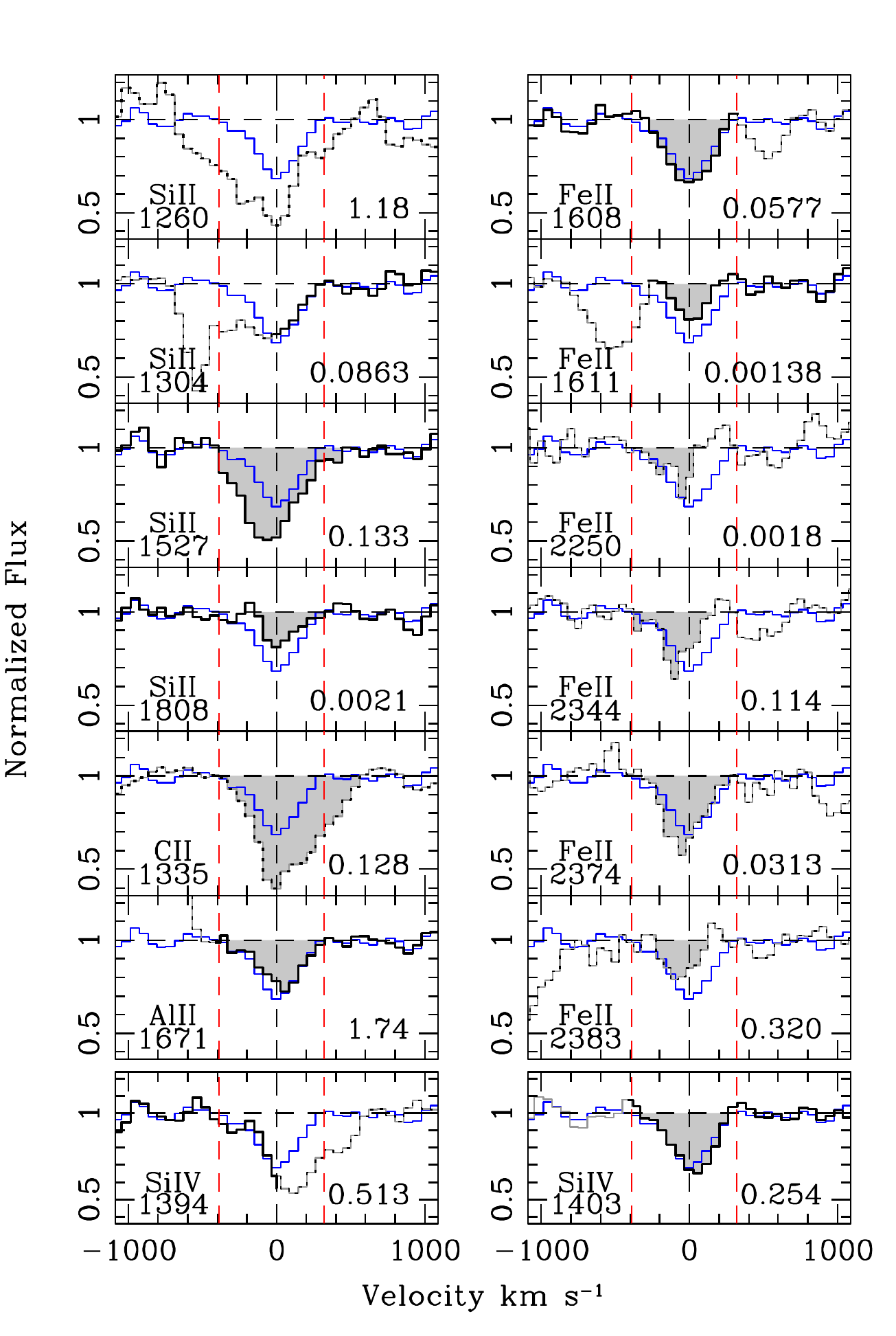}
\caption{Absorption features of both low and high ionization species in SL2S0217.
The blue line represents the average absorption profile, composed of the portions
of the spectra with thick, solid black lines.
The average profile is centered close to 0 km/s and is nearly symmetric, with velocity limits 
of [-392,+319] km/s (red, dashed vertical lines).
In comparison to the $f-$values (listed in the lower right-hand corner for each line),
the strongest lines must be saturated. }
\label{fig7}
\end{center}
\end{figure}

Differences between the average low- and high-ionization line profiles can inform gas properties. 
We determined error-weighted average velocity profiles from the ({\it i}) low-ionization, ({\it ii}) high-ionization,
and ({\it iii}) combined absorption features for comparison.
Given that the near-UV spectrum is significantly more affected by night sky contamination, 
and the Fe~\ii\ profiles may be affected by emission filling (see \S~\ref{sec5.5}), 
absorption features from the red side were not used in the average profile.
From the blue spectrum, the S~\ii\ $+$ Si~\ii\ \W1260 feature was excluded from 
the average due to its blended nature.
Far-UV regions affected by night-sky contamination were also omitted.
The combined average profile, therefore, incorporated the appropriate portions of 
O~\i\ \W1302, Si~\ii\ \W1304, 
Si~\iv\ \W1394, Si~\iv\ \W1403, Si~\ii\ \W1527, Fe~\ii\ \W1608, Fe~\ii\ \W1611, and Al~\ii\ \W1670,
indicated by the black profiles in Figure~\ref{fig7}.

The resulting profiles are nearly identical. 
In Figure~\ref{fig7}, the combined average profile (blue) is overplotted on the 
individual absorption features, where the portions of each line used
in the average are designated by a solid black line.
Apparently the low- and high-ionization gas has similar kinematic 
properties, as all the observed species seem to be nominally consistent 
with the average profile. 
Using a bootstrap Monte Carlo simulation in which the 1$\sigma$ uncertainty was used to perturb 
the average profile and generate 1000 artificial spectra, we measured a flux-weighted centroid of 
$-7.8$ km s$^{-1}$, with limits of $[-392,+319]$ km s$^{-1}$ (designated by red vertical lines in Figure~\ref{fig7}), 
and an equivalent width of 0.80 \AA.
Despite our reticence to interpret the gas kinematics (due to line saturation and low-resolution), 
the average absorption profile of SL2S0217 is clearly uncharacteristic in terms of both velocity 
centroid and range relative to interstellar absorption observed in other $z\sim2$ galaxies 
\citep[e.g.,][]{pettini00, quider10, erb10}. 
Rather, the average profile is characterized by nearly symmetric, narrow absorption 
around the systemic velocity, a signature of little-to-no outflowing gas.  

\begin{figure}
\includegraphics[scale=0.269, trim=24mm 0mm 10mm 20mm, clip]{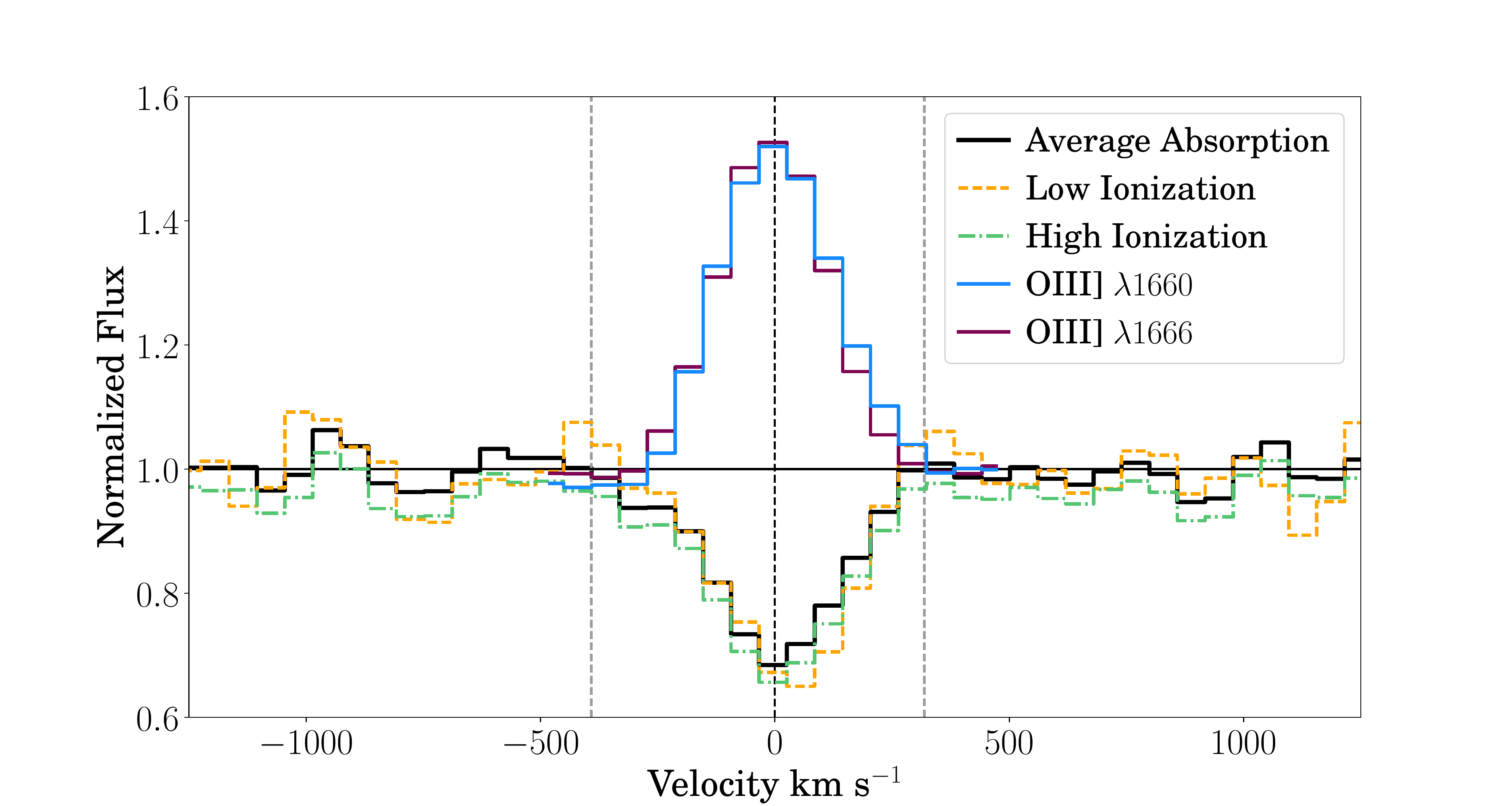}
\caption{Comparison of the emission and absorption velocity profiles in SL2S0217.
The strengths of the emission lines are arbitrarily scaled for visualization. 
An average profile was determined for the low-ionization species by combining the
individual profiles (see Figure~\ref{fig7}), whereas Si~\iv\ \W1403 alone
was used to represent the high-ionization profile.
Remarkably, the emission and absorption profiles are well-matched in terms of line width and velocity centroid. }
\label{fig8}
\end{figure}

In Figure~\ref{fig8} we compare the average velocity profiles of the low- and 
high-ionization absorption lines to the O~\iii] emission features present in the UV spectrum of SL2S0217.
Evidently, the absorption velocity profiles are remarkably similar to those of the emission lines,
whose widths represent the instrumental profile, providing a vivid demonstration of the lack of outflows in SL2S0217.

%----------------------------------------------------------------------------------------------------------------------
%----------------------------------------------------------------------------------------------------------------------

% Figure 7:  Absorption Line Spectra:

%----------------------------------------------------------------------------------------------------------------------

% Figure 8:  Vel Profile Comparison:

%----------------------------------------------------------------------------------------------------------------------
%----------------------------------------------------------------------------------------------------------------------

% Section 5.5:
\subsection{Fine Structure and Resonant Emission Lines}\label{sec5.5}

Fine structure lines, such as Si~\ii* \W\W\W1265,1309,1533, C~\ii* \W1335, 
and Fe~\ii* \W\W2365,2396, are clearly visible and redshifted in other $z\sim2$ emission-line 
galaxies such as BX418 \citep[emission;][]{erb10} and cB58 \citep[absorption;][]{pettini00}, 
but are not detected in the UV spectra of local starbursts \citep[e.g.,][]{schwartz06, leitherer11}.
Additionally, previous studies have found correlations between strong Fe~\ii* features and/or Mg~\ii\ \W\W2796,2803 
resonant emission and galaxies with extreme properties similar to SL2S0217, namely, high sSFRs, 
low stellar masses, blue UV slopes, and large Ly$\alpha$ equivalent widths \citep[e.g.,][]{erb12, jones12, kornei13}.
Based on these properties alone, we might have expected to see Fe fine structure or Mg resonant 
emission in our spectrum of SL2S0217, however, we find no evidence of these features.
On the other hand, from a sample of 184 $z\sim1$ star-forming galaxies, \citet{du17} find that 
Fe~\ii* strength does not vary significantly with stellar mass or color.
While the significance of these trends is still a matter of debate, the fact that we do not significantly 
detect any of the Fe~\ii* or Mg~\ii\ features is interesting.

Previous attempts to reproduce the fine structure emission lines using {\sc cloudy} photoionization models have been 
unsuccessful, suggesting that these lines are not likely produced by nebular photoionization \citep{shapley03, erb10}.
\citet{erb10} measure a consistent redshift of the fine structure lines in the $z = 2.3$ galaxy Q2343-BX418 by 
$\sim200$ km s$^{-1}$ and suggest an origin in the outflowing gas, potentially through resonant scattering, 
but caution that this theory is incomplete as it does not explain the narrowness of the observed lines.
\citet{rubin11} proposed that Fe~\ii* emission, observed at or near the systemic velocity,
arises from photon scattering in outflowing gas, and so may trace the spatial extent of the outflows.
This scenario is supported by the nearly systemic mean velocity centroid for the sample of 96 
$z\sim1-2$ star-forming galaxies analyzed by \citet[$\Delta v_{2626} = -20$ km s$^{-1}$;][]{erb12}.

The absence of Fe~\ii\ fine-structure features in SL2S0217 could then be due, in part, 
to the lack of outflowing gas observed along our line of sight
(see discussion of the velocity structure of the absorption lines in Section~\ref{sec5.4}).
Nevertheless, if scattering in outflows is in fact the production mechanism responsible for Fe~\ii* emission,
then extended emission in transverse outflows, and subsequent slit loses, are possible.
Note, however, that these weak features may simply not be detectable given 
the low resolution and signal-to-noise of the LRIS spectrum.

Galactic outflow models have shown that the overall profiles of the Fe~\ii\ and Mg~\ii\ lines can 
be significantly altered by the effects of photon scattering and re-emitted photons \citep{scarlata15}.
Emission filling, where the re-scattered emission ``fills in" a portion of the absorption profile,
may then be responsible for the bluer velocity centroids and reduced equivalent widths 
of many of the near UV Fe~\ii\ profiles depicted in Figure~\ref{fig7} and listed in Table~\ref{tbl6} \citep[see, also,][]{erb12}.
The effects of re-scattered emission may be even greater for Fe~\ii\ \W2383, whose
only allowed transition after absorption is back to the ground state. 

Mg~\ii\ absorption profiles may be filled by both resonance re-emission and nebular emission \citep[e.g.,][]{henry18}. 
From their composite spectra, \citet{du17} measure decreasing Mg~\ii\ absorption ($-7$ \AA\ $< W_{0,\mbox{\footnotesize Mg~\ii}} < -1$ \AA)
with increasing C~\iii] emission, attributing the effect to nebular emission filling. 
From the scatter in the LRIS spectrum, we estimate the upper limit of the Mg~\ii\ features to 
be $W_{0,\mbox{\footnotesize Mg~\ii}}\sim0.5$ \AA; only a small residual feature is possible, as 
expected for nebular and, possibly, significant re-scattered emission line filling in a target with little-to-no outflows.

%----------------------------------------------------------------------------------------------------------------------
%----------------------------------------------------------------------------------------------------------------------

% TABLE 6
\begin{deluxetable}{cccccc}
\tabletypesize{\scriptsize}
\tablewidth{0pt}
\setlength{\tabcolsep}{1pt}
\tablecaption{Interstellar Absorption Lines }
\tablehead{
\CH{Ion}					% 1
& \CH{\W$_{\mbox{lab}}$}		% 2
& \CH{\textit{f}-value}		% 5
& \CH{$v^a$}				% 6
& \CH{${\Delta v}^b$}		% 7
& \CH{${W_0}^c$} \\
\CH {} & \CH{(\AA)} & \CH{} & \CH{(km s$^{-1})$} & 	\CH{(km s$^{-1})$} & \CH{(\AA)}}		
\startdata
C~\ii		& 1334.5323$^d$ & 0.128	& 93		& -350:+525 	& 1.279$\pm$0.119		\\
Si~\iv	& 1402.7729	& 0.254	& 14		& -314+282 	& 0.511$\pm$0.063		\\
Si~\ii		& 1526.7070	& 0.133	& -37	& -432:+445 	& 1.136$\pm$0.102		\\
Fe~\ii	& 1608.4511	& 0.0577	& 4		& -332:+292 	& 0.643$\pm$0.072		\\
Fe~\ii	& 1611.2005	& 0.00138	& 6		& -197:+201	& 0.237$\pm$0.040		\\
Al~\ii		& 1670.7886	& 1.74	& 19		& -317:+342	& 0.489$\pm$0.088		\\ 
Si~\ii		& 1808.0129	& 0.0021	& 39		& -154:+310 	& 0.271$\pm$0.058		\\
Fe~\ii	& 2249.8768	& 0.0018	& -95	& -299:+61 	& 0.315$\pm$0.079		\\
Fe~\ii	& 2344.2189	& 0.114	& -83	& -323:+149 	& 0.567$\pm$0.074		\\
Fe~\ii	& 2374.4612	& 0.0313	& -27	& -303:+282 	& 0.927$\pm$0.107		\\
Fe~\ii	& 2382.7652	& 0.320	& -63	& -247:+124 	& 0.395$\pm$0.087		\\	
[1.0ex]
Ave.		& 			&		& -8		& -392:+319	& 0.805$\pm$0.050
\enddata
\tablecomments{
Measurements of the cleanest absorption profiles in SL2S0217.
Vacuum wavelengths and \textit{f}-values are taken from \citet{morton03}.
Columns 4 and 5 list the flux-weighted velocity centroids and velocity ranges over which the profiles were 
integrated in order to determined the equivalent widths given in Column 6.
Characteristics of the error-weighted average absorption profile are given in the final row for comparison. \\
$^a$ Velocity relative to $z_{\mbox{sys}}$ = 1.844 (see Table~\ref{tbl2}). \\ 
$^b$ Velocity range used for equivalent width measurements relative to line center (Column 2) at $z_{\mbox{sys}}$ = 1.844. \\
$^c$ Rest-frame equivalent width.  \\
$^d$ The unresolved blend of C~\ii\ \W1334.5323 + C~\ii* \W1335.6627.}
\label{tbl6}
\end{deluxetable}

%----------------------------------------------------------------------------------------------------------------------

% TABLE 7
\begin{deluxetable}{llcccc}
\tabletypesize{\scriptsize}
\tablewidth{0pt}
\setlength{\tabcolsep}{1pt}
\tablecaption{Interstellar Abundances}
\tablehead{
\CH{Ion}										% 1
& \CH{\W (\AA)}								% 2
& \CH{log$(N_{\mbox{\scriptsize HI}}$ cm$^{-2}$)}		% 4
& \CH{log(X/H)}									% 5
& \CH{log(X/H)$_{\odot}^b$}						% 6
& \CH{[X/H]$^c$}}								% 7
\startdata		
H~\i		& 1215	& 20.0$^a$ & -		& -		& - 		\\
Fe~\ii	& 1611	& 15.874	& -4.126	& -4.50	& 0.374 	\\
Si~\ii		& 1808	& 15.649	& -4.351	& -4.49 	& 0.139		
\enddata
\tablecomments{Interstellar abundances estimated from the weakest absorption features,
 assuming the optically thin case. \\
$^a$Value of log$(N_{\mbox{\tiny HI}}$ cm s$^{-1}$) predicted by homogeneous shell models with no expansion velocity,
$b = 10$ km s$^{-1}$, and $\Delta v = 371$ km s$^{-1}$ \citep{verhamme15}. \\
$^b$Solar (meteoric) abundances from the compilation by \citet{asplund09}. \\
$^c$[X/H] = log(X/H) - log(X/H)$_\odot$. }
\label{tbl7}
\end{deluxetable}

%----------------------------------------------------------------------------------------------------------------------
%----------------------------------------------------------------------------------------------------------------------

% SECTION 6
\section{Absorption Line Abundances}\label{sec6}

%----------------------------------------------------------------------------------------------------------------------
%----------------------------------------------------------------------------------------------------------------------

While most of the absorption lines in the UV spectrum of SL2S0217 are likely saturated,
the weakest lines offer the best opportunity to constrain relative abundances. 
Using the integrated equivalent widths from Table~\ref{tbl6}, we estimated the column density
for the weakest lines observed, assuming optically thin conditions (apparent optical depth method).
From \citet{spitzer78}, the linear part of the curve of growth can be characterized as 
\begin{equation}
	{N} = 1.13\times10^{20} \Bigg({\frac{W_{\lambda}}{\lambda^2 f}}\Bigg)\ (\mbox{cm}^{-2}),
\end{equation}
where $N$ is the column density, \W\ is the transition wavelength, and $f$ is the 
oscillator strength (see Table~\ref{tbl6}).
The resulting $N$ values are listed in Table~\ref{tbl7}; however, if any of the lines are 
saturated, the column densities determined here should be considered lower limits. 

Because the Ly$\alpha$ profile is seen purely in emission (see discussion in \S~\ref{sec5.3}), 
we do not have a measure of the total column density of neutral hydrogen. 
As discussed in \S~\ref{sec5.3}, the static, homogeneous radiative transfer shell 
models of \citet{verhamme15} predict a column density of neutral gas of 
$N_{\mbox{\scriptsize HI}} = 10^{20}$ cm$^{-2}$ for the Ly$\alpha$ peak separation in SL2S0217.
Adopting this value as indicative, we subsequently estimated the
element abundances for Si~\ii\ and Fe~\ii, as reported in Table~\ref{tbl7}.

Since Si~\ii\ and Fe~\ii\ both have lines with very weak and similar oscillator strengths,
assuming these lines are unsaturated,
we can estimate the Si~\ii/Fe~\ii\ ratio (independent of $N_{\mbox{\scriptsize HI}}$).
Further, Si~\ii\ and Fe~\ii\ have similar ionization potentials and are the only ionization
states that we observe in absorption for Si and Fe, 
so the ionization correction factor will be small, and, to first order, can be ignored. 
Using our measured values for SL2S0217 in Table~\ref{tbl7}, we find 
log(Si/Fe) $= -0.225\pm0.118$, or $44-77$\% solar.
However, Si is expected, both empirically and theoretically, to be {\it overabundant} relative to Fe in
metal-poor dwarf galaxies \citep[e.g.,][]{tolstoy09}.
This is because Si is predominantly an alpha element ($\alpha$), produced on
relatively short timescales during type II supernovae (SNe; massive stars) explosions, 
whereas Fe is synthesized by SNe Ia (intermediate binaries with mass transfer) and returned much later. 
Since SNe Ia have longer timescales than SNe II, Si and Fe trace different stellar populations and time scales.
Enhanced [$\alpha$/Fe] occurs until SNe Ia begin to contribute to the chemical evolution $10^8-10^9$ 
years after the first episode of star-formation. 

Given the derived properties of SL2S0217, one may have expected a pronounced 
alpha-enhancement in a galaxy experiencing a young burst of star formation.
Instead, the abnormal Si/Fe ratio measured for SL2S0217 could be evidence of an old 
stellar population that has had sufficient time to enrich the gas in Fe. 
Alternatively, the [$\alpha$/Fe]-poor gas could be the result of larger Si than Fe depletion onto dust grains,
although the authors of this work are not aware of any empirical evidence to support this scenario. 
The UV emission line ratios indicate that the Si depletion may be as great as 65\% in SL2S0217 (see \S~\ref{sec7.2.3}).
Unfortunately, we cannot estimate the Fe depletion from emission lines with existing data. 
The Si abundance and dust depletion are further discussed in \S~\ref{sec7.2.3}. 

%----------------------------------------------------------------------------------------------------------------------
%----------------------------------------------------------------------------------------------------------------------

% SECTION 7:
\section{Nebular Emission Line Abundances }\label{sec7}

%----------------------------------------------------------------------------------------------------------------------
%----------------------------------------------------------------------------------------------------------------------

% Section 7.1:
\subsection{Physical Parameters of the ISM }\label{sec7.1}
Directly measuring abundances from nebular emission lines requires knowledge of the 
physical properties of the emitting gas.
While detailed observations of nearby star-forming regions reveal complex nebular 
structures \citep[e.g.,][]{pellegrini12}, simplified \ion{H}{2} region models leverage a spherical 
geometry with three separate ionization volumes.
The advantage of such models is that they account for the fact that the low-, intermediate-, 
and high-ionization zones are governed by different physical properties, and therefore require 
reliable electron temperature and densities for each volume to determine accurate abundances. 
Although the SL2S0217 arc is evidently composed of multiple star-forming regions, they are not 
easily disentangled from the UV spectrum, and so we employ a single \ion{H}{2} region model to 
estimate the nebular gas properties. 
The electron temperature and electron density of SL2S0217 were determined
using the P{\sc y}N{\sc eb} package in {\sc python} \citep{luridiana12, luridiana15}, assuming a 
five-level atom model \citep{derobertis87}. 

%----------------------------------------------------------------------------------------------------------------------

% Section 7.1.1:
\subsubsection{Electron Temperature}\label{sec7.1.1}
Electron temperature is typically determined by observing a temperature-sensitive 
auroral-to-strong-line ratio.
Historically, the [O~\iii] \W4363/[O~\iii] \W5007 emission line ratio has been the 
ideal measure of electron temperature in the high-ionization zone of nebulae.
However, the temperature-sensitive [O~\iii] \W4363 auroral line is not resolved in the {\it HST} grism spectrum.

Fortunately, the electron temperature can also be determined from the O~\iii] \W1666/[O~\iii] \W5007 ratio, 
as is commonly done in high redshift targets where the intrinsically faint optical auroral line is often undetected. 
In the case of SL2S0217, this diagnostic combines space- and ground-based 
observations, potentially introducing flux and aperture mismatching issues, and so this calculation must be done carefully.
Given the excellent flux calibration of the {\it HST} grism spectrum and the match of the  
flux-corrected LRIS continuum to the {\it HST} ACS F606W AB magnitude, 
we used the O~\iii] \W1666/[O~\iii] \W5007 ratio to measure $T_{e,\lambda1666} = 15,400\pm200$ K.
Adopting this $T_{e,\lambda1666}$ measurement for the high ionization zone electron temperature,
the intermediate- and low-ionization zone temperatures ($T_e$ [S~\iii] and $T_e$ [O~\ii] respectively)
were then determined from the theoretical temperature relationships of \citet{garnett92}.
The temperatures used for each ionization zone are listed in Table~\ref{tbl8}.

Interestingly, \citet{brammer12} reported excess emission from the H$\gamma$+[O~\iii] \W4363 
blend in the grism spectrum, which can be attributed to [O~\iii] \W4363 and used to constrain the 
optical electron temperature diagnostic.
We determined the H$\gamma$ flux contribution to the blend using the measured 
H$\beta$ grism flux ($74\pm2\times10^{-17}$ erg s$^{-1}$ cm$^{-2}$) and the Case-B 
H$\gamma$/H$\beta$ value from \citet{hummer87} assuming $T_e =$ 15,000 K and $n_e = 100$ cm$^{-3}$
(conditions representative of SL2S0217; see Section~\ref{sec7.1.2} for $n_e$).
Subtracting this H$\gamma$ value from the blend ($40\pm3\times10^{-17}$ erg s$^{-1}$ cm$^{-2}$; B12), 
the flux corresponding to [O~\iii] \W4363 is $F_{\lambda4363} = 4.9\times10^{-17}$ erg s$^{-1}$ cm$^{-2}$.
We use the standard [O~\iii] \W4363/[O~\iii] \W5007 temperature diagnostic to estimate the 
high-ionization zone electron temperature to be $T_{e,\lambda4363} = 15,700\pm3,200$ K,
in agreement with the combined UV/optical temperature diagnostic.
A higher signal-to-noise and resolution rest-frame optical spectrum would be useful
to more securely measure [O~\iii] \W4363 and compare the two temperature diagnostics. 

%----------------------------------------------------------------------------------------------------------------------

% Section 7.1.2:
\subsubsection{Electron Density}\label{sec7.1.2}
In nearby objects, the gas phase electron density is most commonly determined
from the [S~\ii] and [O~\ii] optical collisionally-excited doublets. 
For SL2S0217, the [O~\ii] lines were not detected and the [S~\ii] lines were outside the
wavelength range covered by the {\it HST} grism spectrum.
Instead, we observe two sets of density-sensitive UV emission-line doublets in our LRIS spectrum: 
Si~\iii] \W\W1883,1892 and C~\iii] \W\W1907,1909.
We calculate densities of $4,500\substack{+1,500 \\ -1,400}$ cm$^{-3}$ and $300\substack{+1,300 \\ -300}$
cm$^{-3}$ for the Si~\iii] and C~\iii] ions respectively\footnotemark[7].
Since the relevant lines are all high S/N and should result purely from nebular emission,
we find no obvious explanation for this discrepancy.

\begin{figure}
\includegraphics[scale=0.375,  trim=0mm 0mm 0mm 0mm, clip]{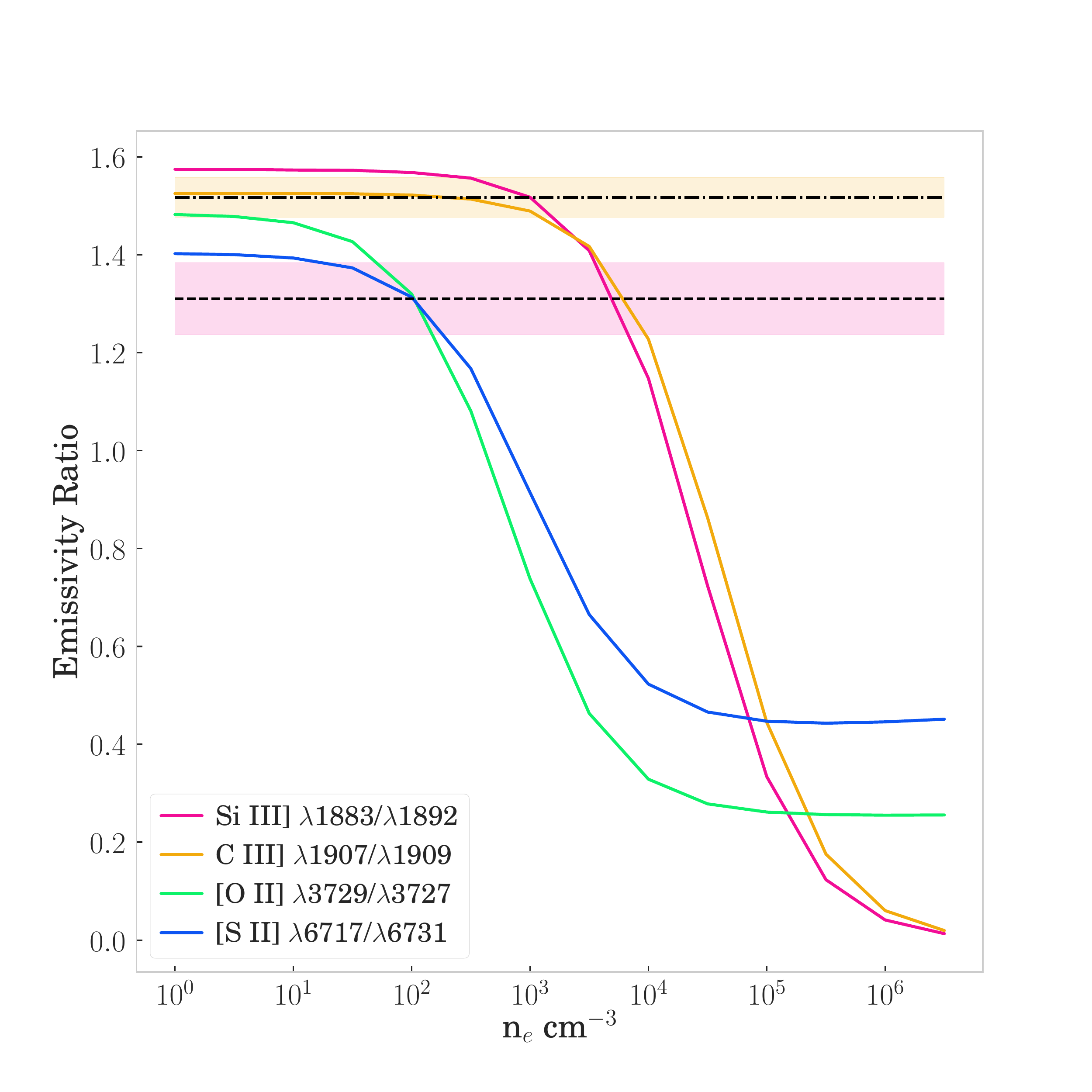}
\caption{
A comparison of common rest-frame UV (Si~\iii] and C~\iii]) and optical ([O~\ii] and [S~\ii])
electron density diagnostic ratios.
Emissivities were calculated using the PyNeb package in Python, assuming an electron 
temperature of 15,000 K. 
The dashed (dashed-dotted) black line and pink (orange) box represent the Si~\iii] (C~\iii])
ratio and uncertainty for SL2S0217. 
Unfortunately, neither C~\iii] or Si~\iii] are sensitive to low densities, so we simply assume
a typical nebular density of $n_e = 100$ cm$^{-3}$.}
\label{fig9}
\end{figure}

The critical densities of C~\iii] and Si~\iii] are both of the order of 10$^4$ cm$^{-3}$,
and so they are generally not useful density diagnostics for $n_e$ values below $\sim10^3$ cm$^{-3}$.
This is evident in Figure~\ref{fig9}, where we plot the emissivity ratios of common rest-frame UV (Si~\iii] and C~\iii]) and 
optical ([O~\ii] and [S~\ii]) electron density diagnostic ratios, assuming a general electron temperature of 15,000 K. 
When measured, the optical line ratios provide constraints below $n_e$ $\sim10^3$ cm$^{-3}$,
however, they probe a different (low-) ionization zone, and indicate significantly lower densities 
relative to the highly-ionized gas in high-redshift star-forming galaxies \citep[e.g.,][]{christensen12, bayliss14, james14}.
Traditionally, the high-ionization zone ($\sim35-55$ eV) is represented by O$^{++}$,
whereas the low-ionization zone ($\sim15-35$ eV) is represented by the O$^+$ or N$^+$ ions,
with the intermediate-zone, based on S$^{++}$, partially overlapping ($\sim23-35$ eV).
By these definitions, the UV Si~\iii] and C~\iii] density diagnostics span the low-to-intermediate and
intermediate-to-high ionization zones respectively. 
Perhaps not surprisingly, the mixed-zone electron densities measured for high-redshift targets, including SL2S0217, 
are significantly larger than the pure low-ionization zone densities typical of \ion{H}{2} regions in nearby galaxies.
Further, \citet{sanders16} find that $z\sim2$ galaxies have mean electron densities that are an order of 
magnitude higher relative to local galaxies at fixed stellar mass.

While the insensitivity of the UV density diagnostics to low densities inhibits the density 
determination for SL2S0217, the C~\iii] ratio is at least in agreement with the low-density 
regime, and so we assumed a standard value of $n_e = 100$ cm$^{-3}$.
Fortunately, the gas temperature and line emissivities are practically independent of the gas
density in the low-density regime, and so this choice of density makes little difference.
In fact, subsequently derived abundance ratios for both $n_e = 100$ cm$^{-3}$ 
(consistent with the $n_e$ from C~\iii]) and $n_e = 4,500$ cm$^{-3}$ (from Si~\iii]), given in Table~\ref{tbl8}, 
show differences that are significantly smaller than the respective uncertainties.
However, detailed studies of these diagnostics must be prioritized to understand whether
targets such as SL2S0217 have abnormally high electron densities/density inhomogeneities, 
the UV diagnostics probe a higher density regime than their optical counterparts, or some other 
effect is at play.

\footnotetext[7]{We note that a preliminary Si~\iii] density determination using TEMDEN in IRAF 
produced unphysical results for SL2S0217. 
We subsequently found that the emissivity ratios for the Si~\iii] density diagnostic approach an unexpected value of 3 
using the default atomic data in IRAF. 
Alternatively, we find that the PyNeb package in Python uses the correct Si~\iii] emissivities.}

%----------------------------------------------------------------------------------------------------------------------

% Figure 9: Density Ratios:

%----------------------------------------------------------------------------------------------------------------------

\subsection{Ionic And Total Abundances}\label{sec7.2}
Due to the number of strong rest-frame UV emission lines observed in the spectrum of 
SL2S0217, we were able to determine relative C, O, N, and Si abundances.  
Using the rest-frame optical grism spectrum, ionic abundances were also determined 
relative to hydrogen:
\begin{equation}
	{\frac{N(X^{i})}{N(H^{+})}\ } = {\frac{I_{\lambda(i)}}{I_{H\beta}}\ } {\frac{j_{H\beta}}{j_{\lambda(i)}}}.
	\label{eq:Nfrac}
\end{equation}
The emissivity coefficients, $j_{\lambda(i)}$, which are functions of both 
temperature and density, were determined using the five-level atom approximation 
and updated atomic data reported in \citet{berg15}. 

Typically, the total oxygen abundance (O/H) in an \ion{H}{2} region is calculated by simply
summing the O$^{+}$/H$^{+}$ and O$^{+2}$/H$^{+}$ ionic abundances, as contributions
from O$^{+3}$/H$^{+}$ (requiring an ionization energy of 54.9 eV) are thought to be negligible.
Although no emission is seen for the strongest UV O~\iv\ transitions ($\lambda\lambda1401,1404$)
in the LRIS spectrum, the presence of He~\ii\ from recombination (requiring an ionization energy of 54.4 eV)
makes this argument less secure.
To account for possible contributions from O$^{+3}$, we measured the 3$\sigma$ upper limit on the flux
at \W\W1401,1404 to be $\leq 3.3\times10^{-18}$ erg s$^{-1}$ cm$^{-2}$.
Comparing this value to the O~\iii] \W\W1661,1666 line strengths, 
we find a marginal oxygen ionization correction factor (ICF) of 1.055.

For SL2S0217 we can measure O$^{+2}$/H$^{+}$ from the [O~\iii] \W\W4959,5007/H$\beta$ 
optical emission line ratio, however the corresponding lines for O$^{+}$/H$^{+}$ are not 
detected in the optical grism spectrum. 
In order to estimate 12+log(O/H) from the optical spectrum, we used the observed [O~\iii] \W5007 
emission and results from photoionization models (see Section~\ref{sec8.1}) 
to predict the [O~\ii] \W3727 emission line flux.
Based on these models, O$^+$/H$^+$ only contributes 2\% to the total oxygen 
abundance (and O$^{0}$ is truly negligible), but such a small contribution is 
expected at the extreme ionization level of SL2S0217.
Therefore, we find SL2S0217 to be an extremely metal poor (EMP; 12+log(O/H) $<7.7$) galaxy,
with an oxygen abundance, corrected for the unseen O$^{+}$/H$^{+}$ and O$^{+3}$/H$^{+}$ contributions,
with 12+log(O/H) $=7.50$. 
The ionic and total oxygen abundances are listed in Table~\ref{tbl8}.

%----------------------------------------------------------------------------------------------------------------------
%----------------------------------------------------------------------------------------------------------------------

% TABLE 8
\begin{deluxetable}{lrl}
\tabletypesize{\scriptsize}
\tablewidth{0pt}
\setlength{\tabcolsep}{3pt}
\tablecaption{Ionic and Relative Nebular Abundances }
\startdata
\hline\hline \\
$n_{e}$ used					& 100 cm$^{-3}$ 						& 4,500 cm$^{-3}$ 			\\
\hline \\
{$T_e$ [O~\iii]}$_{opt}^a$			& \multicolumn{2}{c}{15,700$\pm$3,200 K}							\\
{$T_e$ [O~\iii]}$_{comb}^b$ used	& \multicolumn{2}{c}{15,400$\pm$200 K}								\\
{$T_e$ [S~\iii]}	used				& \multicolumn{2}{c}{14,500$\pm$200 K}								\\
{$T_e$ [O~\ii]}	used				& \multicolumn{2}{c}{13,800$\pm$200 K}								\\
$n_{e}$ C~\iii] measured			& \multicolumn{2}{c}{$300\substack{+1,300 \\ -300}$ cm$^{-3}$	}			\\
$n_{e}$ Si~\iii] measured			& \multicolumn{2}{c}{$4,500\substack{+1,500 \\ -1,400}$ cm$^{-3}$} 			\\
\\
O$^+$/H$^+$$^{a,c}$			& $5.21\times10^{-7}$ 				& $8.77\times10^{-7}$			\\
O$^{+2}$/H$^+$$^a$ 			& $3.14\pm0.12 \times10^{-5}$ 			& $3.13\pm0.12 \times10^{-5}$ 		\\
ICF$^d$						&  \multicolumn{2}{c}{$1.055$}										\\ 
12 + log(O/H)$^a$				& $\geq 7.50^e$					& $\geq 7.51^e$				\\
\\
C$^{+3}$/C$^{+2}$				& $0.86$							& $0.86$						\\
C$^{+2}$/O$^{+2}$				& $0.122\pm0.005$					& $0.124\pm0.005$				\\
log U							&  \multicolumn{2}{c}{$-1.50$}										\\
ICF							&  \multicolumn{2}{c}{$1.27\pm0.27$}								\\ 
log(C/O)						& $-0.81\pm0.09$					& $-0.80\pm0.09$				\\
\\
N$^{+2}$/O$^{+2}$				& $0.033\pm0.062$ 					& $0.029\pm0.054$				\\
log(N/O)						& $-1.48\pm0.46$					& $-1.53\pm0.45$				\\
\\
Si$^{+2}$/C$^{+2}$				& $0.072\pm0.002$					& $0.073\pm0.002$				\\
ICF							&  \multicolumn{2}{c}{2.31$\pm$0.32}								\\
log(Si/C)						& $-0.76\pm0.09$					& $-0.77\pm0.09$				\\
log(Si/O)						& $-1.59\pm0.17$					& $-1.58\pm0.17$							
\enddata
\tablecomments{
Physical conditions and abundances derived for the nebular gas in SL2S0217
assuming either $n_e = 100$ cm$^{-3}$ (consistent with the $n_e$ from C~\iii]) 
or $n_e = 4,500$ cm$^{-3}$ (from Si~\iii]).
Ionic abundances are reported for both the optical and UV species when available.  \\
$^a$Determined from the {\it HST} optical grism spectrum values. \\
$^b$Determined from the combined UV and optical \W1666/\W5007 line ratio. \\
$^c$The [O~\ii] \W3727 emission line intensity was inferred from the observed [O~\iii] \W5007
flux and the [O~\iii] \W5007/[O~\ii] \W3727 ratio from {\sc cloudy} photoionization modeling. \\
$^d$ The oxygen ICF accounts for unseen O$^{+3}$, estimated from the 3$\sigma$ flux upper limit of O~\iv] \W\W1401,1404. \\
$^e$ Because we do not detect [O~\ii], the estimated oxygen abundance is reported as a lower limit.
However, the O$^+$ contribution is expected to be small at such high ionization.}
\label{tbl8}
\end{deluxetable}

%----------------------------------------------------------------------------------------------------------------------

% Section 7.2.1:
\subsubsection{C/O Abundance}\label{sec7.2.1}
Abundance determinations for other elements require ionization correction 
factors to account for their unobserved ionic species. 
The prominence of the high-ionization-potential He~\ii\ and C~\iv\ emission lines observed 
in SL2S0217 indicate that ICFs are particularly important for C. 
Although the relatively narrow emission profiles of the C~\iv\ doublet indicate the dominance of nebular emission, 
it is complicated by potential contributions from a range of stellar features, from pure absorption to P-Cygni profiles. 
To avoid this complication, we used the ICF presented in \citet{berg16} to estimate the 
C$^{+3}$ contribution at log U$=-1.50$ (our best model value; see Section~\ref{sec8.1}).
Despite the strong C~\iv\ emission present in the UV spectrum of SL2S0217, we determine
a modest C ICF = $1.27\pm0.27$.
The C/O abundance is then determined using
\begin{equation}
	{\frac{\mbox{C}}{\mbox{O}} = {\frac{\mbox{C}^{+2}}{\mbox{O}^{+2}\ } {\times{\ \mbox{ICF}}}}}
	\label{eq:CO}
\end{equation}
to be log(C/O) = $-0.81\pm0.03$, which is typical of star-forming dwarf galaxies 
with similar oxygen abundance \citep{garnett95, berg16}.

Carbon and oxygen are produced on different time scales, 
by different mass populations of stars, and so the C/O trend with
metallicity has been studied by many authors with observations for a variety of astrophysical objects,
and using a variety of chemical evolution models and stellar yields.
Oxygen is synthesized by SNe II (massive stars) and returned to the interstellar medium quickly, 
whereas carbon is primarily produced by He burning through the triple-$\alpha$ process,
a reaction that can occur in both massive ($M > 8 M_{\odot}$) and low to intermediate
mass ($1M_{\odot} < M < 8M_{\odot}$) stars.

\begin{figure*}
\begin{center}

\includegraphics[scale=0.95, trim=5mm 10mm 0mm 110mm, clip]{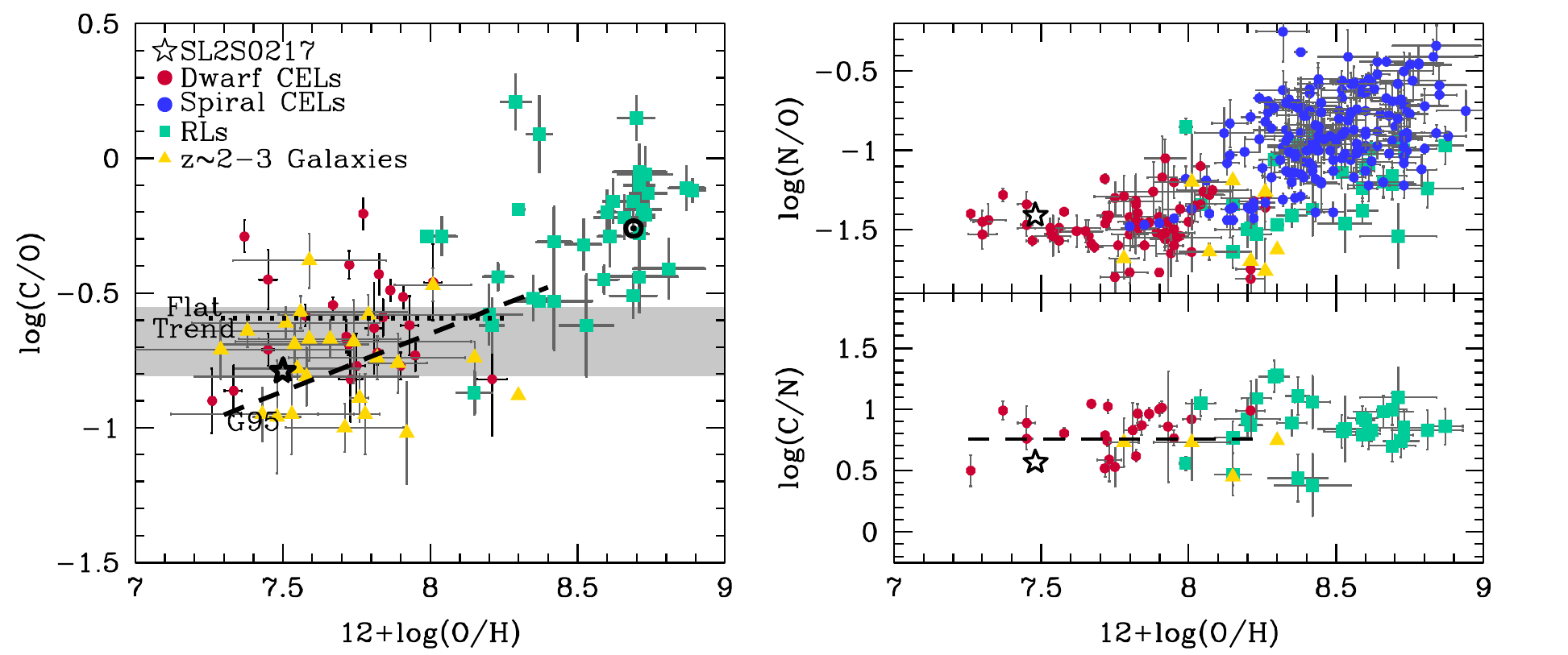}
\caption{
{\it Left:} 
Carbon to oxygen ratio vs. oxygen abundance of SL2S0217 (black star) in comparison to other star forming galaxies.
Comparison data of nearby galaxies is divided into two groups. 
Red circles are metal-poor dwarf galaxies as reported in \citet{berg16} and Berg et al.\ (2018, in prep.), and
green filled squares represent star-forming galaxies with larger oxygen abundances
determined from recombination lines \citep{esteban02, esteban09, esteban14, pilyugin05, garcia-rojas07, lopez-sanchez07}.
The dashed line is the least-squares fit from \citet[][G95]{garnett95} and the dotted line is the 
weighted mean of the significant CEL C/O detections (filled red circles). 
More distant ($z\sim2-3$) galaxies are represented as gold triangles 
\citep{pettini00, fosbury03, erb10, rigby11, christensen12, stark14, bayliss14, james14, vanzella16, steidel16, amorin17, rigby17},
including composite spectra from \cite{steidel16}, \citet{amorin17}, and \citet{rigby17}. 
The background shaded region demonstrates the value for the $z\sim3$ Lyman break galaxy composite spectrum 
determined by \citet{shapley03}. \\
{\it Right:}
Trends in relative nitrogen abundance versus oxygen abundance for galaxies spanning a range in redshift.
Additional direct abundances are plotted for nearby dwarf galaxies \citep{vanzee06a, berg12}
and for nearby spiral galaxies 
(blue points: the CHAOS project - \cite{berg15, croxall15, croxall16}; Berg et al.\ (2018), in prep.).
SL2S0217 appears to have a normal N abundance with respect to both nearby, metal-poor dwarf galaxies
and the few other $z\sim2-3$ galaxies with C/N observations. 
\label{fig10}}
\end{center}
\end{figure*}

Historically, C/O has been difficult to measure, but recently C/O has been measured in 
dozens of nearby, metal-poor, dwarf galaxies using the UV lines \citep[e.g.,][]{berg16, senchyna17}.
Similar to the early work of \citet{garnett95}, these studies find a general trend of increasing 
C/O with oxygen abundance, but with a significant unexplained scatter. 
These data can also be interpreted as a constant trend in C/O at low-metallicities. 
We plot SL2S0217 (open star) on this C/O versus O/H trend in the left panel of Figure~\ref{fig10},
where nearby dwarf galaxies are plotted as filled circles for comparison.
The trend is extended to higher oxygen abundances by incorporating C/O determinations 
from optical recombination lines \citep[filled squares:][]{esteban02, esteban09, esteban14, pilyugin05, garcia-rojas07, lopez-sanchez07}.

Significant UV C and O has also been measured for a handful of $z\sim2-3$ galaxies 
\citep[gold triangles in Figure~\ref{fig10}:][]{pettini00, fosbury03, erb10, christensen12, stark14, bayliss14, james14, vanzella16, steidel16, amorin17, rigby17}, 
amongst which the C/O ratio of SL2S0217 appears to be average.
Interestingly, C/O values for the $z\sim2-3$ galaxies are consistent with the nearby 
dwarf galaxies of similar metallicity, but potentially with a lower average value. 
The distinction between these populations of different ages may 
support the notion put forth by \citet{berg16} that higher C/O values may be 
due to a delayed pseudo-secondary carbon contribution from low- and intermediate-mass stars. 

%----------------------------------------------------------------------------------------------------------------------

% Figure: C/O vs O/H:

%----------------------------------------------------------------------------------------------------------------------

% Section 7.2.2:
\subsubsection{Relative N Abundance}\label{sec7.2.2}
We estimate the N/O ratio by combining the UV O~\iii] \W\W1661,1666 and 
C~\iii] \W\W1907,1909 emission lines and the N~\iii] \W1750 multiplet.
Since the N~\iii] \W1750 feature is only weakly detected in the SL2S0217 LRIS 
spectrum, it is appropriate to think of the following N/O determination as an upper limit. 
Utilizing the fact that the ionization potentials of N$^{++}$ and C$^{++}$ are nearly 
the same (47.448 eV and 47.887 eV respectively), we used the N$^{++}$/C$^{++}$ 
ion ratio to measure a relative N/C enrichment of log(N/C) $= -0.68\pm0.02$.
Then, combining this ratio with the ionization corrected C/O ratio (see Table~\ref{tbl8}),
we inferred a total N/O of log(N/O) $= -1.48\pm0.46$, which is consistent with the range of 
log(N/O) seen for local dwarf galaxies with similar oxygen abundance 
\citep[e.g.,][]{vanzee06a, berg12}.

Relative N/O abundances are well studied in nearby dwarf and spiral galaxies, 
providing a robust sample to compare measurements from more distant targets.
In the upper right panel of Figure~\ref{fig10} we compare the N/O of SL2S0217 to nearby dwarf 
(red points: \cite{vanzee06a, berg12, berg16}; Berg et al.\ 2018, in prep.) and spiral galaxies 
(blue points: the CHAOS project - \cite{berg15, croxall15, croxall16}; Berg et al.\ 2018, in prep.), 
as well as nearby galaxies with oxygen 
abundance measurements from recombination lines (green squares).
SL2S0217 lies amongst the N/O range of both nearby dwarf galaxies and the handful of 
$z\sim2-3$ galaxies with N/O and O/H measurements (gold triangles).
Note that N/O value for SL2S0217 falls on the plateau expected for primary (metallicity-independent) 
N production, as expected for a low oxygen abundance of 12+log(O/H)$\sim7.5$ \citep[see, for example,
the discussion in][]{pettini02a,pettini12}.

In the bottom half of Figure~\ref{fig10} (b) we examine the C/N ratio, 
noting the remarkably flat trend for both nearby and distant galaxies.
\citet{berg16} found that the constant ratio of C/N seen over a large range in O/H for nearby galaxies 
may indicate that carbon is predominantly produced by similar nucleosynthetic mechanisms as nitrogen \citep[also see][]{steidel16}.
However, a larger sample of C/N measurements in distant galaxies is needed to determine whether the 
hypothesis of \citet{berg16} also applies at $z\sim2-3$, or whether different nucleosynthetic 
processes dominate C and N. 

Recently, observations of the standard rest-frame optical emission-line 
[O~\iii]/H$\beta$ versus [N~\ii]/H$\alpha$ diagnostic plot \citep[the BPT diagram;][]{baldwin81}
has revealed a pronounced offset of $\sim2-3$ galaxies from the main sequence 
of local star forming galaxies \citep[e.g.,][]{masters14, steidel14, shapley15}.
Several studies have investigated the physical origin of this offset by analyzing the 
BPT parameter space in terms of other physical properties for large surveys of local galaxies.
Using $\sim100,000$ galaxies from the Sloan Digital Sky Survey Data Release 12, \citet{masters16}
found that the region of the BPT occupied by $\sim2-3$ galaxies corresponds to galaxies with
elevated SFR surface densities, stellar masses, and N/O ratios, concluding that the higher
N/O ratios at fixed [O~\iii]/H$\beta$ are the proximate cause of the offset.
In contrast, the sample of $z\sim2-3$ galaxies plotted in Figure~\ref{fig10} occupy the 
same range in N/O as the local comparison sample. 
SL2S0217, in particular, has an average N/O abundance relative to the local dwarf galaxies,
however, this value is an upper limit estimated from the weak detection of the N~\iii] \W1750 line.
Further, despite the unusual strength of its UV and optical emission lines, the [O~\iii]/H$\beta$ 
and [N~\ii]/H$\alpha$ (inferred from photoionization modeling; \S~\ref{sec8.1.1}) ratios  
of SL2S0217 indicate it may not coincide with the offset region of the BPT. 
Clearly there is still much to understand about the $z\sim2-3$ BPT diagram.

%----------------------------------------------------------------------------------------------------------------------
%----------------------------------------------------------------------------------------------------------------------

% Figure 11:  Si ICF:
\begin{figure}
\includegraphics[scale=0.775, trim=0mm 22mm 0mm 35mm, clip]{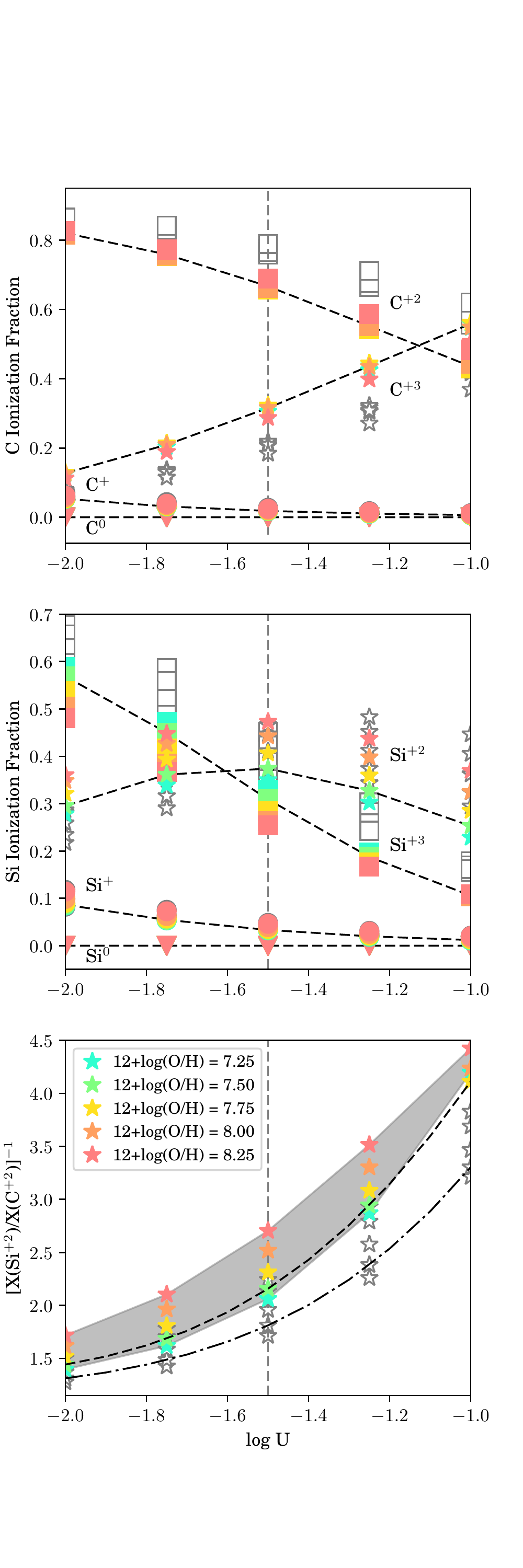}
\caption{
Ionization fraction of C and Si species as a function of ionization parameter from 
{\sc cloudy} models of Z$_{neb} = 0.05$ Z$_{\odot}$ (Z = 0.001) BPASS stellar models. 
Two different ages of instantaneous bursts are shown, demonstrating the limits of the 
parameter space explored: $t=10^{6.4}$ and $10^{7.0}$ yrs.
For the older model, symbols are color coded by the gas-phase oxygen abundance,
whereas the younger model is depicted by gray symbols.
In the top two panels the different ionic species are designated by triangles, circles, 
squares, and stars in order of increasing ionization. 
The bottom panel plots the ionization correction factor (ICF) vs. ionization parameter, where the 
dashed and dashed-dotted curves indicate the ICF for the 12+log(O/H) = 7.5 model at an age of 
$10^{6.4}$ and 10$^{7.0}$ yrs respectively and the gray shading highlights the range spanned
by metallicity.
The gray vertical dashed line marks the predicted ionization parameter for SL2S0217.}
\label{fig11}
\end{figure}

%----------------------------------------------------------------------------------------------------------------------
%----------------------------------------------------------------------------------------------------------------------

% Section 7.2.3:
\subsubsection{Relative Si Abundance}\label{sec7.2.3}
Silicon abundances relative to carbon were determined following the logic presented in \citet{garnett95b}:
The Si~\iii] and C~\iii] doublets arise from comparable energy levels ($\sim$ 6.57 and 6.50 eV respectively),
and therefore the Si/C abundance ratio is relatively insensitive to uncertainties in the electron temperature.
Additionally, Si/C depends little on the assumed density as both Si~\iii] and C~\iii] have very high critical densities, 
meaning their volume emissivities have a similar dependence on density.
Finally, the proximity of the Si~\iii] and C~\iii] emission lines in wavelength 
eliminates strong effects of differential extinction. 

The Si~\iii] and C~\iii] emission line features are extremely prominent in SL2S0217;
we measured their strengths with S/N $>$ 20 (see Figure~\ref{fig5}).
As described in Section~\ref{sec7.1.2}, we assume the ionized gas of SL2S0217 is in the
low density limit ($< 10^3$ cm$^{-3}$), where collisional de-excitation does
not significantly impact the Si$^{+2}$/C$^{+2}$ ratio derived from the UV lines.
Because C$^{+2}$ has a larger ionization potential than Si$^{+2}$ (47.9 and 33.5 eV respectively),
we expect a larger fraction of Si to be in a higher, unobserved ionization state than C, 
and so an ICF is needed to determine the Si/C abundance.
We determine the Si/C element abundance using the equation
\begin{equation}
	{\frac{\mbox{Si}}{\mbox{C} }} = {\frac{\mbox{Si}^{+2}}{\mbox{C}^{+2}}\ } \Bigg[{\frac{X(\mbox{Si}^{+2})}{X(\mbox{C}^{+2})}}\Bigg]^{-1} 
			     			    = {\frac{\mbox{Si}^{+2}}{\mbox{C}^{+2}}\ }{\times{ \mbox{ICF}}},	
\end{equation}
where X(Si$^{+2}$) and X(C$^{+2}$) are the Si$^{+2}$ and C$^{+2}$ volume fractions, respectively.
The ICF is modeled as a function of the ionization parameter using the photoionization code 
{\sc cloudy} \citep{ferland13}, with BPASS models as the input ionizing radiation field (see Section~\ref{sec8.1}).
We sought simple models that reproduced the conditions in SL2S0217 as closely as possible;
see Section~\ref{sec8.1} for a detailed discussion.

The ionization fraction of Si and C species as a function of ionization 
parameter are shown in Figure~\ref{fig11}.
As we will show in Section~\ref{sec8.1.1}, we can reproduce many aspects of 
the observed spectrum using {\sc cloudy} models of a stellar population with 
binaries and an ionization parameter of log U$ = -1.5$; this value is reasonable 
given the very hard ionizing radiation field predicted by the SED modeling of B12.
We estimate the uncertainty in the ICF as the scatter amongst the different
models considered at a given C$^{+2}$ volume fraction.
The resulting Si ICF~$= 2.31\pm$0.32 for 12+log(O/H)~$=7.5$ and 
$t=10^{7.0}$ yrs is significant, correcting the relative Si abundance 
from log(Si/C)$= -1.14$ to log(Si/C)~$= -0.76\pm$0.09.
When combined with the C/O abundance derived for SL2S0217, 
we estimate log(Si/O)~$= -1.59\pm$0.17.
The relevant Si abundance values are reported in Table~\ref{tbl8}.

Due to the need of far-UV spectroscopic observations of nearby galaxies, 
there exists a general paucity of nebular measurements of Si/O to compare to.
Perhaps the best-known study of Si/O comes from \citet{garnett95b}, who measured the 
UV Si~\iii] emission lines for seven extragalactic \ion{H}{2} regions
using the Faint Object Spectrograph on {\it HST}. 
These authors found Si/O to be constant over the observed range in oxygen abundance,
with a weighted mean value of log(Si/O)$=-1.59\pm0.07$. 
Our log(Si/O)$=-1.59\pm0.17$ measurement for SL2S0217 is remarkably consistent with
this average from the \citet{garnett95b} study, but indicates significant Si/O enrichment
or a reduced Si depletion onto dust relative to the $z\sim2.4$ galaxies
of \citet[][log(Si/O) = -1.81$\pm$0.10]{steidel16}. 

Note that our Si/O value does not account for the relative depletion of Si and O onto dust grains.
While one might expect this effect to be small given the low predicted dust reddening in SL2S0217 and the 
probability of grain destruction or erosion in the presence of a hard ionizing radiation field, our Si/O
value is significantly lower than the solar ratio of log(Si/O)$_{\odot} = -1.15$ \citep{jenkins09}.
Many studies argue that both Si and O are predominantly products of massive star nucleosynthesis, 
and so their ratio is expected to vary little with time \citep[e.g.,][]{timmes95}.
Note, however, that some studies find Si production can have a substantial contribution from 
Type Ia SNe outbursts \citep[c.f.,][]{matteucci86, kobayashi06}.
Since oxygen doesn't deplete as strongly, the observed underabundance can likely be 
attributed to the depletion of Si onto dust \citep{steidel16}.

We estimated the logarithmic depletion of Si in the ISM following the prescription laid out in \citet{jenkins09}\footnotemark[8]:
\begin{equation}
	[{\mbox{Si}}/{\mbox{H}}]_{\mbox{\scriptsize gas}} = [\mbox{Si}/\mbox{H}]_{\mbox{\scriptsize obs}} - [\mbox{Si}/\mbox{H}]_{\mbox{\scriptsize ref}}, 
\end{equation}
where, assuming the observed subsolar Si/O ratio is solely due to Si depletion, we use Si/O 
in place of Si/H and use the solar values from \citet{asplund09} for our reference.
This results in a dust depletion of [Si/H]$_{\mbox{\scriptsize gas}} = -0.43$ for SL2S0217;
a value typical with respect to \citet{garnett95b}, but smaller than the value derived by \citet{steidel16}.
From observations of nearby dwarf galaxies we have come to expect [Si/Fe] enrichment
at low [Fe/H] due to early contamination of Si by Type~\ii\ SNe.
However, this exercise indicates that a significant fraction of the Si produced in SL2S0217 is 
missing, perhaps resulting in the subsolar Si/Fe ratio measured from the absorption lines (Section~\ref{sec6}).

\footnotetext[8]{More recently, \citet{jenkins17} updated the elemental depletion equations for 
a given sightline in the Large Magellanic Cloud, including updated parameters for the Si depletion. 
However, this work does not analyze the depletion relationships for 
C, N, or O. Therefore, we have used the \citet{jenkins09} equations for consistency.}
\footnotetext[9]{For dust depletion of D(x) = $[\frac{X}{H}]$, the dust depletion factor
is $\delta(\mbox{x}) = 1 - 10^{\mbox{\tiny D(x)}}$.}

\citet{jenkins09} interpreted the abundances of 17 different elements from more than 100 studies 
of sight lines to 243 different stars to construct a unified representation of depletions that is linearly
dependent on the severity of depletion along a given sight line, $F_{*}$.
Using this model and the inferred depletion value of Si, we estimated the 
depletion of the other significant elements in the SL2S0217 UV spectrum - C, N, and O.  
Using the fit from \citet{jenkins09} for Si, [$X$/H]$_{\mbox{\scriptsize gas}} = B_{\scriptsize X} + A_{\scriptsize X}(F_* - z_{\scriptsize X})$
(coefficients reported in their Table~4), we found a depletion severity of $F_* = 0.182$.
Then, we determine the following dust depletion factors:
$\delta(\mbox{Si}) = 0.628$, $\delta(\mbox{C}) = 0.259$, 
$\delta(\mbox{N}) = 0.222$, and $\delta(\mbox{O}) = 0.111$\footnotemark[9]
([C/H]$_{\mbox{\scriptsize gas}} = -0.130$, [N/H]$_{\mbox{\scriptsize gas}} = -0.109$, 
and [O/H]$_{\mbox{\scriptsize gas}} = -0.051$).
While these depletion values imply a systemically higher relative Si/O abundance ($\Delta \sim0.4$ dex),
the O/H, C/O, and N/O ratios would remain relatively unchanged ($\Delta \sim0.05$ dex).

%----------------------------------------------------------------------------------------------------------------------
%----------------------------------------------------------------------------------------------------------------------

% SECTION 8:
\section{Discussion}\label{sec6}\label{sec8}

%----------------------------------------------------------------------------------------------------------------------
%----------------------------------------------------------------------------------------------------------------------

% Figure 12: Cloudy BPASS models:
\begin{figure*}
\begin{center}
\includegraphics[scale=0.975,trim=10mm 127mm 10mm 10mm,clip]{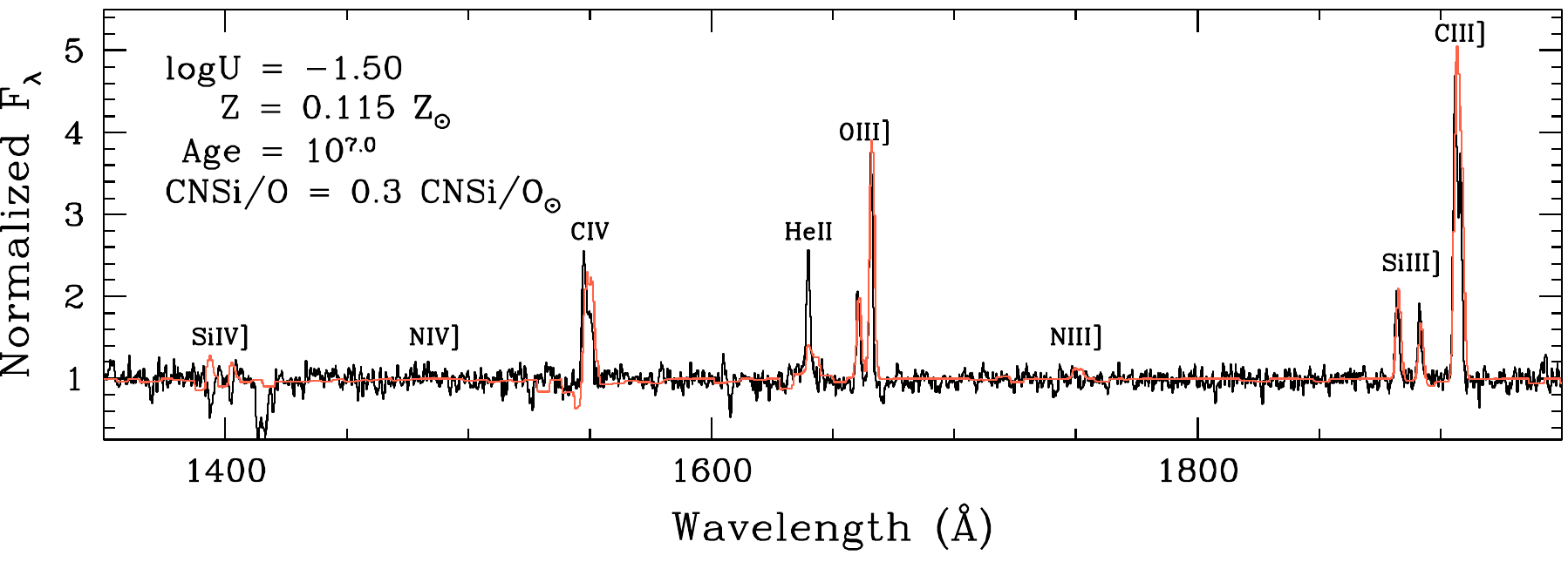}
\caption{
The photoionization model (orange line) best-matched to the blue arm of the UV SL2S0217 spectrum.
SL2S0217 can be described by a high-ionization (log U$ = -1.5$), metal-poor (12+log(O/H) = 7.75;
X/O = 0.3 X/O$_{\odot}$) gas cloud, ionized by a Z = 0.001 star-burst BPASS binary model
at an age of $t=10^{7.0}$ yrs.
All of the emission features, except He~\ii\ which is notoriously difficult to emulate
with pure photoionization models, are reproduced with this model.}
\label{fig12}
\end{center}
\end{figure*}

%----------------------------------------------------------------------------------------------------------------------
%----------------------------------------------------------------------------------------------------------------------

% SECTION 8.1:
\subsection{Photoionization Models}\label{sec8.1}
While the number of objects with detections of extreme UV emission line features continues to grow,
many questions remain regarding their ionizing sources.
To gain insight, we investigated the potential to produce strong UV emission lines 
from an ionizing radiation field powered by stars.
Previous studies have also used photoionization models to examine the factors that regulate 
UV emission lines, focusing on C~\iii] in particular.
\citet{stark14} found that metal-poor ($Z_{neb} < 0.4 Z_{\odot}$), young ($< 50$ Myr) galaxies with
subsolar C/O ratios and large ionization parameters were capable of reproducing the large 
C~\iii] EWs observed for high-redshift galaxies. 
More recently, \citet{jaskot16} created a large grid of photoionization models of star-forming galaxies 
to examine C~\iii] EWs by varying starburst age, metallicity, ionization parameter, C/O abundance, dust 
content, gas density, optical depth, and nebular geometries.  
These authors found that low metallicities and high ionization parameters enhance C~\iii] emission.
Further, the largest C~\iii] EWs were produced by stellar populations that incorporated binary interactions
among massive stars.
Note, however, that models that include stellar rotation have also been shown to extend 
the lifetime (up to $5-7$ Myr) and increase the strength of nebular emission \citep[e.g.,][]{byler17, choi17}.
We focus here on models using binaries alone, as they can produce nebular emission up to $\sim10$ Myr,
but note that likely both rotation and binaries play a role in the observed line strengths.

Our primary goal was to reproduce the strong UV emission line ratios observed for SL2S0217.
Building on the lessons of \citet{jaskot16}, we ran a grid of photoionization models using {\sc cloudy}17.00 in which 
ionization parameter, gas-phase metallicity, and relative C, N, and Si abundances were varied.
For the input ionizing radiation field, we used a suite of models from the new release of ``Binary 
Population and Spectral Synthesis'' \citep[BPASSv2.14;][]{eldridge16, stanway16}, 
with an initial mass function (IMF)  index of $-1.30$ for $0.1-0.5M_{\odot}$ and $-2.35$ for $M > 0.5M_{\odot}$. 

We ran six sets of models depending on three variables: whether 
(1) an IMF upper mass cutoff of 100 or 300 $M_{\odot}$ was used for 
(2) continuous or a burst of star formation, and 
(3) whether or not binary evolution was included. 
Limited parameter ranges were informed by the SED modeling of \citet{brammer12} for the stellar population 
properties and by the UV emission line ratios measured from the LRIS spectrum for the nebular gas properties. 
In particular, \citet{jaskot16} found that C~\iii] EWs as strong as those measured for SL2S0217 ($> 10$ \AA)
only result from a hard radiation field (log U$ > -2$). 
Each model set considered an age range of $10^{6.4}-10^{7.0}$ yrs,
a range in ionization parameter of $-2.0<$ log U$<-1.0$,
with stellar metallicity of Z $=0.001$ (Z$_{\star} = 0.05$ Z$_{\odot}$) to match the estimated gas-phase abundance 
(12+log(O/H) = 7.5 or Z$_{neb} = 0.065$ Z$_{\odot}$). 
The GASS10 solar abundance ratios within {\sc cloudy} were used to initialize the relative gas-phase abundances. 
These abundances were then scaled to cover a range in nebular metallicity ($7.25 <$ 12+log(O/H) $< 8.25$), 
and relative C, N, and Si abundances ($0.1 <$ (X/O)/(X/O)$_{\odot} < 0.5$).
The range in relative N/O, C/O, and Si/O abundances was motivated by our measured values for SL2S0217 and the 
observed values for nearby metal-poor dwarf galaxies \citep[e.g.,][]{berg12, berg16, garnett99}.
The resulting X/O = [0.1,0.3,0.5] X/O$_{\odot}$ model parameters (written as CNSi = [0.1,0.3,0.5] for shorthand) were 
log(C/O) = [$-1.26,-0.78,-0.56$], log(N/O) = [$-1.86,-1.38,-1.16$], and log(Si/O) = [$-2.18,-1.70,-1.48$].

% SECTION 8.1.1:
\subsubsection{Best Model}\label{sec8.1.1}
To determine the best model from our grid, 
we first eliminated sets of models that were obviously poor matches. 
The single-star model sets do not reproduce the UV nebular emission line 
feature strengths \citep[see also,][]{jaskot16}, and so were not investigated further.
For the young ages considered here ($t < 10^{7.2}$ yrs), the continuous star 
formation models are consistent with the burst models, and so only the burst 
models were examined as representative of both types of star formation.
Note, however, that continuous star formation models do produce very different results from bursts for older systems. 
Finally, the models with an IMF cutoff of 300 $M_{\odot}$ do produce more He~\ii\ ionizing photons than the
100 $M_{\odot}$ models, yet they still fall short of producing the He~\ii\ flux observed for SL2S0217, and 
significantly overpredict the Si~\iii]/C~\iii] ratio.
Therefore, we employed an error-weighted $\chi^2$-minimization routine of the grid of 
binary burst models with an IMF cutoff of 100 $M_{\odot}$.
We find we can reproduce most of the features in the UV emission-line spectrum of SL2S0217 
with a model similar to the SED modeling of \citet{brammer12}:
the best model is a high-ionization (log U$ = -1.5$), metal-poor (12+log(O/H) = 7.75; CNSi = 0.3)
gas cloud, ionized by Z = 0.001 star formation with binaries at an age of $t = 10^{7.0}$ yrs.

In Figure~\ref{fig12} we compare the normalized LRIS spectrum of SL2S0217 to 
the output model spectrum at the spectral resolution of the observed spectrum.
The C~\iv, O~\iii], N~\iii], Si~\iii], and C~\iii] emission lines are all relatively well-matched,
while the He~\ii\ emission feature is clearly discrepant. 
Notably, \citet{fosbury03} also observed strong, narrow nebular He~\ii\ and C~\iv\ emission 
from the $z\sim3.4$ Lynx Arc, reportedly due to a young, low-metallicity stellar cluster with 
an extreme ionization parameter of log U$ = -1.0$. 
However, to date, no pure stellar photoionization studies have reported the levels of nebular 
He~\ii\ emission observed in SL2S0217 or the Lynx Arc. 
Interestingly, we do not see the N~\iv] emission line in our SL2S0217 spectrum despite 
the fact that we see strong C~\iv\ emission and C$^{++}$ and N$^{++}$ have similar 
ionization potentials (47.888 eV and 47.445 eV, respectively).
The N~\iv] emission line is also absent from the best model, indicating that N emission is mostly 
sensitive to metallicity and, therefore, not expected in metal-poor galaxies like SL2S0217, 
whereas C~\iv\ is more sensitive to temperature / cooling. 
This fact has motivated the use of the strong-line [N~\ii]/H$\alpha$ metallicity indicator in
nearby galaxies, however, the line ratio is not available from the ground at high redshifts ($z\gtrsim2.5$)
and is highly sensitive to ionization parameter at larger metallicities \citep[e.g.,][]{kewley02}.
Since exceptionally low nitrogen abundances seem to be common amongst distant galaxies 
with similar extreme emission line features \citep[e.g.,][]{bayliss14, james14, smit17}, 
the lack of UV N~\iii] and N~\iv] could help identify additional metal-poor galaxies at high redshifts.
 
\begin{figure}
\includegraphics[scale=0.4, trim=25mm 130mm 0mm 40mm, clip]{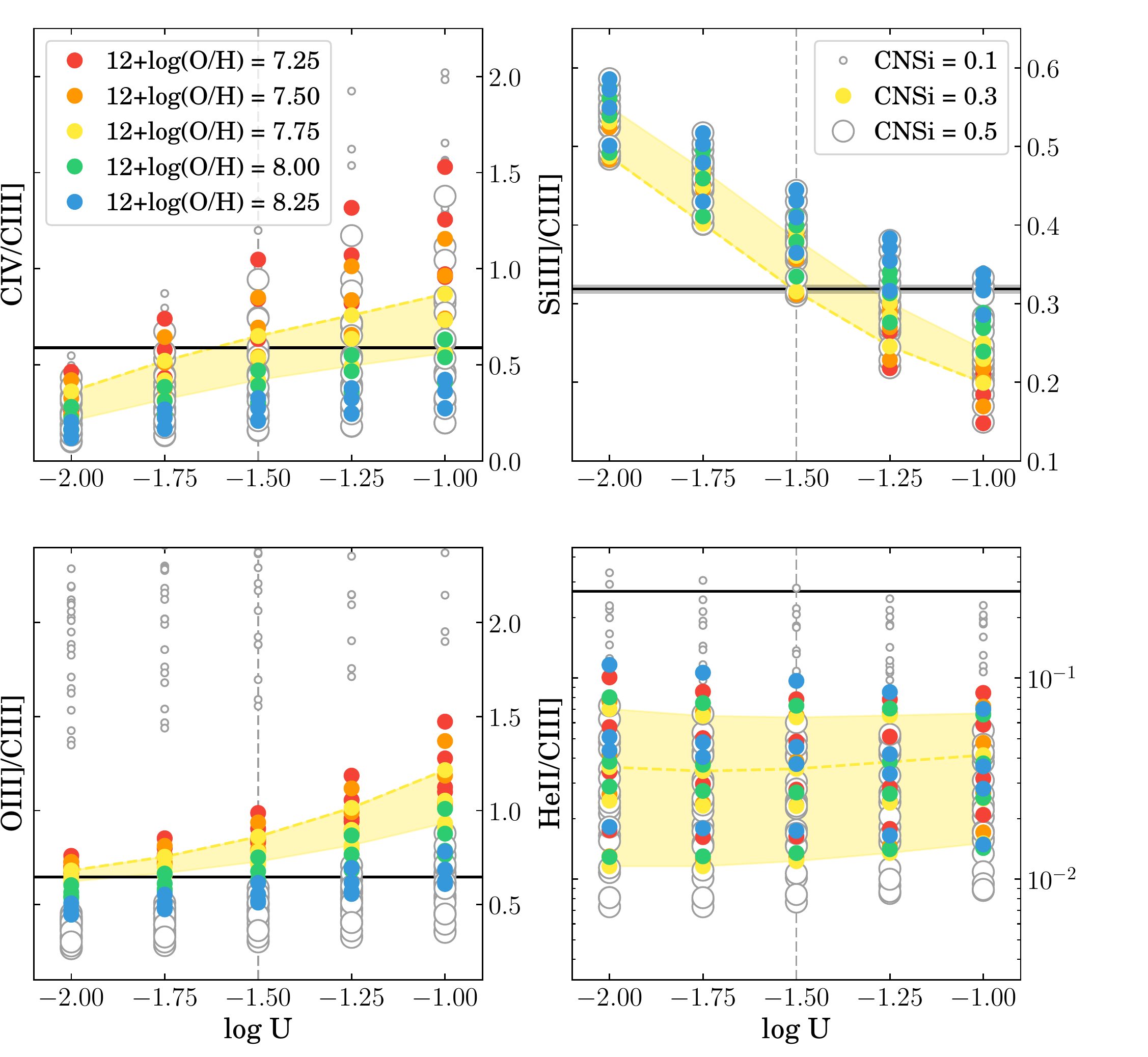}
\caption{
UV emission line ratios from {\sc cloudy} photoionization models using the BPASS Z$_{\star}=0.001$ Z$_{\odot}$ binary models for the ionizing source,
assuming an IMF$_{up}$ = 100 M$_{\odot}$ over a range in age of $10^{6.4}-10^{7.0}$ yrs.
The color coded points represent a range of gas-phase metallicities assumed: 7.25$\le$12+log(O/H)$\le$8.25.
We also explored a range in C/O, N/O, and Si/O abundances relative to their respective solar ratios, 
as indicated by various point sizes (see description in Section~\ref{sec8.1}).
Observed line ratios and corresponding uncertainties for SL2S0217 are plotted as solid black lines and gray extensions, respectively.
For the 12+log(O/H)$=7.75$ models, the change in line ratio as a function of age is shown by the 
spread of the shaded yellow region, where the dashed yellow line marks the upper limit of $10^{7.0}$ yrs. }
\label{fig13}
\end{figure}

The predicted nebular emission line ratios from our photoionization models are plotted relative 
to the ratios measured from the LRIS spectrum of SL2S0217 in Figure~\ref{fig13}.
Different gas-phase metallicity models are indicated by color, where relative C/O, N/O, 
and Si/O abundances increase with point size. 
The statistically best model is indicated by the dashed yellow line (12+log(O/H) = 7.75, 
CNSi = 0.3, $t=10^{7.0}$ yrs), where the corresponding yellow-shaded region extends 
from $t=10^{6.4}-10^{7.0}$ yrs.
Note that nebular He~\ii\ emission peaks before $t=10^{7.0}$ yrs ($\sim10^{6.8}$ yrs) in the models,
and so the dashed yellow line in the He~\ii/C~\iii] plot lies within the yellow-shaded region.
Interestingly, although C~\iv\ also has a very high ionization potential, 
its nebular line strength continues to increase past $t=10^{7.0}$.
Measured line ratios and uncertainties for SL2S0217 are shown as solid black 
horizontal lines and gray-shaded boxes.
Good agreement between model and measurement is found near an ionization parameter 
of log U$ = -1.50$ for all of the UV emission line ratios except He~\ii/C~\iii].

 Just as \citet{stark14}, \citet{berg16}, and \citet{senchyna17} observed strong C~\iii] 
 emission resulting from subsolar C/O in metal-poor systems (Z $\leq0.5\times$Z$_{\odot}$),
 reproducing the observed emission lines in SL2S0217 also requires subsolar C, N, and Si abundance ratios relative to O.
 The low gas metallicity prevents efficient metal cooling by oxygen, resulting in a high electron temperature
 that secures a stable production of the collisionally-excited O, Si, and C nebular emission lines. 
If a portion of the metals are also locked up in dust, 
 C and Si may be under abundant as a result. 
 However, no evidence of dust reddening exists
 and the extreme ionizing radiation field likely destroyed or eroded the dust in the \ion{H}{2} regions
 (but also see discussion of dust depletion in \S~\ref{sec7.2.3}).

%----------------------------------------------------------------------------------------------------------------------

% SECTION 8.2
\subsection{Stellar Ionization}
The rate of ionizing photons needed to power the arc can be estimated from the demagnified H$\beta$ flux.
At the redshift measured in this work, we calculate a luminosity distance of $D_l = 1.444\times10^4$ Mpc \citep{wright06}.
The hydrogen ionizing photon luminosity is then given by
\begin{equation}
	{Q_{\mbox{ion.}}} \approx {\frac{4\pi D^2_l F_{H\beta}}{\mu h \nu_{H\beta}}\ } {\frac{\alpha_B}{\alpha^{eff}_{H\beta}}\ },
\end{equation} 
where $\mu$ is the lensing magnification and $\alpha_B$ and $\alpha^{eff}_{H\beta}$ are the total and effective 
H$\beta$ case-B H recombination coefficients.
We calculate the recombination coefficients for $T_e =1.6\times10^4$ K using the equations of \citet{pequignot91}.
Given the lensing magnification determined here ($\mu = $17.3) and the H$\beta$ flux from the grism spectrum,
we estimate an ionizing photon luminosity of Q$_{ion.} = 7.3\times10^{53}$ photons s$^{-1}$.
The cluster IMF and metallicity is needed in order to then constrain the nature of the ionizing stellar population,
however, this value is still indicative of a large number of massive stars powering the observed spectrum
($> 10^3$ O stars), which can live to greater ages due to rotation and binary evolution.

 %----------------------------------------------------------------------------------------------------------------------

% Figure 13: Cloudy BPASS models:

 %----------------------------------------------------------------------------------------------------------------------

\subsection{Multiple Ionization Sources}
While many of the emission-line features observed in the rest-frame UV spectra of SL2S0217
can be reproduced by photoionization from stars alone, outliers such as He~\ii\
prompted an investigation into additional ionization sources. 
Indeed, the ratios of rest-frame far-UV emission lines provide a means to discern between
nuclear activity and star-formation as the ionizing sources \citep[e.g.,][]{feltre16}.
However, \citet{feltre16} uses the photoionization models of \citet{gutkin16} which are not appropriate 
for our young target as they assume continuous star formation for 100 Myrs and do not 
include the effects of binaries (see Section~\ref{sec8.1.1}). 
Instead, we investigate the effects of incorporating shocks and AGN into the photoionization models presented above. 

%----------------------------------------------------------------------------------------------------------------------
%----------------------------------------------------------------------------------------------------------------------

\begin{figure*}
   \begin{center}
      \begin{tabular}{cc}
             \includegraphics[scale=0.4, trim=25mm 130mm 0mm 40mm, clip]{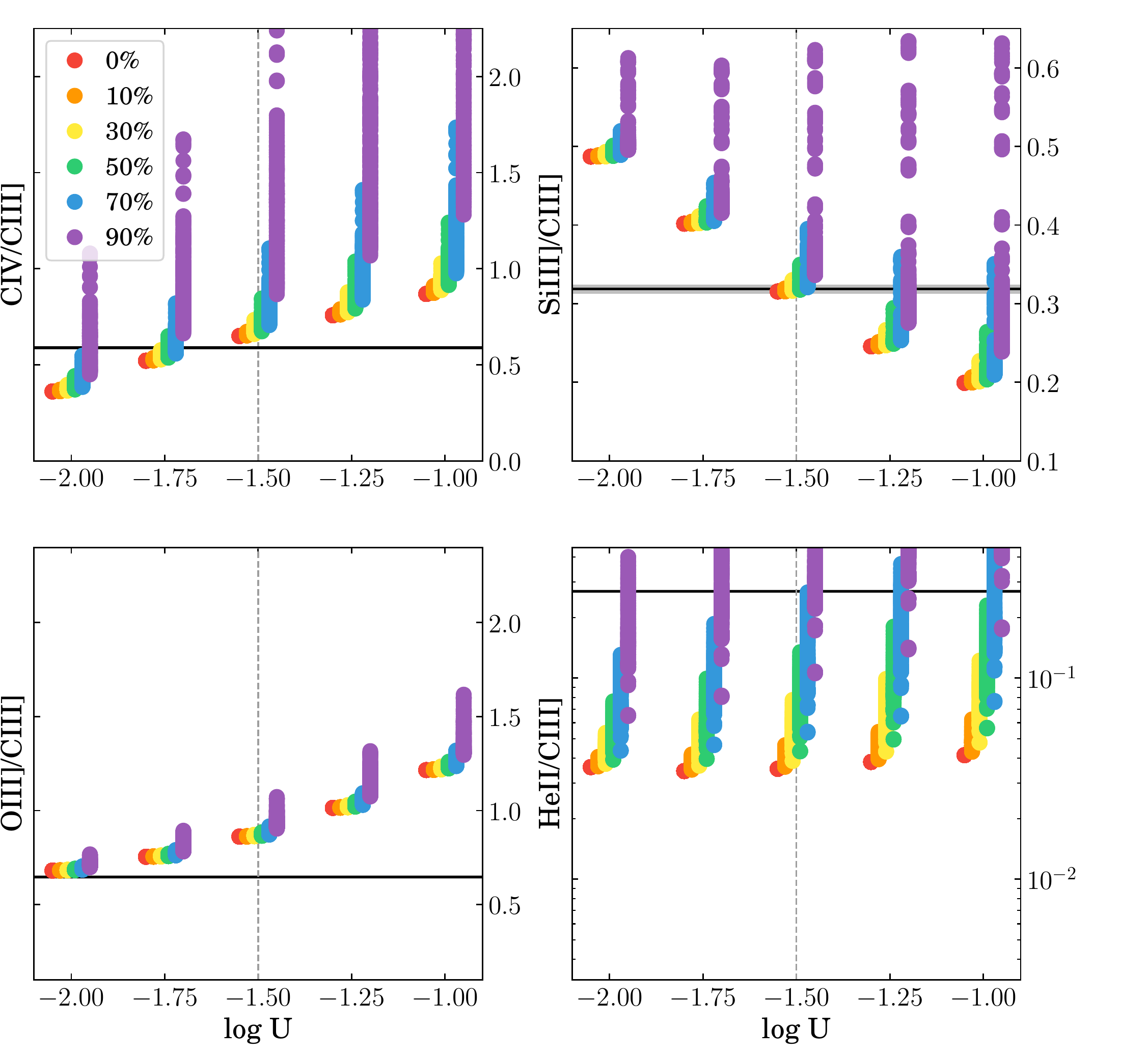}  		
          & \includegraphics[scale=0.4, trim=25mm 130mm 0mm 40mm, clip]{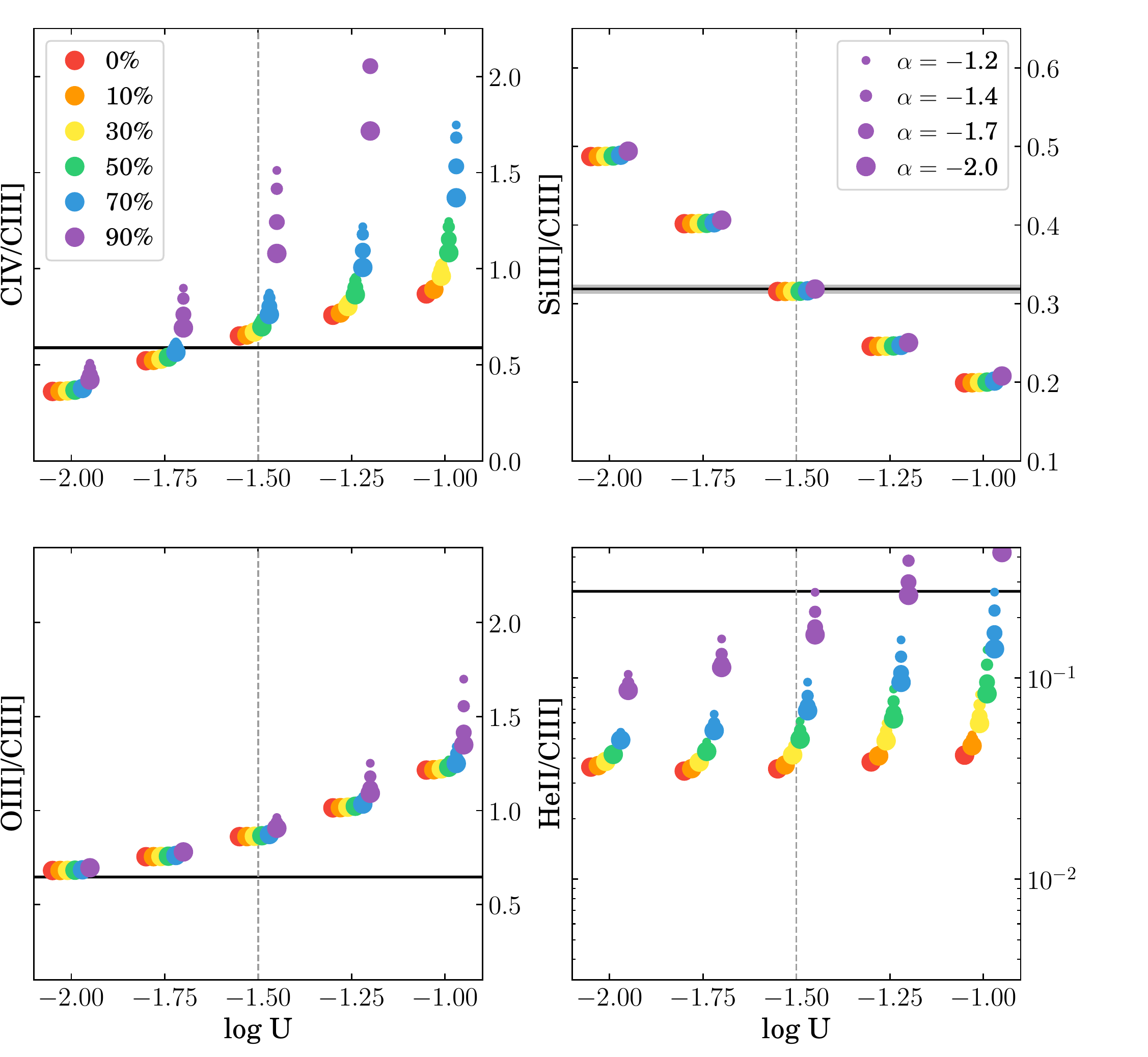}   \\
	{(a) BPASS+Shock Models}	& {(b) BPASS+AGN Models} 
  	\end{tabular}    
	\caption{Cumulative UV emission line ratios from {\sc cloudy} photoionization models plus a (a) shock or (b) AGN contribution,
	while each color, slightly offset for better visualization, indicates a different fractional contribution from shocks or AGN to the total H$\alpha$ emission.
	The input stellar photoionization model is the best model fit to the SL2S0217 spectrum (see Figure~\ref{fig12}). 
	Z$_{\star}\sim0.003$ shock+precursor models from Mappings III \citep{allen08} are used for the shock contribution, 
	where the spread represents a range in velocity of $125-1000$ km s$^{-1}$ and magnetic field strengths of $0.5-10 \mu$G.
	Z$_{\star}=0.25$ Z$_{\odot}$ dust-free AGN models over a range of power-law exponents (symbol size) are taken from \citet{groves04}.
	While the extreme He~\ii/C~\iii] ratio observed for SL2S0217 can be reproduced by models with a significant shock or AGN 
	component, the other line ratios cannot be matched with the same conditions. }
	\label{fig14}
   \end{center} 
\end{figure*}

%----------------------------------------------------------------------------------------------------------------------
%----------------------------------------------------------------------------------------------------------------------

\subsubsection{Shocks}\label{sec8.3.1}
Even in star-forming galaxies, a substantial portion of the nebular emission
may be generated by shocks \citep[e.g.,][]{krabbe14, rich14}.
This fact motivated \citet{jaskot16} to incorporate contributions from shocks
to their pure photoionization models.
In their study, shock+precursor models were taken from Mappings III \citep{allen08}, assuming a metallicity
of $Z\sim0.003$ (the Small Magellanic Cloud models), a velocity range of $125-1000$ km s$^{-1}$,
and magnetic field strengths of $0.5-10 \mu$G.
The authors find that the addition of the shock models result in an increase of both C~\iii] 
and He~\ii\ emission relative to H$\beta$, and, therefore, suggest that nebular He~\ii\ may 
serve as a useful diagnostic of shocks for UV spectra.

In light of these findings, we investigated emission from shock ionization as a source for the
significant He~\ii\ emission seen in SL2S0217.
Following the methodology of \citet{jaskot16}, we scaled the shock emission and added it to the
emission from our photoionization models such that the shocks contribute 10\%, 30\%, 50\%, 70\%, 
or 90\% of the total predicted emission in H$\alpha$ (see their Section 2.3 for a more detailed description).

The emission line ratios for a varying shock contribution to our best photoionization 
model are depicted in Figure~\ref{fig14} (a) relative to those measured for SL2S0217.
These results demonstrate that the addition of shocks to the BPASS photoionization
does not improve the general match of models to the SL2S0217 features. 
While the extreme He~\ii/C~\iii] ratio can be reproduced when the ionizing radiation is dominated by shocks, 
the other observed line ratios are generally over-predicted and cannot be matched with the same conditions.
Further, it is physically unlikely that the gas in SL2S0217 is 90\% shock ionized, and so
we dismiss shocks as a main source of ionization in SL2S0217.

%----------------------------------------------------------------------------------------------------------------------
%----------------------------------------------------------------------------------------------------------------------

\subsubsection{Active Galactic Nuclei}\label{sec8.3.2}
The complex suite of spectral features observed in extreme galaxies such as SL2S0217 may be due to 
a complex combination of ionizing sources, such as AGN emission diluted by stellar radiation. 
For super-massive black holes (SMBH) in massive galaxies, numerous studies have found the SMBH mass 
is well correlated with the velocity dispersion of the host-galaxy bulge
\citep[e.g.,][]{gebhardt00, gultekin09}. 
However, it is unclear whether this well-known scaling relation extends to the galaxies as small
as SL2S0217 \citep[see discussion in][]{kormendy13}.
Black hole masses in dwarf AGNs have been measured in several low-mass star-forming galaxies in the local universe 
\citep[e.g.,][]{reines11, reines14}, with the lowest reported mass being just $\sim$50,000 M$_{\odot}$  \citep{Baldassare15}.
Recently, \citet{baldassare17} found evidence for dwarf AGN in 11 nearby ($z < 0.055$) composite dwarf galaxies.
These authors measured black hole masses ranging from $10^{4.9}-10^{6.1}$ M$_{BH}$/M$_{\odot}$
for galaxies with stellar masses of $10^{8.1}-10^{9.4}$ M$_{\star}$/M$_{\odot}$, on the order of SL2S0217.
While these recent studies of local dwarf galaxies indicate that a non-negligible fraction of low-mass galaxies
might harbor AGN, we caution that we do not know how these results extrapolate to galaxies at high redshifts. 

Given the uncertainties surrounding AGN in low-mass galaxies, 
it is worthwhile to explore the possibility of producing the extreme 
emission-line ratios observed in SL2S0217 with the addition of emission from AGN.
For this purpose, we used the dust-free, low-metallicity (Z$_{\star}=0.25$Z$_{\odot}$), 
narrow-line AGN photoionization models from \citet{groves04}, assuming simple 
power-law radiation fields ($f_{\nu} \propto \nu^{\beta}$) with a range of powers.
As in Section~\ref{sec8.3.1}, we scale the AGN emission such that the AGN 
contribute 10\%, 30\%, 50\%, 70\%, or 90\% of the total emission in H$\alpha$.

Emission line ratios for the best BPASS + AGN models are plotted in 
Figure~\ref{fig14} (b) relative to those measured for SL2S0217.
The resulting shifts in line ratios due to the contribution of emission from AGN look 
very similar to those seen for the BPASS + Shock models.
In either case, the additional hard radiation increases the relative emission from the
highest ionization lines: He~\ii, O~\iii], and C~\iv.
This allows the C~\iv/C~\iii] ratio to be reproduced with small AGN contributions.
Typical narrow-line Type II AGN observations have stronger C~\iv\ than C~\iii] 
emission \citep[e.g.,][]{humphreys08, matsuoka09, hainline11, cassata13},
but our models suggest that the C~\iii] dominant emission in SL2S0217 (C~\iii]/C~\iv$=1.46$)
could be produced by a gas-enshrouded AGN.

On the contrary, Figure~14 shows that the Si~\iii]/C~\iii] ratio is significantly less 
affected by contributions from AGN than from shocks.
This is due, in part, to the fact that the full range of shock velocities are considered at each 
value of log U, such that the maximum contribution from shocks for any ion is always included (near $v\sim175$ km s$^{-1}$ for Si~\iii]).
On the other hand, the AGN contribution to Si~\iii] emission peaks at low ionization parameter, however, it has little effect on the 
total Si~\iii]/H$\alpha$ emission, which is dominated by photoionization. 

An additional difference is seen in the He~\ii/C~\iii] ratio, where the value observed in 
SL2S0217 is only reached by AGN dominated models ($> 90\%$ contribution; purple points), 
yet such models significantly over estimate the C~\iv/C~\iii] and O~\iii]/C~\iii] ratios.
Further, if the strong nebular emission lines in SL2S0217 were powered by a Type II AGN, we would also 
expect to detect several high ionization emission lines that are missing from the rest-frame UV 
spectrum: \ion{N}{5} \W1240, Si~\iv\ \W1403, N~\iv\ \W1486 \citep[e.g.,][]{vandenberk01, hainline11}.
Again, the suite of lines observed for SL2S0217 cannot be simultaneously matched by BPASS photoionization
models with an additional source of hard radiation, in this case, from AGN.

%----------------------------------------------------------------------------------------------------------------------

\subsection{High Ionization Emission Lines}\label{sec8.4}

While we postulate that SL2S0217 is indeed a star-forming galaxy, 
its strong He~\ii\ and C~\iv\ emission is rare and interesting.
In fact, only one percent of UV-selected galaxies at $z\sim3$ show strong nebular
C~\iv\ emission \citep[e.g.,][]{reddy08,hainline11,mainali17}.
In comparison, $z\sim2-3$ LBGs with large Ly$\alpha$ EWs have been reported to measure 
relatively weak C~\iv\ emission with strengths $\le 1$\% of Ly$\alpha$ \citep[e.g.,][]{erb10}.
Notably, significant narrow C~\iv\ and/or He~\ii\ emission are also detected in several 
nearby dwarf galaxies, suggesting that extremely hard ionizing radiation fields may 
be common in low-mass, low-metallicity galaxies \citep[e.g.,][]{kehrig11,berg16}. 
This idea is further supported by several recent studies that have found nebular C~\iv\ and/or 
He~\ii\ in distant metal-poor star-forming galaxies \citep{erb10,christensen12,stark14,debarros16,steidel16,vanzella16}.
Of the few extreme UV emission line galaxies that have been observed at $z\gtrsim2$
\citep[e.g.,][]{erb10,christensen12,bayliss14,stark14,stark15,caminha16,debarros16,vanzella16,smit17},
none show a comparable combination of emission line ratios. 

%----------------------------------------------------------------------------------------------------------------------

\subsubsection{Nebular C~\iv\ \W\W1548,1550}\label{sec8.4.1}
Typically galaxies show C~\iv\ in ({\it i}) absorption from the surrounding ISM or CGM gas
or ({\it ii}) exhibit a P-Cygni profile from the stellar winds of massive stars \citep[e.g.,][]{leitherer01}.
To the contrary, AGN can produce C~\iv\ emission, but line-widths are typically broader than
nebular emission lines (hundreds of km s$^{-1}$).
To date, narrow C~\iv\ has only been observed in a handful of strongly lensed high-redshift galaxies
\citep[e.g.,][]{christensen12, stark14, smit17}, therefore, the significant C~\iv\  
emission detected in SL2S0217 is another interesting piece of this peculiar puzzle.

\begin{figure}
\begin{center}
\includegraphics[scale=0.35, trim=0mm 20mm 0mm 50mm, clip]{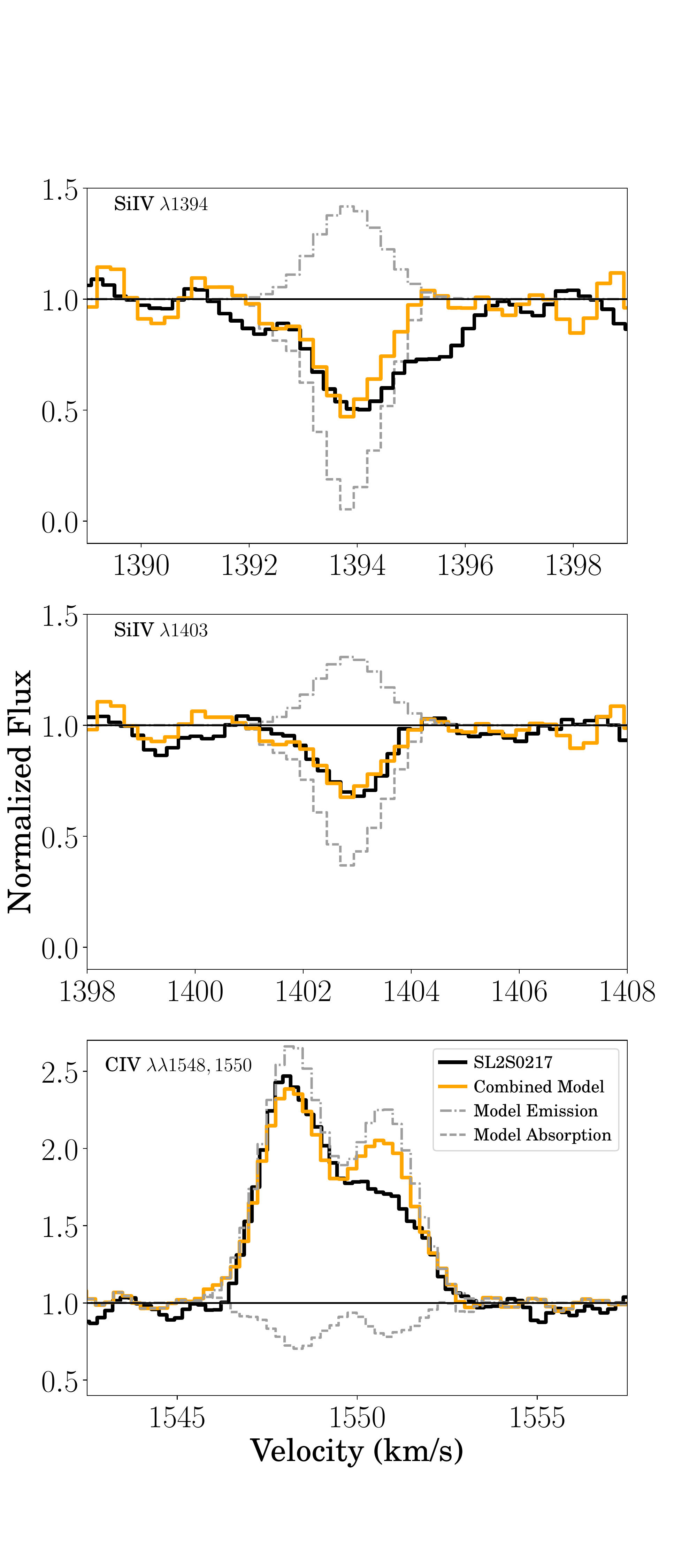}
\caption{
Model to produce the observed Si~\iv\ absorption and C~\iv\ emission features by combining
scaled versions of the average absorption profile and the emission line ratios predicted 
by the photoionization described in Section~\ref{sec8.1.1}.}
\label{fig15}
\end{center}
\end{figure}

Since the C~\iv\ \W\W1548,1550 doublet is expected to be produced as a combination of nebular, stellar,
and ISM/CGM components, we can use a simple combination of absorption and emission profiles to offer a
possible model for the C~\iv\ profile observed in SL2S0217.
Despite the absence of apparent C~\iv\ absorption, if we assume that the absorption in both C~\iv\ and Si~\iv\ 
is due to high ionization clouds in the surrounding ISM of the galaxy, we can use the average absorption profile 
(which is nearly identical to the Si~\iv\ \W1403 profile; see Figure~\ref{fig7}) to model the C~\iv\ absorption.
Although the oscillator strength of Si~\iv\ \W1394 is roughly two times larger than Si~\iv\ \W1403, their
relative absorption profile depths are closer to a factor of 1.5, likely due to saturation.
Similarly, C~\iv\ \W1548 absorption is expected to be twice as strong as C~\iv\ \W1550.
Following the observed model of Si~\iv, we assume relative absorption profile strengths of 
3:2 for both Si~\iv\ \W\W1394,1403 and C~\iv\ \W\W1548,1550. 
The total profile is then produced by co-adding the absorption profile with the emission predicted by the
best photoionization model, $n\times F_{\lambda}$, where $n$ is an arbitrary scaling of the emission
to best-match the observed spectrum of SL2S0217.
The resulting model profiles and their components are plotted 
against the SL2S0217 spectrum in Figure~\ref{fig15}.
These model fits offer {\it a} solution in which the Si~\iv\ and C~\iv\ observed 
profiles could be produced, but we caution that this interpretation cannot be confirmed.

%----------------------------------------------------------------------------------------------------------------------
%----------------------------------------------------------------------------------------------------------------------

%----------------------------------------------------------------------------------------------------------------------
%----------------------------------------------------------------------------------------------------------------------

\subsubsection{Nebular He~\ii\ \W1640}\label{sec8.4.2} 
Arguably the most aberrant feature of the SL2S0217 UV spectrum 
is the impressive nebular He~\ii\ emission strength.
Nebular He~\ii\ emission is produced by recombination from the He$^{+2}$ 
ionization state, and so requires very hard ionizing radiation with photons at 
energies $\ge 54$ eV in order to ionize He$^{++}$.
The interpretation of He~\ii\ emission is further complicated by the fact that it 
can have both nebular and stellar origins.
Stellar He~\ii\ is produced by the stellar winds of massive stars, such as Wolf-Rayet (W-R) stars, and so is 
broadened in relation to the wind velocity \citep[e.g.,][]{brinchmann08, erb10, cassata13}.

\cite{leitherer99} showed that the number of W-R stars relative to O stars
peaks around $10^{6.6}$ yrs, and so stellar winds may no longer be prominently contributing 
to the He~\ii\ emission at a star-burst age of $10^7$ yrs, in line with the fact that we do not observe 
a {\it significant} broad component to the He~\ii\ emission in SL2S0217.
Further, \cite{eldridge09} demonstrated that the WR He~\ii\ \W1640 feature is only prominent in
binary models with stellar metallicities of $Z_{\star} = 0.004$ or $0.008$,
which do not pair well with the gas-phase oxygen abundance of SL2S0217 ($Z_{neb} \sim 0.001$).  
Instead, the He~\ii\ emission in SL2S0217 appears to be nebular in origin.
In the photoionization models of \citet{jaskot16}, the narrow, nebular He~\ii\ emission seen in SL2S0217 
only becomes strong for more metal-rich systems during the W-R phase or when ionized by shocks.
Other sources of high energy photons may be required to produce strong nebular He~\ii, such as AGN, 
exotic Pop III stars, or high-mass X-ray binaries \citep[e.g.,][]{shirazi12,kehrig15}.

As discussed in Section~\ref{sec8.3.1}, we dismiss a significant contribution to the ionizing budget
from shocks as the models significantly overpredict the O~\iii]/C~\iii] budget.
In addition to overshooting O~\iii]/C~\iii], the AGN models (\S~\ref{sec8.3.2}) also produce an excess of C~\iv\ relative to C~\iii]
emission, but this argument is not definitive as some of the emission may be obscured by overlapping C~\iv\ absorption.
However, AGN typically have large He~\ii\ EWs \citep[$\gtrsim$ 8\AA;][]{hainline11, cassata13}, which
are generally greater than those measured for z$\sim$2 galaxies \citep{scarlata09, erb10}
and for SL2S0217 (EW(He~\ii$_n$) = 2.4 \AA\ or EW(He~\ii$_{tot}$) = 2.8 \AA,
where the difference is due to the difficulty in deblending the narrow and wide components).

\begin{figure}
\includegraphics[scale=0.45, trim=20mm 0mm 20mm 20mm, clip]{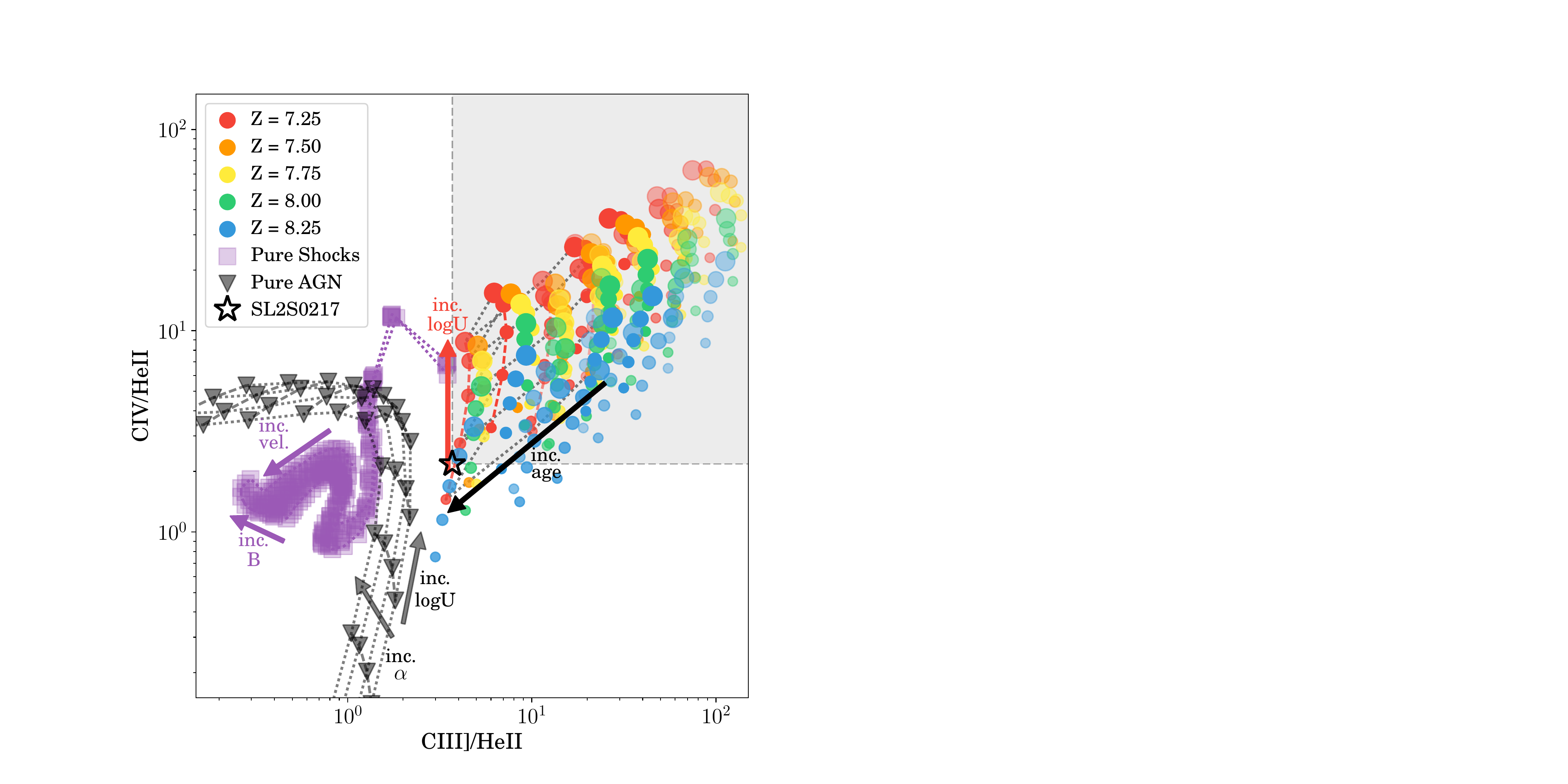}
\caption{
C~\iv/He~\ii\ versus C~\iii]/He~\ii\ emission line diagnostic diagram.
The stellar photoionization models described in Section~\ref{sec8.1} are shown as solid
colored circles, where the metallicity increases with color, ionization parameter increases
with point size, age increases with color saturation. 
As an example, for the CNSi = 0.1 and 12+log(O/H) = 7.25 model, dashed-lines connect points of constant 
age and black dotted-lines connect points of constant ionization parameter. 
Pure shock and AGN models are also shown.
The observed line ratios of SL2S0217 are depicted as the black star, where the gray shaded
region indicates the possible values of SL2S0217 if C~\iv\ or He~\ii\ are not purely nebular emission.}
\label{fig16}
\end{figure}

Photoionization by AGN has also been investigated using diagnostic plots of the C~\iv/He~\ii\ and C~\iii]/He~\ii\ line ratios,
both of which are sensitive to metallicity and ionization parameter \citep{villar-martin97}.
In Figure~\ref{fig16} we plot our stellar photoionization models described in Section~\ref{sec8.1.1}
in comparison to photoionization from pure shocks and AGN for the C~\iv/He~\ii\ versus C~\iii]/He~\ii\ ratios.
The observations for SL2S0217 (black star), are extended to larger values 
(shaded region) in case C~\iv\ is underestimated (due to overlapping C~\iv\ absorption) and/or He~\ii\ is overestimated
(due to a stellar contribution), clearly suggesting a stellar origin of the photoionization budget.
Note, however, that the stellar photoionization models still fail to reproduce the 
absolute strength / EW of the He~\ii\ emission. 
Motivated by the uncertainties of these line ratios, we will present a preferable analysis 
in a forthcoming paper, that allows us to distinguish between sources of ionizing 
radiation using line ratios that are not complicated by stellar and CGM components. 

In observations of star-forming galaxies, He~\ii\ emission is commonly attributed to W-R stars,
which typically show strong and broad ($\Delta$v$>1000$ km/s; \cite{crowther07}) emission lines produced 
by their fast stellar winds.
For example, \citet{chandar04} measured an extreme He~\ii\ \W1640 EW in the local Universe from W-R stars 
in the massive cluster NGC3125-1 with a broad width of $\sim1000$ km s$^{-1}$, and further supported by
strong detections of the N~\iv\ \W1488 and N~{\sc v} \W1720 WN W-R features. 
We therefore rule out W-R stars as the source of the He~\ii\ emission on the account that
we do not observe significant broad He~\ii\ emission in SL2S0217, or other indicative feature, and that the 
W-R phase occurs in a very short and early window that is unlikely the period of our observations.
In binary star models, where the effects of mixing and mass transfer extends the life of these massive stars,
He~\ii\ emission peaks $\sim50$ Myrs after a burst of Z$_{\star}= 0.004$ star formation.
However, the He~\ii\ emission from W-Rs depends on the metallicity, and the lower metallicity 
stellar spectra representative of SL2S0217 are not capable of reproducing the observed He~\ii\
(see Figure~\ref{fig13}).

Theoretically, gravitational cooling radiation from gas accretion is another mechanism that can produce
He~\ii\ (as well as Ly$\alpha$) emission profiles like those in SL2S0217.
In these models, infalling gas onto dark matter can be heated up to temperatures of $T\sim10^5$ K,
and then is cooled by H and He line emission \citep{keres05, yang06}.
It has been suggested that Ly$\alpha$ halos 
\citep[Ly$\alpha$ ``blobs" that show extended Ly$\alpha$ emission on scales of tens of kpc or more, e.g.,][]{caminha16,  patricio16} 
are the observational signature of this cooling radiation in forming galaxies \citep{haiman00, fardal01}. 
While inflowing gas would also help explain the strong blue peak in the Ly$\alpha$ profile, 
it is unlikely that SL2S0217 is a Ly$\alpha$ halo.
Our observations and lensing model show a small galaxy with Ly$\alpha$ emission coming from
three dense star-forming regions, where the Ly$\alpha$ emission extends beyond the continuum 
on much smaller spatial scales (as detected in narrow-band {\it HST} imaging; Erb et al.\ 2018, in prep.).

The current arsenal of stellar population models in the literature are only able to reproduce 
strong He~\ii\ emission lines with pure photoionization for extreme stellar populations. 
Due to its high ionization potential, He~\ii\ emission has been used as a probe to find Pop III stars \citep[e.g.,][]{visbal15}.
A top-heavy IMF in extremely low metallicity regions can produce large He~\ii\ line luminosities \citep{tumlinson00, schaerer03, yoon12}.
However, the He~\ii\ EW is very sensitive to metallicity, and so requires Z $< 10^{-5}$ for this model.
\citet{steidel16} predicted that \ion{H}{2} regions with normal oxygen abundances can exhibit extreme UV features 
if they are powered by stars that are O/Fe enhanced relative to solar.
Such a scenario would be expected if the ISM was enriched by core-collapse SNe, and then the best-matching stellar
model would actually have a metallicity several times lower than the gas-phase oxygen abundance.
Since stellar spectra are primarily dependent on the total opacity, which is largely modulated by the Fe abundance \citep[e.g.,][]{rix04}, 
much harder ionizing radiation would be produced than is expected from the nebular metallicity.
Alternatively, \citet{tornatore07} predicted that zero-metallicity regions can survive in low-density regions around 
large over densities, allowing Pop III stars to form, until as recently as $z\sim2$.

Other studies do not require large modifications to relative abundances or metal-free stars to produce narrow He~\ii\ emission. 
\citet{kudritzki02} showed that certain O stars are indeed hot enough to ionize He~\ii\ and \citet{brinchmann08}
suggested that at low metallicities nebular He~\ii\ is predominantly produced by O stars where optically 
thin winds can be penetrated by ionizing photons. 
The energy production of low-metallicity massive stars may also be increased if they are fast rotators \citep{meynet07}.
Further support is provided by the study of \citet{kehrig15}, who found that the He~\ii-ionization budget of IZw~18 can
only be produced by models of either low-metallicity super-massive O stars or rotating metal-free stars.
However, the details of very low-metallicity O stars are still widely debated, and further
complicated by the fact that some He~\ii\ nebulae don't appear to have surviving O or W-R stars \citep{kehrig11}.
\citet{grafener15} prosed an alternative scenario for the origin of narrow He~\ii\ emission in which metal-poor, very 
massive stars form strong but slow W-R-type stellar winds due to their proximity to the Eddington limit.
Post-AGB stars from an older stellar population may also produce He~\ii\ emission.
The combined radiation field of post-AGB stars will dominate the ionizing spectrum about 100 Myrs after a 
burst of star formation and would be strong enough to ionize He~\ii\ \citep{binette94}.
However, the He~\ii\ luminosities produced by post-AGB stars are insufficient to explain the 
emission line strength seen for SL2S0217 \citep{cassata13}.

It seems likely that very massive stars play a significant role in the He~\ii\ line strength.
Unfortunately, the nature of metal-poor massive stars remains poorly understood and therefore
very massive stars are often withheld from current population synthesis models. 
Theoretical work on rotation and binary evolution of massive stars \citep[e.g.,][]{eldridge09, demink14} 
has produced a vast range in the predicted lifetimes and energy output of low-metallicity massive stars.
Further, these models make simplifying assumptions about the rotation velocities and distribution of binary
separations that are poorly motivated.
In the future, more sophisticated models of metal-poor massive stars may be able to explain the strength
of high ionization nebular lines observed for SL2S0217.
However, since narrow He~\ii\ emission is more common at $z\sim3$ \citep{cassata13} than at $z\sim0$ \citep{kehrig11},
it is possible that the stellar populations that produce He~\ii\ ionizing radiation are more common at higher redshifts
and are fundamentally different than typical populations in the local Universe.

%----------------------------------------------------------------------------------------------------------------------
%----------------------------------------------------------------------------------------------------------------------

%----------------------------------------------------------------------------------------------------------------------
%----------------------------------------------------------------------------------------------------------------------

\section{CONCLUSIONS}
We present the rest-frame UV Keck/LRIS spectrum of SL2SJ021737-051329 (SL2S0217),
a lensed galaxy magnified by a factor of roughly 17 by a massive galaxy at z = 0.65. 
SL2S0217 is particularly interesting given its very low mass (M $< 10^9$ M$_{\odot}$), 
low estimated metallically (Z$_{\star}\sim$ 1/20 Z$_{\odot}$), and extreme star-forming conditions
that produce strong nebular UV emission lines in the absence of any apparent outflows. 
Because these characteristics are more common at higher redshifts \citep[e.g.,][]{brinchmann08, stark14} 
and have been shown to be correlated with escaping ionizing radiation \citep[e.g.,][]{verhamme17}, 
we present SL2S0217 as a template to study the extreme conditions that may be responsible
for the reionization-era.  

In our analysis of the UV spectrum of SL2S0217 we found the Ly$\alpha$, interstellar absorption, 
and nebular emission features to be notable in the following ways:

1. The Ly$\alpha$ feature is observed purely in emission that is double-peaked and centered near the systemic velocity.
The dominant blue peak and small peak separation indicate little-to-no outflowing neutral gas along the line of sight.
Given the large Ly$\alpha$ EW, we measure an unexpectedly low Ly$\alpha$ escape fraction from the 
$L\substack{obs \\ Ly\alpha}$/$L\substack{int \\ Ly\alpha}$ ratio of 7\%.

2. Nearly all of the observed absorption features are saturated due to optically thick ionized gas. 
However, we were able to use the weakest Si~\ii\ and Fe~\ii\ lines to estimate a relative abundance of 
log(Si/Fe) = $-0.225\pm0.118$, or $44-77$\%\ solar, where this underabundance can be accounted for
by Si depletion onto dust grains. 
Similar to the Ly$\alpha$ emission, both the low- and high-ionization absorption features show 
profiles with nearly systemic velocity centers and small velocity ranges, indicating very little or no 
outflowing ionized gas along the sightline to the lensed galaxy.

3. Large equivalent widths are measured for the C~\iv\ \W\W1548,1550, He~\ii\ \W 1640, 
O~\iii] \W\W1661,1666, Si~\iii] \W\W1883,1892, and C~\iii] \W\W1907,1909 nebular emission lines. 
In the case of C~\iii] and C~\iv, these line strengths are similar to those observed for 
the extreme star-forming galaxies at higher redshifts \citep[$z>7$; e.g.,][]{stark15, stark16}.

With the exception of He~\ii, the relative emission line strengths can be reproduced by ionization from
a very high ionization, low metallicity starburst with binaries.
We rule out large contributions from AGN and shocks to the photoionization budget, 
suggesting that the emission features requiring the hardest radiation field likely result from 
extreme stellar populations that are beyond the capabilities of current models. 
We, therefore, argue that the combination of features observed for SL2S0217 are the 
result of an extreme episode of star formation in a young galaxy along a sightline that does
not cross any outflowing gas.

We have studied the extreme spectral features of SL2S0217, finding both its
absorption and emission features to be distinct from the general population of $z\sim2-3$ galaxies. 
In particular, our models show that the emission line strengths of SL2S0217 require very hard ionizing radiation,
and so it adds to the small population of galaxies with UV emission lines powered by the hard
ionizing stellar spectra that may be responsible for the reionization era.
Therefore, due to its optimal properties and leveraged by its lensed nature, SL2S0217 is
an ideal template to study the extreme conditions that are important for reionization 
and thought to be more common in the early Universe.

%----------------------------------------------------------------------------------------------------------------------
%----------------------------------------------------------------------------------------------------------------------

\acknowledgements
This work evolved over the course of the project thanks to many creative and insightful 
conversations with colleagues, especially John Chisholm, Anne Jaskot, and Claudia Scarlata.
We extend our gratitude to the Lorentz Center for useful discussions during the 
``Characterizing Galaxies with Spectroscopy with a view for JWST" workshop.
Additionally, we are grateful to the referee for valuable comments that improved and clarified the paper.  
We wish to recognize and acknowledge the very significant cultural role and reverence 
that the summit of Mauna Kea has always had within the indigenous Hawaiian community. 
We are most fortunate to have the opportunity to conduct observations from this mountain.

DAB and DKE are supported by the US National Science Foundation 
through the Faculty Early Career Development (CAREER) Program, grant AST-1255591.
Models in this work were computed using the computer cluster, Nemo, at 
The Leonard E Parker Center for Gravitation, Cosmology and Astrophysics is supported by NASA, 
the National Science Foundation, UW-Milwaukee College of Letters and Science, and UW-Milwaukee Graduate School. 
Any opinions, findings, and conclusions or recommendations expressed in this material are those of the author(s) 
and do not necessarily reflect the views of these organizations. 

\clearpage

\bibliography{SL2S_v9_arxiv}{}

\clearpage

\end{document}